\documentclass[aps,prd,a4paper,twocolumn]{revtex4}

\usepackage{graphicx}
\usepackage{bm}
\usepackage{amsfonts}
\usepackage{amsmath}

\newcommand{\muas}[0]{\hbox{\rm $\mu$as}}
\newcommand{\ve}[1]{\mbox{\boldmath $#1$}}

\begin{document}

\title{Light propagation in the gravitational field of $N$ arbitrarily moving bodies in the 1.5PN approximation for high-precision astrometry}

\author{Sven Zschocke}

\affiliation{Institute of Planetary Geodesy - Lohrmann Observatory,
Dresden Technical University, Helmholtzstrasse 10, D-01069 Dresden, Germany}

\begin{abstract}
High-precision astrometry on sub-micro-arcsecond level in angular resolution requires accurate determination of the trajectory  
of a light-signal from the celestial light source through the gravitational field of the Solar system toward the observer.  
In this investigation the light trajectory in the gravitational field of $N$ moving bodies is determined in the  
1.5 post-Newtonian approximation. In the approach presented two specific issues of particular importance are accounted for:   

({\bf 1}) According to the recommendations of International Astronomical Union, the metric of  
the Solar system is expressed in terms of intrinsic mass-multipoles and intrinsic spin-multipoles   
of the massive bodies, allowing for arbitrary shape, inner structure and rotational motion of the
massive bodies of the Solar system. 
 
({\bf 2}) The Solar system bodies move along arbitrary worldlines which can later be specified by Solar system ephemeris.  

The presented analytical solution for light trajectory is a primary requirement
for extremely high-precision astrometry on sub-micro-arcsecond level of accuracy
and associated massive computations in astrometric data reduction. 
An estimation of the numerical magnitude for time delay and light deflection of the leading multipoles is given.  

\end{abstract}


\maketitle


\section{Introduction}\label{Section0}  

A substantial advancement in astrometric measurements has been achieved by the astrometry mission {\it Hipparcos} 
(launch: 8 August 1989) of European Space Agency (ESA), which has reached an accuracy of a milli-arcsecond (${\rm mas}$) in determining the angular positions 
of about $10^5$ stars \cite{Hipparcos1,Hipparcos2}.
Meanwhile, the state-of-the-art angular observations have finally arrived at the level of a few micro-acrseconds ($\muas$) \cite{Kovalevsky,History_Astrometry}.
Especially, the stunning progress in astrometry has proceeded with the ESA mission {\it Gaia} \cite{GAIA} (launch: 19 December 2013) which  
aims at an all-sky survey of more than $10^9$ stars of our galaxy and targets angular accuracy of up to  
a few $\muas$ for bright stars in the final catalog scheduled for publication in 2022.   

In view of these advancements it becomes obvious that future astrometry is going to force into the exciting areas of sub-$\muas$ or even nano-arcsecond (nas)  
level of accuracy. To step up efforts toward sub-\muas-astrometry is of fundamental importance in astro\-nomy and astrophysics. 
For example, an accuracy of about $10\,{\rm nas}$ in angular resolution would allow for direct measurement of 
trigonometric parallaxes of stars belonging to galaxies of the Local Group which spans a diameter of about $10^7$ light-years, that means   
would enable to determine spatial distances of extra-galactic objects independently of dynamical models of the Universe. 
Moreover, also extremely high-precision tests of relativity, detection of dark-matter distributions within or outside of our galaxy, 
determination of stellar and galactic kinematics, and finally even the discovery of one-Earth-mass exoplanets in the habitable zone of nearby Sun-like stars  
would be possible by means of sub-$\muas$-astrometry. Recently, there are several mission proposals in this respect.  
For instance, the mission NEAT \cite{NEAT1,NEAT2} has been proposed to ESA which intends to reach a precision of about $50\,{\rm nas}$ 
in angular resolution for being able to detect Earth-like exoplanets surrounding stars in the stellar neighborhood of the Sun.
Further space missions like ASTROD \cite{Astrod1,Astrod2}, LATOR \cite{Lator1,Lator2}, ODYSSEY \cite{Odyssey}, SAGAS \cite{Sagas}, or TIPO \cite{TIPO}  
have been proposed to ESA which imply the determination of light trajectory through the Solar system on sub-\muas $\;$ or even at ${\rm nas}$ level of accuracy.    
Also earth-bound telescopes are under consideration which aim at angular resolutions of about $10\,{\rm nas}$ \cite{nas_telescopes}.  

But, although in view of the recent impressive achievements, the step from $\muas$-astrometry toward sub-$\muas$-level or even nas-level of accuracy in angular 
resolution will surely be a long-term goal in the astronomical science. This is because the envisaged advancement toward space-based ${\rm nas}$-astrometry 
implies many subtle effects and new kind of challenges which have not been encountered before:  
What kind of optical technology would allow for nas-astrometry?  
Is it technologically possible to measure the velocity of spacecraft (observer) with sufficient accuracy  
allowing for a precise determination of aberrational effects? How accurate do we have to determine the ephemeris of the  
Solar system bodies and could such precise ephemeris be provided? Is it possible to model accurate enough the influence of
interstellar medium on light propagation? How strong is the effect of gravitational waves on light propagation on nano-arcsecond level?  
How is it possible to account for the gravitational light deflection caused by massive bodies located outside the Solar system?  

Each of these and many other problems have to be clarified before ${\rm nas}$-astrometry becomes feasible. But certainly, the fundamental assignment  
in astrometry remains to trace a lightray observed in the Solar system back to the celestial light source. The importance of this fact has also been  
underlined recently by the ESA-Senior-Survey-Committee (SSC) in response of the selection of science themes for the L2 and L3 launch opportunities,  
where it has been stated that "{\it SSC recommends that proper modeling tools, most notably the availability of a General Relativistic framework able  
to model photon trajectories to the accuracy required should be given the proper attention to prove feasibility}" of high-precision astrometry \cite{SSC}.  
According to this, the primary effort in any astrometrical framework concerns the precise description of the light trajectory, that is to say the determination  
of the spatial coordinates of a light-signal as function of coordinate time, $\ve{x}\left(t\right)$, in some global coordinate system.  
Accordingly, the principal purpose of this investigation is the determination of the trajectory of a light-signal propagating through the Solar system.   
In the following four subsections it will be enlightened how one has to proceed in order to arrive that goal:  
(A) the theory of light propagation, (B) the post-Newtonian expansion, (C) the state-of-the-art, and (D) the primary objective of this investigation.  

\subsection{Theory of light propagation}\label{SS_1}  

The determination of spatial coordinates of the light\-ray takes the most simple form in the flat Minkowskian space-time and  
assuming a Cartesian coordinate system which covers the entire space, implying the metric tensor $\eta_{\alpha\beta}= {\rm diag}\left(-1,+1,+1,+1\right)$.  
Suppose the light-signal is emitted at some initial time $t_0$ by a light source  
located at some space-point $\ve{x}_0$, then the light trajectory is simply given by a straight line which is also called unperturbed light trajectory,  
\begin{eqnarray}
\ve{x}_{\rm N}\left(t\right) &=& \ve{x}_0 + c \left(t-t_0\right) \ve{\sigma}\,, 
\label{Introduction_1}
\end{eqnarray}

\noindent
where the unit-vector $\ve{\sigma}$ determines the direction of light-propagation and the sublabel "${\rm N}$" denotes Newtonian approximation.  

In general relativity the four-dimensional space-time in the presence of matter is curved, that means is described by a semi-Riemannian manifold with 
non-vanishing curvature tensor rather than a flat Minkowskian space-time, and a light trajectory is no longer a straight line but   
propagates along a so-called null geodesic, which generalizes the concept of a straight light trajectory.   
The four-coordinates $x^{\alpha}\left(\lambda\right)$ of a light trajectory depend on some affine curve-parameter $\lambda$, and are determined by the  
geodesic equation \cite{MTW,Brumberg1991},  
\begin{subequations}
\begin{equation}
\frac{d^2 x^{\alpha}\left(\lambda\right)}{d \lambda^2}
+ \Gamma^{\alpha}_{\mu\nu}\,\frac{d x^{\mu}\left(\lambda\right)}{d \lambda}\,
\frac{d x^{\nu}\left(\lambda\right)}{d \lambda} = 0\,,
\label{Geodetic_Equation}
\end{equation}
\begin{equation}
g_{\alpha\beta}\,\frac{d x^{\alpha}\left(\lambda\right)}{d \lambda}\,\frac{d x^{\beta}\left(\lambda\right)}{d \lambda} = 0\,,
\label{Geodetic_Equation1}
\end{equation}
\end{subequations}

\noindent
where (\ref{Geodetic_Equation}) represents the geodesic equation, while the isotropic condition (\ref{Geodetic_Equation1}) is an additional constraint 
for a null geodesic, a term which refers to the fact that the invariant line element vanishes, 
$ds^2 = d x_{\alpha}\left(\lambda\right)\,d x^{\alpha}\left(\lambda\right) = 0$, at any point along the light trajectory.  
The Christoffel symbols in (\ref{Geodetic_Equation}) are related to the metric of curved space-time as follows:  
\begin{eqnarray}
\Gamma^{\alpha}_{\mu\nu} &=& \frac{1}{2}\,g^{\alpha\beta}
\left(\frac{\partial g_{\beta\mu}}{\partial x^{\nu}}
+ \frac{\partial g_{\beta\nu}}{\partial x^{\mu}}
- \frac{\partial g_{\mu\nu}}{\partial x^{\beta}}\right),
\label{Christoffel_Symbols}
\end{eqnarray}

\noindent
where $g^{\alpha\beta}$ and $g_{\alpha\beta}$ are the contravariant and covariant components of the metric tensor, respectively, where the 
metric signature $\left(-,+,+,+\right)$.  
The geodesic equation (\ref{Geodetic_Equation}) represents a second-order differential equation, hence an unique solution implies the need of two initial values 
for the lightray:  
\begin{eqnarray}
x^{\alpha}\left(\lambda\right)\bigg|_{\lambda=\lambda_0} &&\,,
\label{Initial_Value_A}
\\
\nonumber\\
\frac{d x^{\alpha}\left(\lambda\right)}{d \lambda}\bigg|_{\lambda=\lambda_0} &&\,.  
\label{Initial_Value_B}
\end{eqnarray}

\noindent 
The equations in (\ref{Geodetic_Equation}) - (\ref{Geodetic_Equation1}) are valid in any reference system. 
But in practical astrometry one is necessarily enforced to specify the reference systems for concrete observational data. 
In line with the recommendations of the International Astronomical Union (IAU) \cite{IAU_Resolution1,IAU_Resolution2}, 
the Barycentric Celestial Reference System (BCRS) with coordinates $\left(ct,\ve{x}\right)$ is the standard global chart to be used in modern-day astrometry, 
where $t$ is the BCRS coordinate-time and $\ve{x}$ are Cartesian-like spatial coordinates from the barycenter of the Solar system to some field-point.  
Consequently, it becomes much preferable to exploit the freedom in the choice of scalar curve-parameter $\lambda$ and to rewrite the affinely parametrized 
geodesic equation (\ref{Geodetic_Equation})  
and the isotropic condition (\ref{Geodetic_Equation1}) in terms of BCRS coordinate-time \cite{MTW,Brumberg1991,Kopeikin_Efroimsky_Kaplan}:  
\begin{subequations}
\begin{equation}
\frac{d^2 x^{\alpha}\left(t\right)}{c^2 dt^2}  
+ \Gamma^{\alpha}_{\mu\nu} \frac{d x^{\mu}\left(t\right)}{c dt} 
\frac{d x^{\nu}\left(t\right)}{c dt}   
= \Gamma^{0}_{\mu\nu} \frac{d x^{\mu}\left(t\right)}{c dt} \frac{d x^{\nu}\left(t\right)}{c dt} \frac{d x^{\alpha}\left(t\right)}{c dt},   
\label{Geodetic_Equation2}
\end{equation}
\begin{equation}
g_{\alpha\beta}\,\frac{d x^{\alpha}\left(t\right)}{c dt}\,\frac{d x^{\beta}\left(t\right)}{c dt} = 0\,. 
\label{Geodetic_Equation3}
\end{equation}
\end{subequations}

\noindent 
The zeroth component in (\ref{Geodetic_Equation2}) does not carry any new information because it vanishes identically.   
In order to determine the solution of (\ref{Geodetic_Equation2}) it is advantageous to transform the initial conditions 
in (\ref{Initial_Value_A}) - (\ref{Initial_Value_B}) into initial-boundary conditions \cite{Brumberg1991}: 
\begin{eqnarray}
\ve{x}_0 &=& \ve{x}\left(t\right)\bigg|_{t=t_0}\,,
\label{Introduction_6}
\\
\nonumber\\
\ve{\sigma} &=& \frac{d \ve{x}\left(t\right)}{d ct}\bigg|_{t = - \infty}\,,
\label{Introduction_7}
\end{eqnarray}

\noindent
with (\ref{Introduction_6}) being the position of the light source at the moment $t_0$ of emission of the light-signal and (\ref{Introduction_7})  
being the unit-direction of the lightray at past-null infinity. Then, the exact solution of (\ref{Geodetic_Equation2}) for the light trajectory  
from the light source through the Solar system toward the observer can be written as follows,  
\begin{eqnarray}
\ve{x}\left(t\right) &=& \ve{x}_0 + c \left(t-t_0\right) \ve{\sigma} + \Delta \ve{x}\left(t,t_0\right)\,,
\label{Introduction_5}
\end{eqnarray}

\noindent
where the term $\Delta \ve{x}\left(t,t_0\right)$ denotes gravitational corrections to the unperturbed light trajectory (\ref{Introduction_1}).  

\subsection{Post-Newtonian expansion}\label{SS_2}  

The correction terms $\Delta \ve{x}\left(t,t_0\right)$ in Eq.~(\ref{Introduction_5}) are highly complicated expressions which cannot be determined exactly 
and one has to resort on approximation schemes. Such an approximation scheme is provided by the post-Newtonian expansion of the metric of Solar system, which 
represents an expansion in terms of inverse power of the speed of light, up to terms of the order ${\cal O} \left(c^{-5}\right)$ given by: 
\begin{equation}
g_{\alpha \beta} = \eta_{\alpha \beta} + h^{(2)}_{\alpha\beta} + h^{(3)}_{\alpha\beta} + h^{(4)}_{\alpha\beta} + {\cal O} \left(c^{-5}\right),
\label{post_Newtonian_metric_C}
\end{equation}

 \noindent
where $h^{(n)}_{\alpha\beta} = {\cal O} \left(c^{-n}\right)$ with $n=2,3,4$. The justification of such an expansion is based on the fact that the gravitational  
fields in the Solar system are weak, $\left(G\,M_A\right) / \left(c^2\,P_A\right) \ll 1$, as well as the velocities of the Solar system bodies are slow, 
$v_A / c \ll 1$, where $M_A$, $P_A$, and $v_A$ means mass, radius, and velocity, respectively, of some massive body $A$. For these reasons the 
post-Newtonian expansion is also called weak-field slow-motion expansion. As outlined in \cite{MTW,Poisson_Will,Kopeikin_Efroimsky_Kaplan,Expansion_2PN}, 
such an expansion is valid inside the near-zone of the Solar system, $\left|\ve{x}\right| \ll \lambda_{\rm gr}$, where $\lambda_{\rm gr} \sim 10^{17}\,{\rm meter}$ 
is a characteristic wavelength of gravitational waves emitted by the Solar system. The near-zone of Solar system is so
large that it still contains all Solar system bodies and even encompasses the nearest stars of the stellar neighborhood of the Sun. 

Inserting the expansion (\ref{post_Newtonian_metric_C}) into (\ref{Geodetic_Equation2}) yields the geodesic equations for lightrays up to terms of the order  
${\cal O}\left(c^{-5}\right)$. Accordingly, the expansion of the metric in (\ref{post_Newtonian_metric_C}) inherits a corresponding expansion of the lightray,  
that means the corrections to the unperturbed lightray can formally be written as follows: 
\begin{eqnarray}
\Delta \ve{x} &=& \Delta \ve{x}_{\rm 1PN} + \Delta \ve{x}_{\rm 1.5PN} + \Delta \ve{x}_{\rm 2PN} + {\cal O}\left(c^{-5}\right),  
\label{Light_Trajectory_2PN}
\end{eqnarray}

\noindent
where $\Delta \ve{x}_{\rm 1PN} = {\cal O}\left(c^{-2}\right)$ are 1PN corrections, $\Delta \ve{x}_{\rm 1.5PN} = {\cal O}\left(c^{-3}\right)$ are  
1.5PN corrections, and $\Delta \ve{x}_{\rm 2PN} = {\cal O}\left(c^{-4}\right)$ are 2PN corrections to the unperturbed lightray.  
In view of the fact that the post-Newtonian expansion of the metric (\ref{post_Newtonian_metric_C}) is only valid within the near-zone of the Solar system, the  
post-Newtonian expansion of the lightray (\ref{Light_Trajectory_2PN}) allows for near-zone astrometry, in particular for reduction of astrometric observations 
of all Solar system objects. The unique interpretation of astrometrical data of far objects, like stars or quasars, 
is the subject of far-zone astrometry and necessitates the determination of light trajectory outside the near-zone of the Solar system. That  
especially means, the light trajectory in the near-zone has to be aligned with the light trajectory in the far-zone by means of a so-called matching procedure 
as described in detail in \cite{KlionerKopeikin1992,Will_2003,Kopeikin_Efroimsky_Kaplan} which, however, will not be a topic of this investigation.

\subsection{State-of-the-art in the theory of light propagation}\label{SS_3}  

A brief survey about the present status in the theory of light propagation in the gravitational field of massive bodies has recently been presented  
\cite{Zschocke_1PN}. Here we will summarize and update that survey. In particular, we will restrict our review on those investigations where the explicit  
time-dependence of the photon's spatial coordinate, $\ve{x}\left(t\right)$, has been determined, a prerequisite for interpreting real astrometrical observations.

\subsubsection{Monopoles at rest}\label{SSS_1}  

The case of light propagation in the Schwarzschild metric, i.e. in the gravitational field of one spherically symmetric massive body at rest, 
\begin{eqnarray}
\ve{x}_A\left(t\right) &=& \ve{x}_A\,, 
\label{worldline_introduction_A}
\end{eqnarray}

\noindent
where $\ve{x}_A = {\rm const}$ is the constant position of the body, is the  
most simple case and has been determined long time ago in 1PN approximation, e.g. \cite{Brumberg1987,Brumberg1991,KlionerKopeikin1992,Klioner2003a,Zschocke_1PN}. 
The solution for the light trajectory is given by Eq.~(\ref{1PN_Solution_A}).  
Besides its simplicity, the determination of the photon's spatial coordinate in the Schwarzschild-field is the initial point in the theory of 
light propagation in astrometry.

\subsubsection{Monopoles in motion}\label{SSS_2}

In reality, the bodies $A=1,...,N$ of the Solar system move along their time-like worldlines $\ve{x}_A\left(t\right)$ and for todays extremely high-precision  
in astrometric measurements the gravitational field of some Solar system body can not any longer be treated as static and spherically symmetric.  
In a first approximation, the motion of one massive body $A$ can be considered as translational motion with constant velocity $\ve{v}_A$:  
\begin{eqnarray}
\ve{x}_A\left(t\right) &=& \ve{x}_A + \ve{v}_A \left(t - t_A\right),  
\label{worldline_introduction_B}
\end{eqnarray}

\noindent
where $\ve{x}_A = \ve{x}_A\left(t_A\right)$ and $\ve{v}_A = \ve{v}_A\left(t_A\right)$ are the spatial position and velocity of the body $A$  
at some initial time-moment $t_A$. The light-trajectory in the field of one massive body in translational motion has completely been solved in 1PN approximation  
in \cite{Klioner1989}. This solution has later been rederived by means of a suitable Lorentz transformation \cite{Klioner2003b}. Following a suggestion  
in \cite{Hellings1986}, in the investigation \cite{KlionerKopeikin1992} it has been shown that the free parameter $t_A$ in Eq.~(\ref{worldline_introduction_B}) 
should be chosen as the time-moment of closest approach (given by Eqs.~(\ref{time_of_closest_approach_t_0_moving_body}) and 
(\ref{time_of_closest_approach_t_1_moving_body})) between the massive body $A$ and the photon in order to minimize  
the residual effects caused by the approximation of the real motion by a translational motion of the massive body.  
With the aid of advanced integration methods, originally introduced in \cite{Kopeikin1997} and further developed in \cite{KopeikinSchaefer1999_Gwinn_Eubanks},  
a rigorous solution for the trajectory of a light-signal through the gravitational field of an arbitrarily moving body has thoroughly been solved  
in \cite{KopeikinSchaefer1999} in the first post-Minkowskian approximation. The first post-Minkowskian approximation takes into account all terms proportional  
to the gravitational constant and especially all terms to any power in $v_A/c$, hence  
the body can even be in ultra-relativistic motion and, therefore, the post-Minkowskian approximation is often called {\it weak-field approximation} opposite 
to the post-Newtonian approximation which is called {\it weak-field slow-motion approximation}. Comparing the solution in \cite{KopeikinSchaefer1999} with  
\cite{Klioner1989,Klioner2003b}, it has been demonstrated in \cite{KlionerPeip2003} that the simpler solution for the light-trajectory in the field of 
a uniformly moving body is actually sufficient for high-precision astrometry on sub-\muas-level provided the free parameter $t_A$ is chosen either as  
time-moment of closest approach or as retarded time-moment (given by Eq.~(\ref{Retarded_Time})) between the photon and the position of the massive body.  
All these results agree with our investigation in \cite{Zschocke_1PN} for the case of bodies in slow-motion.  

\subsubsection{Spin-dipoles at rest}\label{SSS_3}

The light trajectory in the gravitational field of one body at rest having spin-dipole $\ve{S}_A = {\rm const}$ has first been solved 
in \cite{Klioner1991} and later confirmed in \cite{KlionerKopeikin1992}. 
The magnitude of light deflection due to the rotational motion of Solar system bodies has been determined in \cite{Klioner1991,Klioner2003a} and 
turns out to be significant for astrometry on sub-\muas-level of accuracy.

\subsubsection{Spin-dipoles in motion}\label{SSS_4}

In \cite{Deng_2015} an explicit solution for the light-trajectory in the field of $N$ uniformly moving bodies with intrinsic spin 
has been obtained.  
A comprehensive solution in 1PM approximation for the light-trajectory in the field of 
$N$ arbitrarily moving bodies with individual spin-structure has been derived in \cite{KopeikinMashhoon2002} using the already mentioned 
advanced integration methods originally developed in \cite{Kopeikin1997,KopeikinSchaefer1999_Gwinn_Eubanks}.

\subsubsection{Mass-quadrupoles at rest}\label{SSS_5}

The solution for the light-trajectory in the field of mass-quadrupoles at rest in 1PN approximation was given in \cite{Klioner1991} and later 
in \cite{KlionerKopeikin1992,Klioner2003a,Zschocke_1PN}. Especially, in \cite{Klioner1991} the magnitude of light deflection caused by the mass-quadrupole 
structure of Solar system bodies has been determined, where it was figured out that astrometry on \muas-level of accuracy is able to detect this  
light deflection effect. In fact, the light deflection due to the quadrupole-structure of Jupiter is presently under investigation  
by the ESA astrometry-mission Gaia \cite{GAIA}.

\subsubsection{Mass-quadrupoles in motion}\label{SSS_6}  

The light trajectory in the field of $N$ arbitrarily slowly-moving bodies with quadrupole structure has been determined in \cite{Zschocke_1PN}. Recently,  
the light-trajectory in the field of $N$ uniformly moving bodies with mass-quadrupole structure has also been obtained in \cite{Deng_2015} by integrating the  
geodesic equations for the lightray. Another interesting approach has been found in \cite{Hees_Bertone_Poncin_Lafitte_2014a},  
which is based on the Time Transfer Function (TTF) which avoids to solve the geodesic equations and hence circumvents some of its involved peculiarities.

\subsubsection{Higher mass-multipoles and spin-multipoles at rest}\label{SSS_7}

A fruitful and systematic approach which allows to integrate analytically the geodesic equations in 1.5 approximation in the field of one body at rest having full 
time-independent mass-multipoles $M^A_L$ and spin-multipoles $S^A_L$ to any order in the multi-index $L$ has been introduced in \cite{Kopeikin1997}.  

The advanced integration method in \cite{Kopeikin1997} has been developed further in \cite{KopeikinSchaefer1999_Gwinn_Eubanks} for the case of time-dependent  
mass-multipoles $M^A_L\left(t\right)$ and spin-multipoles $S^A_L\left(t\right)$ in 1PM approximation. Using this advanced approach the analytical solution in  
1PM approximation for the light-trajectory in the field of one massive body at rest with the full set of time-dependent multipoles has been determined in  
\cite{KopeikinKorobkovPolnarev2006,KopeikinKorobkov2005}. One comment should be in order at this stage. Namely, it is of course possible to interpret 
the Solar system just as one global massive body $A$ which consists of many individual small massive bodies. But then the solution 
in \cite{KopeikinKorobkovPolnarev2006,KopeikinKorobkov2005} has to be interpreted as still expressed in terms of  
global multipoles $m_L\left(t\right)$ and $s_L\left(t\right)$ which characterize the entire multipole structure of the Solar system as a whole. 
However, physically meaningful multipoles can only be defined in the local reference system of each individual massive body.  
This important issue will later be further considered in some more detail.   

Another approach is based on the solution for the TTF and its spatial derivative. A corresponding multipole decomposition of the TTF has been applied 
in \cite{Poncin_Lafitte_Teyssandier_2008} in order to determine the coordinate travel time  
and the light deflection of a lightray in the gravitational field of one axisymmetric body at rest expressed in terms of mass-multipoles $M_L^A$.

\subsubsection{Higher mass-multipoles in uniform motion}\label{SSS_8}

In \cite{Hees_Bertone_Poncin_Lafitte_2014a} the TTF approach in 1PM approximation has been applied for the case of light propagation in the field of 
one axisymmetric body in uniform motion. Especially, an expression for the TTF and its spatial derivative is obtained for this case, which allows to determine 
astrometric observables like the coordinate travel time of the lightray, the direction of an incident lightray, and the gauge-invariant angle between the 
direction of two incoming photons. A similar investigation has been done in \cite{Soffel_Han}, where the TTF approach has been used in order to determine  
the coordinate travel time of a lightray in the field of one slowly and uniformly moving extended body with full mass-multipole and spin-multipole structure.

\subsubsection{2PN light propagation in the field of monopoles}\label{SSS_9}  

Light propagation 2PN approximation is not on the scope of the presented investigation, but for reasons of completeness some results obtained in 2PN 
approximation will briefly be mentioned, not only because of its relevance for future high-precision astrometry on sub-\muas-level of accuracy but also for 
its importance in todays high-precision astrometry on \muas-level.  

An important progress has been made in \cite{Brumberg1987,Brumberg1991}, where an analytical solution of the light-trajectory in 2PN approximation has been  
determined with explicit time-dependence of the photon's spatial coordinates by solving the null geodesic equations. This solution has later been confirmed  
by several progressing and ongoing investigations \cite{KlionerKopeikin1992,Brugmann,Klioner_Zschocke,Deng_Xie,Deng_2015}, and has also been determined 
in this investigation, see Eqs.~(\ref{2PN_Solution_Brumberg_2}) - (\ref{Vectorial_Function_B2}).  
Furthermore, in \cite{Deng_Xie} the time-derivative of the light trajectory in the field of two pointlike bodies at rest has been obtained, 
allowing to determine the light deflection in such a system. An important new result of this investigation is the fact that the 2PN two-body effect 
in the Solar system is less than $0.1\,{\rm nas}$ which considerably simplifies future analytical investigations for high-precision astrometry 
on sub-\muas-level of accuracy.  

In \cite{Poncin_Lafitte_Teyssandier_2004,Teyssandier,Hees_Bertone_Poncin_Lafitte_2014b} the general formalism of how to determine the 
TTF and its derivatives has been extended up to the second post-Newtonian (2PN) and second post-Minkowskian (2PM) order, that means 
including all terms to order ${\cal O}\left(G^2\right)$.  
The formalism has finally been specified for the case of light propagation in the gravitational field of one spherically symmetric body at rest 
where the 2PM and 2PN approximations become identically. Especially, explicit expressions for the coordinate travel time of lightray,  
for the direction of the lightray, and for the angular separation between two incident lightrays have been obtained.    

Finally, we also mention another approach which is based on the eikonal concept \cite{Ashby_Bertotti}, where the light trajectory in 2PN approximation 
in the field of one spherically symmetric body at rest has also been derived. The results of this work completely agree with \cite{Klioner_Zschocke}.

\subsection{Primary objective of this investigation}\label{SS_4} 

According to the survey given above about the present situation in the theory of light propagation, thus far there is no analytical solution available  
for the light trajectory in the field of arbitrarily moving extended bodies in 1.5PN approximation which, however, is of decisive importance in future  
high-precision astrometric measurements on sub-\muas-level of accuracy and its foreseen involved massive computations, see also \cite{Zschocke_1PN}. 
In respect thereof, two important aspects must carefully be treated: 

({\bf 1}) The metric perturbations in the exterior of the massive bodies can be decomposed in terms of global mass-multipoles $m_L$ 
and global spin-multipoles $s_L$ \cite{Thorne,Blanchet_Damour1,Blanchet_Damour2,Multipole_Damour_2}:  
\begin{eqnarray}   
h^{\left(n\right)}_{\alpha\beta} &=& h^{\left(n\right)}_{\alpha\beta} \left(m_L,s_L\right), \quad n=2,3,...\,.  
\end{eqnarray}   
 
\noindent
These global mass and spin multipoles describe the gravitational field of the Solar system as a whole.  
However, from the theory of relativistic reference systems it is clear that physically meaningful
multipole moments of a massive body $A$ have to be defined in the body's local reference system $\left(c T_A, \ve{X}_A\right)$ 
tied to that body under consideration. Such multipoles are called intrinsic mass-multipoles $M_L^A$ and intrinsic spin-multipoles $S_L^A$.  
Then the question arises about how to express the global BCRS metric in terms of such intrinsic multipoles, 
that is to say how to determine the global metric perturbations: 
\begin{eqnarray}
h^{\left(n\right)}_{\alpha\beta} &=& h^{\left(n\right)}_{\alpha\beta} \left(M^A_L,S^A_L\right), \quad n=2,3,...\,.
\label{Global_Metric_B} 
\end{eqnarray}

\noindent
Such a framework has been elaborated by the approach of {\it Damour-Soffel-Xu} (DSX) \cite{DSX1,DSX2,DSX3,DSX4} and within the {\it Brumberg-Kopeikin} (BK)  
formalisms \cite{Brumberg1991,BK1,Reference_System1,BK2,BK3}, both of which became a part of the IAU resolutions \cite{IAU_Resolution1,IAU_Resolution2}.  

({\bf 2}) The second issue concerns the motion of the massive Solar system bodies. While in first approximation these bodies orbit the barycenter  
of the Solar system along ellipse-shaped trajectories, in reality their orbital motion $\ve{x}_A\left(t\right)$ is highly complicated due to the 
mutual interactions among these bodies. The worldlines of all massive bodies can be concretized by Solar system ephemeris \cite{JPL} at any stage of the 
calculations.   
One might prefer to series expand these worldlines as follows:
\begin{eqnarray}
\ve{x}_A\left(t\right) &=& \ve{x}_A + \frac{\ve{v}_A}{1!}\left(t - t_A\right)
+ \frac{\ve{a}_A}{2!}\left(t-t_A\right)^2 + {\cal O}\left(\dot{a}_A\right),
\nonumber\\
\label{worldline_introduction}
\end{eqnarray}

\noindent
where $\ve{x}_A = \ve{x}_A\left(t_A\right)$, $\ve{v}_A = \ve{v}_A\left(t_A\right)$ and $\ve{a}_A = \ve{a}_A\left(t_A\right)$
are the position, velocity and acceleration of body $A$ at some time-moment $t_A$. However, such an approach is problematic mainly for two reasons:
\begin{enumerate}
\item[(i)] all terms of the infinite series expansion (\ref{worldline_introduction}) contribute on 1PN or 1.5PN level, because 
the expansion in (\ref{worldline_introduction}) is not performed with respect to the inverse powers of the speed of light. 
\item[(ii)] the time-moment $t_A$ remains an open parameter as long as no additional arguments are introduced,  
which would uniquely allow to identify that parameter with the time of closest approach or with the retarded time.  
\end{enumerate}

\noindent
These both aspects ({\bf 1}) and ({\bf 2}) enforce to determine the light trajectory $\ve{x}\left(t\right)$ of a light-signal from 
the celestial light source toward the observer as function of intrinsic multipoles $M^A_L$ and $S^A_L$  
as well as function of the arbitrary worldlines $\ve{x}_A\left(t\right)$ of these massive Solar system bodies.  

In a previous investigation \cite{Zschocke_1PN} a solution for the light trajectory in 1PN approximation in the gravitational field
of $N$ massive bodies in arbitrary motion and expressed in terms of their intrinsic multipoles has been obtained:  
\begin{eqnarray}
\ve{x}\left(t\right) &=& \ve{x}_0 + c \left(t-t_0\right) \ve{\sigma} + \Delta\ve{x}_{\rm 1PN} + {\cal O}\left(c^{-3}\right). 
\label{Introduction_A}
\end{eqnarray}

\noindent 
However, as outlined in more detail in \cite{Zschocke_1PN}, such 1PN solution is not sufficient for astrometry on sub-\muas-level of accuracy. 
For instance, the rotational motion of the massive bodies cannot be taken into account in 1PN approximation.  
However, the impact of the spin-dipole structure of the massive bodies on light deflection 
amounts to be about $0.7$ \muas$\,$, $0.2$ \muas$\,$, and $0.04$ \muas $\,$ for a grazing lightray at Sun, Jupiter, and Saturn,  
respectively \cite{Klioner2003a,Klioner1991}. Moreover, also higher spin-multipoles have a significant impact on sub-\muas-level  
\cite{Jan-Meichsner_Diploma_Thesis,Zschocke_1PN}. Furthermore, in 1PN approximation there are no terms proportional to $\displaystyle \frac{v_A}{c}\,M_{ab}$ 
where $M_{ab}$ is the mass-quadrupole term. Already a straightforward estimate reveals that such terms become relevant on sub-\muas-level of 
accuracy \cite{Zschocke_1PN}, see also Table \ref{Table3}. 
In order to scrutinize the impact of such terms one is necessarily enforced to determine the 1.5PN solution for the light trajectory.  

In view of these facts, the primary goal of this investigation is to determine a solution for the light trajectory in 1.5PN 
approximation, which includes all terms up to the order ${\cal O}\left(c^{-4}\right)$, where   
both of the important aspects ({\bf 1}) and $({\bf 2})$ addressed above are fully taken into account:  
\begin{eqnarray}
\ve{x}\left(t\right) \!\!&=& \!\! \ve{x}_0 + c \left(t-t_0\right) \ve{\sigma} + \Delta\ve{x}_{\rm 1PN} + \Delta\ve{x}_{\rm 1.5PN} + {\cal O}\left(c^{-4}\right).
\nonumber\\
\label{Introduction_B}
\end{eqnarray}

\noindent  
Especially, the massive bodies of the Solar system are allowed to move along  
arbitrary worldlines $\ve{x}_A\left(t\right)$ and they are having arbitrary shape and inner structure and rotational motion, given in terms 
of time-dependent intrinsic mass-multipoles $M^A_L\left(t\right)$ and spin-multipoles $S^A_L\left(t\right)$, in accordance with the IAU recommendations  
\cite{IAU_Resolution1,IAU_Resolution2} and the theory of relativistic reference systems \cite{DSX1,DSX2,DSX3,DSX4,Brumberg1991,BK1,Reference_System1,BK2,BK3}.  
The given solution for the light trajectory is considered as a further step towards
a consistent model of general-relativistic theory of light propagation in the gravitational field of the Solar system, which finally aims
at accuracies on ${\rm sub}$-\muas-level and even on nas-level.
 
The article is organized as follows: 
In section \ref{Section2} the geodesic equation in 1.5PN approximation is considered. 
A compendium of the DSX framework is presented in section \ref{Section3}.  
The transformation of geodesic equation in terms of new variables, which are more efficient than the standard parametrization, is given in section \ref{Section4}.  
The first and second integration of geodesic equation is determined in section \ref{First_Integration} and section \ref{Second_Integration}, respectively.  
The important case of light-propagation in the gravitational field of moving spin-dipoles is investigated in section \ref{Section5}. 
Finally, the expressions for the observables of time delay and light deflection are obtained in section \ref{Observable_Effects_Time_Delay} and 
\ref{Observable_Effects_Light_Deflection}. Especially, numerical values for the impact of the leading mass-multipoles and spin-multipoles on 
time delay and light deflection are given in Table~\ref{Table2} and Table~\ref{Table3}, respectively. 
A summary and outlook can be found in section \ref{Summary_Outlook}.  
The used notations and conventions and further details and several checks of the calculations are shifted into appendix.

\section{Geodesic equation in 1.5PN approximation}\label{Section2}

The Solar system is composed of $N$ arbitrarily shaped, rotating and deformable massive bodies which move under the influence of their mutual gravitational 
interaction among their common barycenter. It is clear, that the  
metric of such a highly complicated $N$-body system is not known in its exact form and can only be determined within an approximative scheme.  
In view of the weak gravitational fields and slow motions of the bodies, the metric tensor of the Solar system in the BCRS coordinate system 
$x^{\mu} = \left(ct,\ve{x}\right)$ can be expanded in terms of inverse powers in the light-velocity, called post-Newtonian expansion \cite{MTW}:  
\begin{eqnarray}
g_{\alpha\beta}\left(t,\ve{x}\right) &=& \eta_{\alpha\beta} 
+ h^{(2)}_{\alpha\beta}\left(t,\ve{x}\right) + h^{(3)}_{\alpha\beta}\left(t,\ve{x}\right)  
+ {\cal O} \left(c^{-4}\right),
\nonumber\\
\label{metric_perturbation_pN}
\end{eqnarray}

\noindent
where $\eta_{\alpha\beta}$ is the metric tensor of flat Minkowski space-time and the metric perturbations are of the order  
$h^{(2)}_{\alpha\beta} = {\cal O} \left(c^{-2}\right)$ and $h^{(3)}_{\alpha\beta} = {\cal O} \left(c^{-3}\right)$, cf. Eq.~(\ref{post_Newtonian_metric_C}).  
Inserting (\ref{metric_perturbation_pN}) into (\ref{Geodetic_Equation2}) yields  
the geodesic equation in 1.5PN approximation, which in terms of global coordinate time reads 
\cite{Brumberg1991,KlionerPeip2003,KlionerKopeikin1992,KopeikinSchaefer1999_Gwinn_Eubanks,KopeikinKorobkovPolnarev2006,KopeikinKorobkov2005}:  
\begin{eqnarray}
\frac{\ddot{x}^i \left(t\right)}{c^2} &=& \frac{1}{2}\,h_{00,i}^{(2)}  
- h_{00,j}^{(2)} \frac{\dot{x}^i\left(t\right)}{c}\frac{\dot{x}^j\left(t\right)}{c}
- h_{ij,k}^{(2)}\,\frac{\dot{x}^j\left(t\right)}{c}\frac{\dot{x}^k\left(t\right)}{c}
\nonumber\\
\nonumber\\
&& \hspace{-1.0cm} + \frac{1}{2}\,h_{jk,i}^{(2)}\,\frac{\dot{x}^j\left(t\right)}{c}\frac{\dot{x}^k\left(t\right)}{c}
- \frac{1}{2}\,h_{00,0}^{(2)}\,\frac{\dot{x}^i\left(t\right)}{c}
- h_{ij,0}^{(2)} \frac{\dot{x}^j\left(t\right)}{c}
\nonumber\\
\nonumber\\
&& \hspace{-1.0cm} + \frac{1}{2}\,h_{jk,0}^{(2)} \frac{\dot{x}^i\left(t\right)}{c}
\frac{\dot{x}^j\left(t\right)}{c}\frac{\dot{x}^k\left(t\right)}{c}
- h_{0i,j}^{(3)} \frac{\dot{x}^j\left(t\right)}{c}
+ h_{0j,i}^{(3)} \frac{\dot{x}^j\left(t\right)}{c}
\nonumber\\
\nonumber\\
&& \hspace{-1.0cm} - h_{0j,k}^{(3)}\frac{\dot{x}^i\left(t\right)}{c}\frac{\dot{x}^j\left(t\right)}{c}\frac{\dot{x}^k\left(t\right)}{c}  
+ {\cal O}\left(c^{-4}\right),  
\label{geodesic_equation_1}
\end{eqnarray}

\noindent
where a dot means total time-derivative. Note that the constraint in (\ref{Geodetic_Equation3}) results 
in $\displaystyle \frac{\dot{\ve{x}}\left(t\right) \cdot \dot{\ve{x}}\left(t\right)}{c^2} = 1 + {\cal O}\left(c^{-2}\right)$, hence   
will not change the form of geodesic equation in 1.5PN approximation in (\ref{geodesic_equation_1}). 
 
In (\ref{geodesic_equation_1}) we have taken into account that in general $h_{0i}^{(2)}=h_{00}^{(3)}=h_{ij}^{(3)} = 0$  
and $h_{0i,0}^{(3)} = {\cal O}\left(c^{-4}\right)$. The metric perturbations in (\ref{metric_perturbation_pN}) are functions of the field-points $(t,\ve{x})$, 
while in the geodesic equation (\ref{geodesic_equation_1}) the metric perturbations are of relevance at the coordinates of the photon $\ve{x}\left(t\right)$.  
Consequently, the derivatives in (\ref{geodesic_equation_1}) are taken along the lightray:
\begin{eqnarray}
h_{\alpha \beta, \mu}^{(n)} &=& \frac{\partial h_{\alpha \beta}^{(n)}\left(t,\ve{x}\right)}{\partial x^{\mu}}
\Bigg|_{\ve{x}=\ve{x}\mbox{\normalsize $\left(t\right)$}}\,,\quad n=2,3\,.  
\label{geodesic_equation_3}
\end{eqnarray}

\noindent
In order to find an unique solution of the geodesic equation in (\ref{geodesic_equation_1}), so-called mixed
initial-boundary conditions can be imposed, which have extensively been used in the literature, e.g.
\cite{Brumberg1991,KlionerKopeikin1992,Klioner_Zschocke,Kopeikin1997,KopeikinSchaefer1999_Gwinn_Eubanks,Brumberg1987,KopeikinKorobkovPolnarev2006}: 
\begin{eqnarray}
\ve{x}_0 &=& \ve{x}\left(t_0\right),
\label{Initial_Boundary_Condition_1}
\\
\nonumber\\
\ve{\sigma} &=& \lim_{t \rightarrow - \infty}\, \frac{\dot{\ve{x}}\left(t\right)}{c}\,.
\label{Initial_Boundary_Condition_2}
\end{eqnarray}

\noindent
The first condition (\ref{Initial_Boundary_Condition_1}) defines the spatial coordinates of the photon at the moment
$t_0$ of emission of light. The second condition (\ref{Initial_Boundary_Condition_2}) defines the unit-direction
$\left(\ve{\sigma}\cdot\ve{\sigma} = 1\right)$ of the lightray at past null infinity, that means the
unit-tangent vector along the light path in the infinite past hence at infinite spatial distance from the origin of the global coordinate system.

In the flat space-time there is no gravitational field, $h_{\alpha \beta}^{(n)} = 0$,  
hence the geodesic equation (\ref{geodesic_equation_1}) simplifies to the form $\ddot{\ve{x}}\left(t\right) = 0$, having the solution  
\begin{eqnarray}
\ve{x}\left(t\right) &=& \ve{x}_0 + c \left(t - t_0\right) \ve{\sigma} + {\cal O}\left(c^{-2}\right), 
\label{Unperturbed_Light_Trajectory}
\end{eqnarray}
 
\noindent
which is nothing else than just the unperturbed light trajectory in Eq.~(\ref{Introduction_1}).

The exact light trajectory $\ve{x}\left(t\right)$ in (\ref{Introduction_5}) deviates from the Newtonian approximation in (\ref{Unperturbed_Light_Trajectory}) 
by terms of the order ${\cal O}\left(c^{-2}\right)$, that means $\ve{x}\left(t\right)=\ve{x}_{\rm N}\left(t\right) + {\cal O}\left(c^{-2}\right)$.  
Accordingly, in (\ref{geodesic_equation_1}) we may replace $\dot{\ve{x}}\left(t\right)$ by its Newtonian
approximation, $\dot{\ve{x}}_{\rm N} = c\,\ve{\sigma}$, and (\ref{geodesic_equation_1}) simplifies as follows:  
\begin{eqnarray}
\frac{\ddot{x}^i \left(t\right)}{c^2} &=&
\frac{1}{2}\,h_{00,i}^{(2)}  
- h_{00,j}^{(2)}\,\sigma^i\,\sigma^j - h_{ij,k}^{(2)}\,\sigma^j\,\sigma^k
\nonumber\\
\nonumber\\
&& \hspace{-1.5cm} + \frac{1}{2}\,h_{jk,i}^{(2)}\,\sigma^j\,\sigma^k
- \frac{1}{2}\,h_{00,0}^{(2)}\,\sigma^i\, 
- h_{ij,0}^{(2)}\,\sigma^j
+ \frac{1}{2}\,h_{jk,0}^{(2)}\,\sigma^i\,\sigma^j\,\sigma^k  
\nonumber\\
\nonumber\\
&& \hspace{-1.5cm} - h_{0i,j}^{(3)}\,\sigma^j
+ h_{0j,i}^{(3)}\,\sigma^j
- h_{0j,k}^{(3)}\,\sigma^i\,\sigma^j\,\sigma^k
+ {\cal O}\left(c^{-4}\right),  
\label{geodesic_equation_5}
\end{eqnarray}

\noindent
which agrees with Eq.~(3) in \cite{KopeikinSchaefer1999_Gwinn_Eubanks}; recall $h^{(3)}_{0i,0} = {\cal O}\left(c^{-4}\right)$.  
Furthermore, in 1.5PN approximation the metric perturbations in (\ref{geodesic_equation_5}) can be taken at the  
spatial coordinates of the unperturbed lightray. That means, in (\ref{geodesic_equation_5}) one has first to perform the 
differentiations with respect to BCRS coordinates $x^{\mu}=\left(ct,\ve{x}\right)$ and afterwards to insert the unperturbed lightray: 
\begin{eqnarray}
h_{\alpha \beta, \mu}^{(n)} &=& \frac{\partial h_{\alpha \beta}^{(n)}\left(t,\ve{x}\right)}{\partial x^{\mu}}
\Bigg|_{\ve{x}=\ve{x}_{\rm N}\mbox{\normalsize $\left(t\right)$}}\,.
\label{transformed_geodesic_equation_15}
\end{eqnarray}

\noindent
In this investigation we will determine the solution of the geodesic equation (\ref{geodesic_equation_5}) in 1.5PN approximation, which  
can formally be written as follows (cf. Eq.~(\ref{Introduction_B})): 
\begin{eqnarray}
\ve{x}\left(t\right) &\!\!=\!\!& \ve{x}_0 + c\,\ve{\sigma}\left(t - t_0\right) 
\nonumber\\
\nonumber\\
&& \hspace{-1.0cm} + \Delta \ve{x}_{\rm 1PN}\left(t,t_0\right) + \Delta \ve{x}_{\rm 1.5PN} \left(t,t_0\right) + {\cal O}\left(c^{-4}\right).  
\label{light_trajectory_B}
\end{eqnarray}

\noindent
The 1PN corrections $\Delta \ve{x}_{\rm 1PN}\left(t,t_0\right)$ in Eq.~(\ref{light_trajectory_B}) are terms of the order ${\cal O}\left(c^{-2}\right)$ and 
have already been determined in our recent analysis \cite{Zschocke_1PN}. Here, the primary goal is the determination of the 
1.5PN corrections $\Delta \ve{x}_{\rm 1.5PN}\left(t,t_0\right)$ in Eq.~(\ref{light_trajectory_B}) which are terms of the order ${\cal O}\left(c^{-3}\right)$.

\section{Compendium of DSX framework}\label{Section3}

The DSX framework represents a well-established formalism in the general-relativistic celestial mechanics of a $N$-body system of arbitrarily shaped,  
rotating and deformable bodies, and has been introduced and thoroughly formulated in \cite{DSX1,DSX2,DSX3,DSX4}.  
The original intension of DSX was the description of the dynamics of $N$ massive bodies, that is the equations of motion in celestial mechanics for  
$N$ extended bodies under the influence of their mutual gravitational interaction.  

The basic assumption is to introduce $N+1$ reference systems: one global chart (BCRS) with coordinates $x^{\mu} = \left(ct,\ve{x}\right)$  
having its origin of the spatial axes at the barycenter of the Solar system, and $N$ local charts with coordinates $X_A^{\alpha} = \left(cT_A,\ve{X}_A\right)$,  
one for each individual body $A=1,...,N$ and having their origins at the barycenter of these massive bodies and co-moving with them.  
The local coordinate systems are tied to each individual massive body and are defined very similar to the 
Geocentric Celestial Reference System (GCRS) which is in use for the Earth and, therefore, they are called GCRS-like reference systems.  
A central result of the DSX approach is the form of the global metric $g_{\mu\nu}$ of BCRS and the form of the local metric $G_{\alpha\beta}^A$ for 
each GCRS-like system, and the coordinate transformation among all these reference systems $\left(ct,\ve{x}\right) \leftrightarrow \left(cT_A,\ve{X}_A\right)$. 
Another central achievement in the DSX formalism is the decomposition of the global metric in terms of intrinsic mass-multipoles $M^A_L$ and intrinsic 
spin-multipoles $S^A_L$.  
In this section we will present a compendium of the DSX theory, which has become a basic part of IAU resolution B1.3 (2000) \cite{IAU_Resolution1} 
and which are of upmost relevance for our own considerations aiming at applications of the DSX approach in the astrometrical science.

\subsection{BCRS}\label{BCRS} 

The harmonic BCRS coordinates are denoted by $x^{\mu}=\left(ct,x^i\right)$, where $t={\rm TCB}$ is the BCRS coordinate time;  
about a practical synchronization of a set of clocks distributed somewhere in the Solar system we refer to \cite{Synchronization1}.  
The origin of the spatial axes of BCRS is located at the barycenter of the Solar system and cover the entire three-dimensional space and can therefore  
be used to model light trajectories from distant celestial objects to the observer. The IAU Resolution B2 (2006) \cite{IAU_Resolution2} recommends  
the spatial axes of BCRS to be oriented according to the spatial axes of the International Celestial Reference System (ICRS) \cite{ICRS}.
Furthermore, according to IAU resolution B1.3 (2000) \cite{IAU_Resolution1} the Solar system is assumed to be isolated and the space-time  
is asymptotically flat, that means the BCRS metric $g_{\mu\nu}\left(t,\ve{x}\right)$ at infinity reads:  
\begin{eqnarray}
\lim_{r \rightarrow \infty}\,
g_{\mu\nu}\left(t,\ve{x}\right) &=& \eta_{\mu\nu}\,, 
\label{boundary_condition_global_metric}
\end{eqnarray}

\noindent
where $r=\left|\ve{x}\right|$. The BCRS is completely characterized by the form of its metric tensor which, however, is not known in its exact form. 
According to the geodesic equation in 1.5PN approximation (\ref{geodesic_equation_5}), for our intentions  
the metric is required to be known only up to terms of the order ${\cal O}\left(c^{-4}\right)$, which are  
given by \cite{IAU_Resolution1}:  
\begin{eqnarray}
g_{00}\left(t,\ve{x}\right) &=& - 1 + \frac{2\,w\left(t,\ve{x}\right)}{c^2} + {\cal O}\left(c^{-4}\right),  
\label{BCRS_1}
\\
\nonumber\\
g_{0i}\left(t,\ve{x}\right) &=& - \frac{4\,w^i\left(t,\ve{x}\right)}{c^3} + {\cal O}\left(c^{-5}\right),
\label{BCRS_2}
\\
\nonumber\\
g_{ij}\left(t,\ve{x}\right) &=& \left( 1 + \frac{2\,w\left(t,\ve{x}\right)}{c^2}\right) \delta_{ij} + {\cal O}\left(c^{-4}\right).  
\label{BCRS_3}
\end{eqnarray}

\noindent
The gravitational potentials in (\ref{BCRS_1}) - (\ref{BCRS_3}) are given by the integrals  
\begin{eqnarray}
w\left(t,\ve{x}\right) &=& \frac{G}{c^2} \int d^3 x^{\prime}\;
\frac{t^{00}\left(t,\ve{x}^{\prime}\right)}{\left|\ve{x} - \ve{x}^{\prime}\right|} + {\cal O}\left(c^{-2}\right),  
\label{BCRS_4}
\\
\nonumber\\
w^i\left(t,\ve{x}\right) &=& \frac{G}{c} \int d^3 x^{\prime}\;
\frac{t^{0i}\left(t,\ve{x}^{\prime}\right)}{\left|\ve{x} - \ve{x}^{\prime}\right|} + {\cal O}\left(c^{-2}\right),  
\label{BCRS_5}
\end{eqnarray}

\noindent
where the integrals in (\ref{BCRS_4}) and (\ref{BCRS_5}) run over the entire Solar system, and $t^{\mu\nu}$  
is the energy-momentum tensor of the Solar system in global BCRS
coordinates; recall the components of energy-momentum tensor scale as follows:
$t^{00} = {\cal O}\left(c^2\right), t^{0i} = {\cal O}\left(c^1\right), t^{ij} = {\cal O}\left(c^0\right)$. 

The global gravitational potentials in (\ref{BCRS_4}) - (\ref{BCRS_5}) admit an expansion in terms of  
global Blanchet-Damour (BD) mass-multipoles and spin-multipoles, $m_L$ and $s_L$,   
\cite{Thorne,Blanchet_Damour1,Blanchet_Damour2,KopeikinSchaefer1999_Gwinn_Eubanks}, which characterize  
the multipole structure of the Solar system as a whole,   
\begin{eqnarray}
w\left(t,\ve{x}\right) &=& G \sum\limits_{l=0}^{\infty} \frac{\left(-1\right)^l}{l!}\,m_{\langle L\rangle}\left(t\right)\,
\partial_{\langle L\rangle}\,\frac{1}{r} + {\cal O}\left(c^{-2}\right), 
\label{global_multipole_expansion1}
\\
\nonumber\\
w^i\left(t,\ve{x}\right) &=& - G\,\sum\limits_{l = 2}^{\infty} \frac{\left(-1\right)^l}{l!}
\dot{m}_{\langle i L-1 \rangle}\left(t\right)\;\partial_{\langle L-1 \rangle}\,\frac{1}{r}
\nonumber\\
\nonumber\\
&& \hspace{-1.5cm} - G\,\sum\limits_{l = 1}^{\infty} \frac{\left(-1\right)^l\;l}{\left(l + 1\right)!}\, 
\epsilon_{iab}\,s_{\langle b L-1 \rangle} \left(t\right)  
\partial_{\langle a L-1 \rangle}\,\frac{1}{r} + {\cal O}\left(c^{-2}\right). 
\nonumber\\
\label{global_multipole_expansion2}
\end{eqnarray}

\noindent  
The global mass-multipoles and global spin-multipoles in (\ref{global_multipole_expansion1}) - (\ref{global_multipole_expansion2}) are Cartesian symmetric 
and trace-free (STF) tensors, and up to order ${\cal O}\left(c^{-2}\right)$ given by (cf. Eqs.~(2.34a) and (2.34b) in \cite{Blanchet_Damour1}):
\begin{eqnarray}
m_{\langle L \rangle}\left(t\right) &=& \underset{L}{\rm STF} \int d^3x\,x_L\, \frac{t^{00}\left(t,\ve{x}\right)}{c^2}\,, 
\label{global_mass_multipoles}
\\
\nonumber\\
s_{\langle L \rangle}\left(t\right) &=& \underset{L}{\rm STF} \int d^3x\,\epsilon_{a b c_l}\,x_{a L-1}\,\frac{t^{0b}\left(t,\ve{x}\right)}{c}\,,  
\label{global_spin_multipoles}
\end{eqnarray}

\noindent
where $m_0 = {\rm const.}$ is the mass of the entire Solar system, the mass-dipole $m_i =0$ because the origin of BCRS is located at the barycenter of the 
Solar system, and the spin-dipole $s_i = {\rm const}$ describes the spin of the entire Solar system which safely can be assumed to be time-independent.  
The spatial derivative operator in (\ref{global_multipole_expansion1}) - (\ref{global_multipole_expansion2}) is defined by  
\begin{eqnarray}
\partial_{\langle L \rangle} &=& \underset{i_1...i_l}{\rm STF}\;\frac{\partial}{\partial x^{i_1}}\,.\,.\,.\,\frac{\partial}{\partial x^{i_l}}\,,
\label{Spatial_Derivative_BCRS} 
\end{eqnarray}

\noindent
and a dot means derivative with respect to global coordinate-time.
The expansion in (\ref{global_multipole_expansion1}) - (\ref{global_multipole_expansion2}) has two specific features,  
which do not allow a straightforward application in our investigations:

(1) The expansion in (\ref{global_multipole_expansion1}) - (\ref{global_multipole_expansion2}) is valid outside a sphere which encloses the  
$N$-body system \cite{Thorne,Blanchet_Damour1,Blanchet_Damour2,Multipole_Damour_2,Zschocke_Multipole_Expansion}. It is quite obvious that for modeling  
of light propagation through the Solar system we need to have a metric which is valid in spatial domains between these $N$ massive bodies.  

(2) It has already been underlined in the introductory section that according to the theory of reference systems  
\cite{IAU_Resolution1,Brumberg1991,BK1,Reference_System1,BK2,BK3,DSX1,DSX2,DSX3,DSX4} physically meaningful  
multipole moments of some massive body $A$ have to be defined in the local reference system $\left(cT_A, \ve{X}_A\right)$ tied to that body and co-moving with it.

For these reasons, the global gravitational potentials in (\ref{global_multipole_expansion1}) - (\ref{global_multipole_expansion2})  
must have to be expressed by intrinsic mass-multipoles $M^A_L$ and intrinsic spin-multipoles $S^A_L$,  
which are defined in the local reference system $\left(cT_A,\ve{X}_A\right)$ of each individual massive body A. 
The prototype of all these GCRS-like coordinate systems is the GCRS especially designed for the Earth and which will be considered now.  

\subsection{GCRS}\label{GCRS} 

For the description of physical problems nearby the Earth the GCRS is the appropriate reference system.   
The harmonic GCRS coordinates are denoted by $X^{\alpha}=\left(cT,X^i\right)$, where $T={\rm TCG}$ is the GCRS coordinate time.  
According to IAU resolution B1.3 (2000) \cite{IAU_Resolution1}, the origin of the spatial axes of GCRS is located at the center-of-mass  
of the Earth and co-moving with it. The spatial axes of GCRS are kinematically non-rotating with respect to the BCRS, that means 
the GCRS is a space-fixed reference system and is not a local inertial system.  
The GCRS is completely characterized by the form of its metric tensor, up to order ${\cal O}\left(c^{-4}\right)$ given by
\cite{IAU_Resolution1,DSX1,DSX2},
\begin{eqnarray}
G_{00}\left(T,\ve{X}\right) &=& - 1 + \frac{2\,W\left(T,\ve{X}\right)}{c^2} + {\cal O}\left(c^{-4}\right),
\label{GCRS_1}
\\
\nonumber\\
G_{0i}\left(T,\ve{X}\right) &=& - \frac{4\,W^i\left(T,\ve{X}\right)}{c^3} + {\cal O}\left(c^{-5}\right),
\label{GCRS_2}
\\
\nonumber\\
G_{ij}\left(T,\ve{X}\right) &=& \left(1 + \frac{2\,W\left(T,\ve{X}\right)}{c^2}\right)\delta_{ij} + {\cal O}\left(c^{-4}\right).
\label{GCRS_3}
\end{eqnarray}

\noindent
The gravitational potentials in (\ref{GCRS_1}) - (\ref{GCRS_3}) can uniquely be separated into two components: a local component,  
$\left(W_{\rm loc}, W^i_{\rm loc}\right)$ which originates from the body $A$ itself and an external component, 
$\left(W_{\rm ext}, W^i_{\rm ext}\right)$, which is associated with inertial effects and 
tidal forces \cite{IAU_Resolution1,DSX1,DSX2}:  
\begin{eqnarray}
W\left(T,\ve{X}\right) &=& W_{\rm loc}\left(T,\ve{X}\right) + W_{\rm ext}\left(T,\ve{X}\right), 
\label{GCRS_4}
\\
\nonumber\\
W^i\left(T,\ve{X}\right) &=& W^i_{\rm loc}\left(T,\ve{X}\right) + W^i_{\rm ext}\left(T,\ve{X}\right).
\label{GCRS_5}
\end{eqnarray}

\noindent
Explicit expressions for the external potentials are given in \cite{DSX1,DSX2}, while  
the local potentials are defined by the following integrals:
\begin{eqnarray}
W_{\rm loc}\left(T,\ve{X}\right) &=&
\frac{G}{c^2} \int_{V_E} d^3 X^{\prime}\;
\frac{T^{00}\left(T,\ve{X}^{\prime}\right)}{\left|\ve{X} - \ve{X}^{\prime}\right|} + {\cal O}\left(c^{-2}\right),
\nonumber\\
\label{GCRS_6}
\\
\nonumber\\
W^i_{\rm loc}\left(T,\ve{X}\right) &=&
\frac{G}{c} \int_{V_E} d^3 X^{\prime}\;
\frac{T^{0i}\left(T,\ve{X}^{\prime}\right)}{\left|\ve{X} - \ve{X}^{\prime}\right|} + {\cal O}\left(c^{-2}\right),
\nonumber\\
\label{GCRS_7}
\end{eqnarray}

\noindent
where the integrations run over the entire volume of the Earth, and where $T^{\mu\nu}$ are the components of  
the energy-momentum tensor in GCRS coordinates; recall the components of energy-momentum tensor scale as follows:  
$T^{00} = {\cal O}\left(c^2\right), T^{0i} = {\cal O}\left(c^1\right), T^{ij} = {\cal O}\left(c^0\right)$. The  
local potentials (\ref{GCRS_6}) - (\ref{GCRS_7}) generated by the Earth can be expanded into a series of STF multipole 
moments, $M_L$ and $S_L$. In the harmonic skeletonized gauge they are given by  
\cite{IAU_Resolution1,Thorne,Blanchet_Damour1,Blanchet_Damour2,Multipole_Damour_2,DSX1}:  
\begin{eqnarray}
W_{\rm loc} \left(T,\ve{X}\right) &=&
G \sum\limits_{l=0}^{\infty} \frac{\left(-1\right)^l}{l!} M_{\langle L \rangle}\left(T\right) {\cal D}_{\langle L \rangle} \frac{1}{R} 
+ {\cal O}\left(c^{-2}\right),  
\nonumber\\
\label{BCRS_10}
\\
\nonumber\\
W^i_{\rm loc} \left(T,\ve{X}\right) &=&
- G \sum\limits_{l=1}^{\infty} \frac{\left(-1\right)^l}{l!} \dot{M}_{\langle i L-1 \rangle}\left(T\right)\,
{\cal D}_{\langle L-1 \rangle} \frac{1}{R}  
\nonumber\\
\nonumber\\
&& \hspace{-2.2cm} - G \sum\limits_{l=1}^{\infty} \frac{\left(-1\right)^l\;l}{\left(l+1\right)!}\, 
\epsilon_{iab}\,S_{\langle b L-1 \rangle}\left(T\right) {\cal D}_{\langle a L-1 \rangle} \frac{1}{R} + {\cal O}\left(c^{-2}\right),  
\nonumber\\
\label{BCRS_11}
\end{eqnarray}

\noindent
where $R =\left|\ve{X}\right|$ is the spatial distance from the origin of GCRS to some field point outside the Earth, and  
\begin{eqnarray}
{\cal D}_{\langle L \rangle} &=& \underset{a_1...a_l}{\rm STF}\;\frac{\partial}{\partial X^{a_1}} \,.\,.\,.\,\frac{\partial}{\partial X^{a_l}}\,,   
\label{Spatial_Derivative_GCRS} 
\end{eqnarray}

\noindent
and a dot in (\ref{BCRS_11}) denotes a derivative with respect to GCRS coordinate time $T$.  

The intrinsic STF multipoles in (\ref{BCRS_10}) and (\ref{BCRS_11}) can be approximated by their Newtonian expressions, that means up 
to terms of the order ${\cal O}\left(c^{-2}\right)$ they are given by:
\begin{eqnarray}
M_{\langle L \rangle} \left(T\right) &=& \underset{L}{\rm STF} \int_{V_E} d^3 X\;X_L\;\frac{T^{00}\left(T,\ve{X}\right)}{c^2}\,, 
\label{local_Newtonian_Mass_Multipole}
\\
\nonumber\\
S_{\langle L \rangle}\left(T\right) &=& \underset{L}{\rm STF} \int_{V_E} d^3 X \epsilon_{a b c_l}
X_{a L-1} \frac{T^{0b}\left(T,\ve{X}\right)}{c} ,
\label{local_Newtonian_Spin_Multipole}
\end{eqnarray}

\noindent
where the integration runs over the volume of the Earth, and $T^{\alpha\beta}$ is the energy-momentum tensor in the local system of the Earth.  

The intrinsic mass-monopole term $M = {\rm const}$ in (\ref{local_Newtonian_Mass_Multipole}) is the Newtonian mass of the Earth.  
Actually, the mass-dipole vanishes, $M_{i}=0$, because the origin of the GCRS is assumed to be located at the barycenter of the Earth, 
but in real measurements of celestial mechanics the center-of-mass of massive Solar system bodies can not be determined exactly,  
so it is meaningful to keep this term and to assume $M_{i} = {\rm const.}$ in general.  
The spin-dipole $S_i\left(T\right)$ of the Earth is not constant but time-dependent due to inner forces of the Earth and due to the gravitational interaction 
of the Earth with other massive bodies.

\subsection{Metric of Solar system in terms of intrinsic multipoles}\label{Solar_System_Metric}

In order to describe the light trajectory through the Solar system, one needs to introduce one global chart (BCRS) $x^{\mu} = \left(ct,\ve{x}\right)$ 
but expressed in terms of intrinsic multipoles, $M^A_L$ and $S^A_L$, of each massive body $A=1,...,N$. For being able to define the multipole  
structure of each individual body in a physically meaningful manner, the DSX formalism \cite{DSX1,DSX2} introduces $N$ local GCRS-like reference systems 
$X_A^{\alpha} = \left(c T_A, \ve{X}_A\right)$, each one very similar to the GCRS in Eqs.~(\ref{GCRS_1}) - (\ref{GCRS_3}). 
These $N+1$ coordinate systems are linked with each other via coordinate-transformations.  
The DSX theory \cite{DSX1,DSX2,DSX3,DSX4} provides the theoretical framework for such an approach, and has                              
originally been established for celestial mechanics and for deriving the equations of motion of a system of $N$ massive bodies with full
multipole structure. Consequently, one central result of DSX theory are the coordinate transformations among these reference systems,  
which are given by 
\begin{eqnarray}
x^{\mu} &=& x^{\mu}_A\left(T_A\right) + e^{\mu}_{a}\left(T_A\right) X^a_A + {\cal O} \left(c^{-2}\right),
\label{coordinate_transformation_1}
\end{eqnarray}

\noindent
where $x^{\mu}_A\left(T_A\right)$ is the worldline of body $A$ in BCRS coordinates. 
The inverse coordinate transformations could be found in \cite{IAU_Resolution1}, but is not of relevance here for our purposes. 
The tetrads $e^{\mu}_{a}$ along the worldline of this body are explicitly given by (cf. Eqs.~(2.16) in \cite{DSX1}):  
\begin{eqnarray}
e_a^0\left(T_A\right) &=& \frac{\dot{x}_A^a\left(T_A\right)}{c} + {\cal O}\left(c^{-3}\right),
\label{tetrade_2}
\\
\nonumber\\
e_a^i\left(T_A\right) &=& \delta_{ai} + {\cal O}\left(c^{-2}\right),
\label{tetrade_3}
\end{eqnarray}

\noindent
where in (\ref{tetrade_2}) a dot means derivative with respect to the local coordinate time of body $A$. That means, $\dot{\ve{x}}_A\left(T_A\right)$ is  
the three-velocity of body $A$ in the global system and given in terms of the body's local coordinate time $T_A$, which could easily be transformed into terms 
of global BCRS coordinate-time.

The contravariant components of the BCRS metric tensor $g^{\mu\nu}\left(t,\ve{x}\right)$ in Eqs.~(\ref{BCRS_1}) - (\ref{BCRS_3}) and the 
contravariant components of the metric tensor $G^{\alpha\beta}\left(T_A,\ve{X}_A\right)$ in Eqs.~(\ref{GCRS_1}) - (\ref{GCRS_3}) in the local GCRS-like 
coordinate system of body $A$ are related via the following transformation:  
\begin{eqnarray}
g^{\mu\nu}\left(t,\ve{x}\right) &=& \frac{\partial x^{\mu}}{\partial X_A^{\alpha}}\,\frac{\partial x^{\nu}}{\partial X_A^{\beta}}\, 
G^{\alpha\beta}\left(T_A,\ve{X}_A\right). 
\label{Tensor_transformation}
\end{eqnarray}

\noindent 
Using (\ref{coordinate_transformation_1}) in virtue of (\ref{Tensor_transformation}), the global potentials $\left(w,w^i\right)$ in (\ref{BCRS_4}) - (\ref{BCRS_5})  
can be expressed in terms of intrinsic STF multipoles $M_L^A$ and $S_L^A$ as follows \cite{DSX1,DSX2,IAU_Resolution1}:  
\begin{eqnarray}
w\left(t,\ve{x}\right) &=& \sum\limits_{A=1}^{N}\, w_A\left(t,\ve{x}\right),   
\label{global_metric_potentials_1}
\\
\nonumber\\
w_A\left(t,\ve{x}\right) &=& G \sum\limits_{l = 0}^{\infty} \frac{\left(-1\right)^l}{l!}
M_{\langle L \rangle}^A\left(T_A\right)\;{\cal D}^A_{\langle L \rangle} \frac{1}{R_A} + {\cal O}\left(c^{-2}\right),
\nonumber\\
\label{global_metric_potentials_2}
\\
\nonumber\\
\nonumber\\
w^i\left(t,\ve{x}\right) &=& \sum\limits_{A=1}^{N}\, w_A^i\left(t,\ve{x}\right),  
\label{global_metric_potentials_4}
\\
\nonumber\\
w_A^i\left(t,\ve{x}\right) &=&  
- G\,\sum\limits_{l=1}^{\infty} \frac{\left(-1\right)^l}{l!}\;\dot{M}^A_{\langle i L-1 \rangle}\left(T_A\right)\,
{\cal D}^A_{\langle L-1 \rangle} \frac{1}{R_A}
\nonumber\\
\nonumber\\
&& \hspace{-1.5cm} - G\,\sum\limits_{l=1}^{\infty} \frac{\left(-1\right)^l}{l!}
\,\frac{l}{l+1} \epsilon_{iab}\,S^A_{\langle b L-1 \rangle}\left(T_A\right)\;{\cal D}^A_{\langle a L-1 \rangle}\,\frac{1}{R_A}  
\nonumber\\
\nonumber\\
&& \hspace{-1.5cm} + G\,v_A^i\left(T_A\right) \sum\limits_{l = 0}^{\infty} \frac{\left(-1\right)^l}{l!}
M_{\langle L \rangle}^A\left(T_A\right)\;{\cal D}^A_{\langle L\rangle} \frac{1}{R_A} + {\cal O}\left(c^{-2}\right).  
\nonumber\\
\label{global_metric_potentials_3}
\end{eqnarray}

\noindent
In (\ref{global_metric_potentials_1}) and (\ref{global_metric_potentials_4}) the sum runs over all bodies of the $N$-body system, 
$R_A =\left|\ve{X}_A\right|$ is the spatial distance from the origin of local coordinate system
to some field point located outside the massive body, and 
\begin{eqnarray}
{\cal D}^A_{\langle L\rangle}=\underset{a_1...a_l}{\rm STF}\;\frac{\partial}{\partial X^{a_1}_A}\,.\,.\,.\,\frac{\partial}{\partial X_A^{a_l}}\,.
\label{Spatial_Derivative_GCRS_A}
\end{eqnarray}

\noindent
The local mass-multipoles and spin-multipoles of some massive body $A$ in Newtonian approximation, i.e. up to terms of the  
order ${\cal O}\left(c^{-2}\right)$, are given by (cf. Eqs.~(\ref{local_Newtonian_Mass_Multipole}) - (\ref{local_Newtonian_Spin_Multipole}):
\begin{eqnarray}
M^A_{\langle L \rangle}\left(T_A\right) &=& \underset{L}{\rm STF} \int_{V_A} d^3 X_A X^A_L\,\frac{T^{00}_A\left(T_A,\ve{X}_A\right)}{c^2}\,,  
\label{local_Newtonian_Mass_Multipole_A}
\\
\nonumber\\
S^A_{\langle L \rangle}\left(T_A\right) &=& \underset{L}{\rm STF} \int_{V_A} d^3 X_A\,\epsilon_{a b c_l}  
X^A_{a L-1} \frac{T_A^{0b}\left(T_A,\ve{X}_A\right)}{c}\,,
\nonumber\\
\label{local_Newtonian_Spin_Multipole_A}
\end{eqnarray}

\noindent
where the integration runs over the spatial volume of massive body A, and $T_A^{\alpha\beta}$ is the energy-momentum tensor of body $A$  
in the local coordinate system of that body.

Finally, using coordinate transformations (\ref{coordinate_transformation_1}), the spatial derivatives in (\ref{global_metric_potentials_2}) 
and (\ref{global_metric_potentials_3}) must be transformed into the BCRS, where they read as follows:  
\begin{eqnarray}
\frac{\partial}{\partial cT_A} &=& \frac{\partial}{\partial c t}
+ \frac{v_A^a\left(T_A\right)}{c} \frac{\partial}{\partial x^a} + {\cal O}\left(c^{-2}\right),
\label{coordinate_transformation_6}
\\
\nonumber\\
\frac{\partial}{\partial X_A^a} &=& \frac{\partial}{\partial x^a}
+ \frac{v_A^a\left(T_A\right)}{c} \frac{\partial}{\partial ct} + {\cal O}\left(c^{-2}\right),  
\label{coordinate_transformation_7}
\end{eqnarray}

\noindent
where the second term in (\ref{coordinate_transformation_6}) as well as in (\ref{coordinate_transformation_7}) generate terms which are  
beyond the order of ${\cal O}\left(c^{-4}\right)$ in the global metric, that means these terms will actually not contribute in the final results for 
the light trajectory. 
From (\ref{coordinate_transformation_1}) follows the relation \cite{IAU_Resolution1,Kopeikin_Efroimsky_Kaplan,DSX1,DSX2}:
\begin{eqnarray}
R_A &=& \left| \ve{x} - \ve{x}_A\left(t\right)\right| + {\cal O} \left(c^{-2}\right),   
\label{coordinate_relation_5} 
\end{eqnarray}

\noindent
where we recall that some massive body $A$ moves along the arbitrary worldline $\ve{x}_A\left(t\right)$, which can later be concretized by Solar system 
ephemeris \cite{JPL} at any stage of the calculations.  
Because of the fact that the BCRS coordinate-time and the coordinate time $T_A$ of local system of body $A$ are related as 
follows \cite{IAU_Resolution1,Kopeikin_Efroimsky_Kaplan,DSX1,DSX2},  
\begin{eqnarray}
T_A &=& t + {\cal O} \left(c^{-2}\right),  
\label{coordinate_relation_10}
\end{eqnarray}

\noindent
we obtain for the time-dependence of the intrinsic multipoles the following relation:  
\begin{eqnarray}
M_{\langle L \rangle}^A\left(T_A\right) &=& M_{\langle L \rangle}^A\left(t\right) + {\cal O}\left(c^{-2}\right), 
\label{coordinate_relation_15}
\\
\nonumber\\
S_{\langle L \rangle}^A\left(T_A\right) &=& S_{\langle L \rangle}^A\left(t\right) + {\cal O}\left(c^{-2}\right),  
\label{coordinate_relation_20}
\end{eqnarray}

\noindent
that means the neglected terms in (\ref{coordinate_relation_15}) and (\ref{coordinate_relation_20}) are beyond 1.5PN approximation for lightrays.  

Summarizing the conclusions in Eqs.~(\ref{coordinate_transformation_1}) - (\ref{coordinate_relation_20}), the metric perturbation in the near-zone 
of the Solar system and expressed in terms of local multipoles is given by: 
\begin{eqnarray}
h^{\left(2\right)}_{00}\left(t,\ve{x}\right) &=& \sum\limits_{A=1}^{N} h^{\left(2\right) A}_{00}\left(t,\ve{x}\right),  
\label{global_metric_perturbation_A}
\\
\nonumber\\
h^{\left(2\right) A}_{00}\left(t,\ve{x}\right) &=& \frac{2\,G}{c^2}
\sum\limits_{l = 0}^{\infty} \frac{\left(-1\right)^l}{l!}
M_{\langle L \rangle}^A \left(t\right)\;\partial_{\langle L \rangle}\,\frac{1}{r_A\left(t\right)}\,,
\label{global_metric_perturbation_B1}
\\
\nonumber\\
\nonumber\\
h^{\left(2\right)}_{ij}\left(t,\ve{x}\right) &=& h^{\left(2\right)}_{00}\left(t,\ve{x}\right)\,\delta_{ij}\,,
\label{global_metric_perturbation_B3}
\\
\nonumber\\
\nonumber\\
h^{\left(3\right)}_{0i}\left(t,\ve{x}\right) &=& \sum\limits_{A=1}^{N} h^{\left(3\right) A}_{0i}\left(t,\ve{x}\right),  
\label{global_metric_perturbation_B2}
\\
\nonumber\\
h^{\left(3\right) A}_{0i}\left(t,\ve{x}\right) &=& 
\frac{4\,G}{c^3}\,\sum\limits_{l=1}^{\infty} \frac{\left(-1\right)^l}{l!}\;\dot{M}^A_{\langle i L-1 \rangle}\left(t\right)\,
\partial_{\langle L-1 \rangle} \frac{1}{r_A\left(t\right)}
\nonumber\\
\nonumber\\
&& \hspace{-2.0cm} + \frac{4\,G}{c^3}\,\sum\limits_{l=1}^{\infty} \frac{\left(-1\right)^l\;l}{\left(l + 1 \right)!}\, 
\epsilon_{iab}\,S^A_{\langle b L-1 \rangle} \left(t\right)\;\partial_{\langle a L-1 \rangle}\,\frac{1}{r_A\left(t\right)}
\nonumber\\
\nonumber\\
&& \hspace{-2.0cm} - \frac{4\,G}{c^3}\,v_A^i\left(t\right) \sum\limits_{l = 0}^{\infty} \frac{\left(-1\right)^l}{l!}
M_{\langle L \rangle}^A \left(t\right)\;\partial_{\langle L \rangle}\frac{1}{r_A\left(t\right)}\,, 
\label{global_metric_perturbation_C}
\end{eqnarray}

\noindent
where the summation in (\ref{global_metric_perturbation_A}) and (\ref{global_metric_perturbation_B2}) runs over all massive bodies of the Solar system, while  
the metric perturbations caused by one individual body are given by (\ref{global_metric_perturbation_B1}) and
(\ref{global_metric_perturbation_C}). 
The dot in the first term of expression (\ref{global_metric_perturbation_C}) means here derivative with respect to 
global BCRS coordinate time, and the spatial derivatives in (\ref{global_metric_perturbation_A}) - (\ref{global_metric_perturbation_C}) are derivatives  
in the global system and given by
\begin{eqnarray}
\partial_{\langle L \rangle}=\underset{i_1...i_l}{\rm STF}\;\frac{\partial}{\partial x^{i_1}}\,...\,\frac{\partial}{\partial x^{i_l}}\,, 
\label{spatial_derivative}
\end{eqnarray}

\noindent
and 
\begin{eqnarray}
r_A\left(t\right) &=& \left|\ve{x} - \ve{x}_A\left(t\right)\right|\,,
\label{spatial_distance_1}
\end{eqnarray}

\noindent
is the distance between some field-point with spatial coordinate $\ve{x}$ and the spatial position $\ve{x}_A\left(t\right)$ of massive body $A$ in the global
reference system at BCRS time $t$. The metric perturbations in (\ref{global_metric_perturbation_A}) - (\ref{global_metric_perturbation_C}) have to be implemented 
into the geodesic equation in (\ref{geodesic_equation_1}) and, therefore, the field-point $\ve{x}$ in (\ref{spatial_distance_1}) will be identified with the 
photons position $\ve{x}\left(t\right)$ according to Eq.~(\ref{geodesic_equation_3}).  
In view of this fact we will use the same notation for the distance in (\ref{spatial_distance_1}) and for the absolute value of (\ref{notation_5a}).   

Before going further, we underline the absence of terms proportional to $\displaystyle \frac{{v}_A^2}{c^2}\,M_L^A$,  
$\displaystyle \frac{{v}_A}{c}\,\dot{M}_L^A$,   
and $\displaystyle \frac{{v}_A}{c}\,S_L^A$ or higher time-derivatives  
of the multipoles like $\ddot{M}_L^A$ or $\dot{S}_L^A$ in the DSX metric tensor (\ref{global_metric_perturbation_A}) - (\ref{global_metric_perturbation_C}).  
Such terms are of the order ${\cal O}\left(c^{-4}\right)$ in the metric, that means they are beyond the 1.5PN approximation for lightrays.

\section{Transformation of geodesic equation}\label{Section4}

As it has been discussed above, instead of (\ref{geodesic_equation_1}) we actually may consider the simpler form of geodesic equation
in (\ref{geodesic_equation_5}), which is integrated along the unperturbed light trajectory (\ref{Introduction_1}). That means, according
to Eq.~(\ref{transformed_geodesic_equation_15}), the field-point $\ve{x}$ in Eq.~(\ref{spatial_distance_1}) can be approximated by the unperturbed
photon-trajectory $\ve{x}_{\rm N}\left(t\right)$ in (\ref{Introduction_1}), so that we get the following expression for the vector pointing  
from the center of massive body $A$ toward the spatial position of the photon at time $t$:  
\begin{eqnarray}
\ve{r}^{\rm N}_A\left(t\right) &=& \ve{x}_{\rm N}\left(t\right) - \ve{x}_A\left(t\right)\,, 
\label{spatial_distance_4}
\\
\nonumber\\
r^{\rm N}_A\left(t\right) &=& \left|\ve{x}_{\rm N}\left(t\right) - \ve{x}_A\left(t\right)\right|\,,
\label{spatial_distance_5}
\end{eqnarray}

\noindent
where the unperturbed lightray is given by Eq.~(\ref{Introduction_1}) or Eq.~(\ref{Unperturbed_Light_Trajectory}). 
It especially means, that all derivatives in geodesic equation
(\ref{geodesic_equation_5}) and in the metric perturbations in (\ref{global_metric_perturbation_A}) - (\ref{global_metric_perturbation_C}) act on
the unperturbed lightray. In view of this important fact, it is highly effective to embark on a strategy, where all expressions in the geodesic equation
(\ref{geodesic_equation_5}) are expressed in terms of new parameters which fully characterize the unperturbed light trajectory from the very beginning of
the integration procedure. This strategy especially implies, that we will transform the spatial derivatives in (\ref{spatial_derivative}), the derivatives
in the geodesic equation (\ref{geodesic_equation_5}), the distance in (\ref{spatial_distance_5}) and the time-argument of the multipoles
in terms of these new parameters.
\begin{figure}[!ht]
\begin{center}
\includegraphics[scale=0.15]{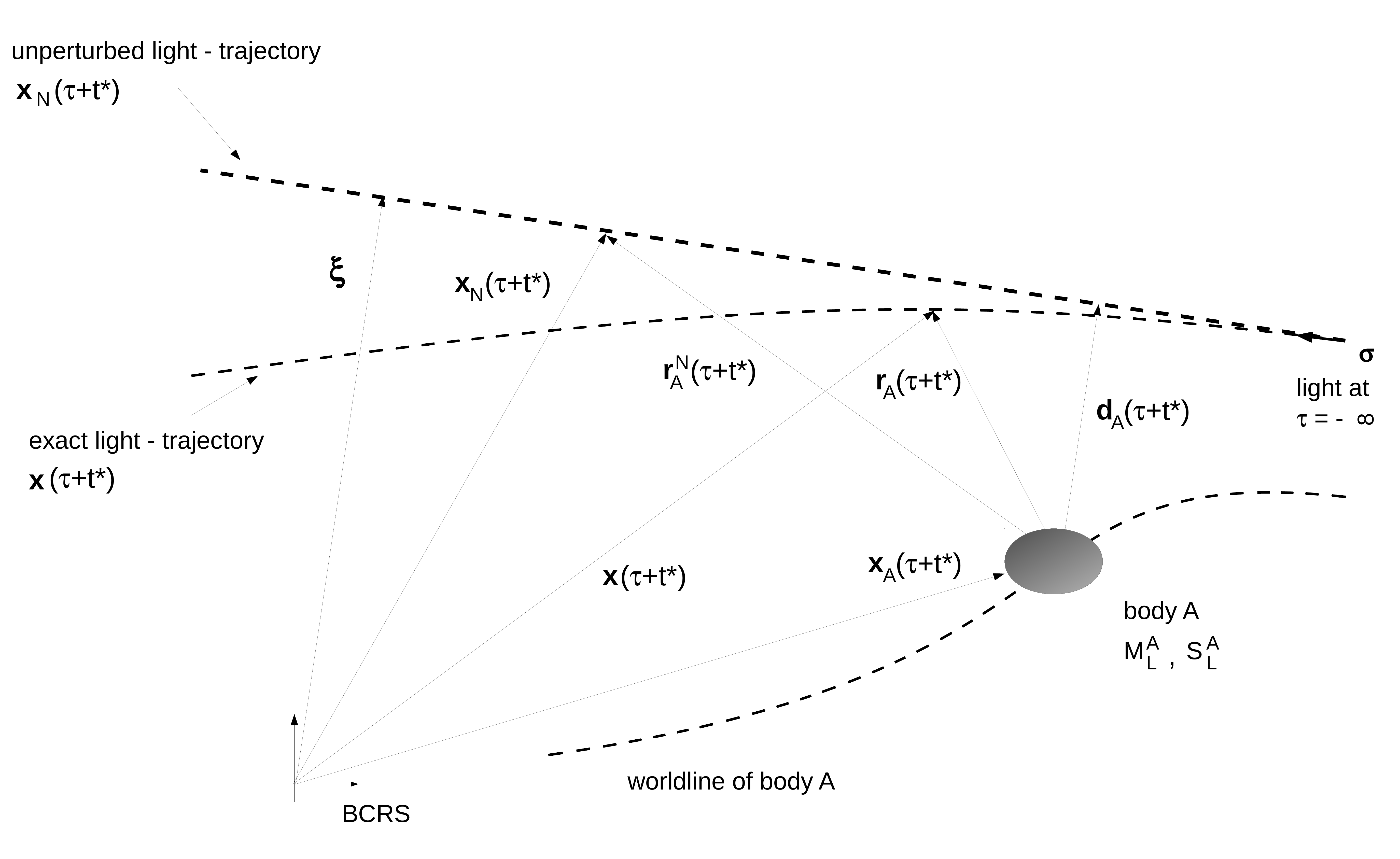}
\end{center}
\caption{A geometrical representation of the light trajectory through the Solar system (only one massive body $A$ of the $N$-body Solar system is depicted) 
in terms of the new variables $\ve{\xi}$ and $\tau$. The impact vector $\ve{\xi}$ is defined by Eq.~(\ref{variable_2})  
and points from the origin of global system to the point of closest approach of the
unperturbed lightray to that origin, and is time-independent. The
impact vector $\ve{d}_A\left(\tau+t^{\ast}\right)$ is defined by Eq.~(\ref{First_Integration_22})
and points from the origin of local system of body $A$ toward the point of closest approach
of unperturbed lightray to that origin, and is time-dependent due to the motion of the body.
Furthermore, $\ve{x} \left(\tau+t^{\ast}\right)$ is the global spatial coordinate of the photon
of the light trajectory, while $\ve{x}_{\rm N}\left(\tau+t^{\ast}\right)$ is the unperturbed lightray. The
worldline of massive body $A$ in the global system is given by $\ve{x}_A\left(\tau+t^{\ast}\right)$, and
$\ve{r}_A\left(\tau+t^{\ast}\right)$ points from the origin of local system toward the exact photon's position, while
$\ve{r}^{\rm N}_A\left(\tau+t^{\ast}\right)$ points from the origin of local system toward the unperturbed lightray.}
\label{Diagram1}
\end{figure}

The problem and the need for introducing new variables is namely the following. The variables $t$ and $\ve{x}$ are field variables
of the gravitational field and, therefore, they are of course independent of each other. But since the integration of geodesic equation
proceeds along the lightray (cf. Eq.~(\ref{geodesic_equation_3})) these field variables have to be replaced by the photon trajectory, $\ve{x}\left(t\right)$,
and then these variables become dependent on each other. A drastical simplification is achieved in view of Eq.~(\ref{transformed_geodesic_equation_15})  
which states that the geodesic equation in (\ref{geodesic_equation_5}) can be integrated along the unperturbed lightray.   
Therefore, we are looking for new time-variable and spatial-variable, which
fully parametrize the unperturbed lightray and which are independent on each other. In this way the integration of geodesic equation becomes feasible.
Just for that reason, the following independent variables $\tau$ and $\ve{\xi}$ have been introduced
in \cite{KopeikinSchaefer1999_Gwinn_Eubanks,KopeikinKorobkovPolnarev2006,KopeikinKorobkov2005}:
\begin{eqnarray}
c\,\tau &=& \ve{\sigma}\cdot\ve{x}_{\rm N}\left(t\right)\,,\quad
c\,\tau_0 = \ve{\sigma}\cdot\ve{x}_{\rm N}\left(t_0\right),
\label{variable_1}
\\
\nonumber\\
\xi^i &=& P^i_j\,x_{\rm N}^j\left(t\right),
\label{variable_2}
\end{eqnarray}

\noindent
where $P^i_j$ is the operator of projection onto the plane perpendicular to vector $\ve{\sigma}$,
\begin{eqnarray}
P^{ij} &=& \delta_{i j} - \sigma^i\,\sigma^j\,,
\label{variable_3}
\end{eqnarray}

\noindent
where the covariant and contravariant positions of spatial indices is insignificant: $P^i_j = P_{ij} = P^{ij}$.
According to (\ref{variable_2}), the three-vector $\ve{\xi}$ is the impact vector of the unperturbed lightray,
see also Eq.~(\ref{notation_2}). Especially, $\ve{\xi}$ is time-independent and directed from the origin of global coordinate system
toward the point of closest approach of the unperturbed light trajectory and the absolute value is denoted by $d = \left|\ve{\xi}\right|$.
For a graphical elucidation see Fig.~\ref{Diagram1}.

Another important parameter is the time of closest approach of unperturbed lightray to the origin of the global coordinate system, defined by
\begin{eqnarray}
t^{\ast} &=& t_0 - \frac{\ve{\sigma}\cdot\ve{x}_0}{c}\,,
\label{time_of_closest_approach}
\end{eqnarray}

\noindent
which differs from (\ref{time_of_closest_approach_t_0_moving_body}) which is the time of closest approach of the lightray to the origin of the local
coordinate system of some massive body A. Notice that $d t = d \tau$ for the total differentials, because $t^{\ast}$ is a constant for each particular lightray, 
and $\tau = t - t^{\ast}$ and $\tau_0 = t_0 - t^{\ast}$.   
With the aid of these new variables $\ve{\xi}$ and $\tau$, the mixed initial-boundary conditions
(\ref{Initial_Boundary_Condition_1}) and (\ref{Initial_Boundary_Condition_2}) take the form
\begin{eqnarray}
\ve{x}_0 &=& \ve{x}\left(\tau_0 + t^{\ast}\right),
\label{Transformed_Initial_Boundary_Condition_1}
\\
\nonumber\\
\ve{\sigma} &=& \lim_{\tau \rightarrow - \infty}\, \frac{\dot{\ve{x}}\left(\tau + t^{\ast}\right)}{c}\,,
\label{Transformed_Initial_Boundary_Condition_2}
\end{eqnarray}

\noindent
where a dot means derivative with respect to variable $\tau$. In the new variables the interpretation of these initial-boundary
conditions remains the same: the first condition (\ref{Transformed_Initial_Boundary_Condition_1}) defines the spatial coordinates
of the photon at the moment of emission of light, while the second condition (\ref{Transformed_Initial_Boundary_Condition_2})
defines the unit-direction ($\ve{\sigma}\cdot\ve{\sigma} = 1$) at infinite past and infinite distance from the origin of
global coordinate system, that means at the so-called past null infinity.

The unperturbed lightray in (\ref{Unperturbed_Light_Trajectory}) transforms as follows
\cite{KopeikinSchaefer1999_Gwinn_Eubanks,KopeikinSchaefer1999,KopeikinKorobkovPolnarev2006,KopeikinKorobkov2005,Zschocke_1PN}:
\begin{eqnarray}
\ve{x}_{\rm N}\left(\tau+t^{\ast}\right) &=& \ve{\xi} + c\,\tau\,\ve{\sigma}\,,
\label{variable_5}
\end{eqnarray}

\noindent
while its derivative with respect to variable $\tau$ reads $\dot{\ve{x}}_{\rm N}\left(\tau+t^{\ast}\right)=c\,\ve{\sigma}$.
The vector pointing from the spatial position of the arbitrarily moving body toward the unperturbed lightray in these new variables transforms as follows:
\begin{eqnarray}
\ve{r}^{\rm N}_A\left(\tau + t^{\ast}\right) &=& \ve{\xi} + c\,\tau\,\ve{\sigma} - \ve{x}_A\left(\tau + t^{\ast}\right),
\label{notation_15}
\end{eqnarray}

\noindent
with the absolute value $r^{\rm N}_A\left(\tau + t^{\ast}\right)=\left|\ve{r}^{\rm N}_A\left(\tau + t^{\ast}\right)\right|$, and
the impact parameter in (\ref{notation_6}) for arbitrarily moving bodies in these new variables reads:
\begin{eqnarray}
\ve{d}_A \left(\tau + t^{\ast}\right) &=& \ve{\sigma} \times \left(\ve{r}^{\rm N}_A\left(\tau + t^{\ast}\right) \times \ve{\sigma}\right),
\label{First_Integration_22}
\end{eqnarray}

\noindent
with the absolute value $d_A\left(\tau + t^{\ast}\right)=\left|\ve{d}_A\left(\tau + t^{\ast}\right)\right|$.

In virtue of Eqs.~(\ref{variable_1}) and (\ref{variable_2}) two new variables, $\tau$ and $\ve{\xi}$, have been introduced
and in addition the auxiliary variable $t^{\ast}$ by Eq.~(\ref{time_of_closest_approach}). As next, the partial derivatives with respect to
space and time in the geodesic equation (\ref{geodesic_equation_5}) have to be expressed in terms of these new variables. In the
pioneering investigations in \cite{KopeikinSchaefer1999_Gwinn_Eubanks,KopeikinKorobkovPolnarev2006,KopeikinKorobkov2005} it has been shown by chain rule
that these partial derivatives transform in the following way:
\begin{eqnarray}
&& \hspace{-0.5cm} \frac{\partial h_{\alpha \beta}^{(n)}\left(t,\ve{x}\right)}{\partial x^i}
\Bigg|_{\ve{x}=\ve{x}_{\rm N}\mbox{\normalsize $\left(t\right)$}}
\nonumber\\
&& \hspace{-0.5cm} = \left(P^{ij} \frac{\partial}{\partial \xi^j} + \sigma^i\,\frac{\partial}{\partial c\,\tau}
- \sigma^i \frac{\partial}{\partial c t^{\ast}}\right)
h_{\alpha \beta}^{(n)}\left(\tau + t^{\ast},\ve{\xi} + c \tau\,\ve{\sigma}\right),
\nonumber\\
\label{Transformation_Derivative_2A}
\\
\nonumber\\
&& \hspace{-0.5cm} \frac{\partial h_{\alpha \beta}^{(n)}\left(t,\ve{x}\right)}{\partial c t}
\Bigg|_{\ve{x}=\ve{x}_{\rm N}\mbox{\normalsize $\left(t\right)$}}
= \frac{\partial}{\partial c t^{\ast}}\,
h_{\alpha \beta}^{(n)}\left(\tau + t^{\ast},\ve{\xi} + c \tau\,\ve{\sigma}\right).
\label{Transformation_Derivative_2B}
\end{eqnarray}

\noindent
Two remarks are in order to interpret these relations correctly.

First, we notice that the explicit time-dependence of the metric tensor, $h_{\alpha \beta}^{(n)}\left(t,\ve{x}\right)$, is caused by the time-dependence of 
the multipoles $M_L^A\left(t\right), S_L^A\left(t\right)$ as well as by the motions of the massive bodies $\ve{x}_A\left(t\right)$. Therefore,  
the partial time-derivative on the l.h.s. in (\ref{Transformation_Derivative_2B}) acts on the multipoles 
as well as on the worldlines of the massive bodies. For the same reason, the time-derivatives on the r.h.s. in (\ref{Transformation_Derivative_2A}) 
and (\ref{Transformation_Derivative_2B}) act on the multipoles, the worldlines of the massive bodies and on the unperturbed lightray. 
The unperturbed lightray in (\ref{variable_5}) does, however, not depend on variable $t^{\ast}$. 
 
Second, it should be realized, that in the left-hand side in (\ref{Transformation_Derivative_2A})
and (\ref{Transformation_Derivative_2B}) one has first to perform the differentiations
and afterwards the field-point $\ve{x}$ has to be substituted by the unperturbed lightray
$\ve{x}_{\rm N}\left(t\right) = \ve{x}_0 + c\,\ve{\sigma}\left(t-t_0\right)$.
Opposite, in the right-hand side in (\ref{Transformation_Derivative_2A}) and (\ref{Transformation_Derivative_2B})
one has first to substitute $t^{\ast}+\tau$ and
$\ve{x}_{\rm N}\left(\tau+t^{\ast}\right) = \ve{\xi} + c\,\tau\,\ve{\sigma}$
and afterwards to perform the differentiations.
By means of these relations (\ref{Transformation_Derivative_2A}) and (\ref{Transformation_Derivative_2B}),
the geodesic equation in 1.5PN approximation in (\ref{geodesic_equation_5}) transforms as follows:
\begin{eqnarray}
\frac{\ddot{x}^i \left(\tau+t^{\ast}\right)}{c^2} &=&
+ \frac{1}{2}\,P^{ij}\,\frac{\partial}{\partial \xi^j}\,h_{00}^{(2)}
- \frac{1}{2}\,\sigma^i\,\frac{\partial}{\partial c \tau}\,h_{00}^{(2)}
\nonumber\\
\nonumber\\
&& \hspace{-2.5cm} + \frac{1}{2} \sigma^k \sigma^l P^{ij} \frac{\partial}{\partial \xi^j} h_{kl}^{(2)}
 + \frac{1}{2} \sigma^i \sigma^j \sigma^k \frac{\partial}{\partial c \tau} h_{jk}^{(2)}
- \sigma^j \frac{\partial}{\partial c \tau} h_{ij}^{(2)}
\nonumber\\
\nonumber\\
&& \hspace{-2.5cm} - \frac{\partial}{\partial c \tau} h_{0i}^{(3)}
+ \sigma^j P^{ik} \frac{\partial}{\partial \xi^k}\,h_{0j}^{(3)}
+ {\cal O}\left(c^{-4}\right),
\label{transformed_geodesic_equation_A}
\end{eqnarray}

\noindent
which agrees with Eq.~(36) in \cite{KopeikinSchaefer1999_Gwinn_Eubanks} and Eq.~(19) in \cite{KopeikinSchaefer1999}; note that $P_{ab}\,\sigma^b = 0$.
The double-dot on the left-hand side in (\ref{transformed_geodesic_equation_A}) means twice of the total differential with respect to the new variable
$\tau$. Subject to relation (\ref{global_metric_perturbation_B3}), the geodesic equation in (\ref{transformed_geodesic_equation_A}) simplifies further:
\begin{eqnarray}
\frac{\ddot{x}^i \left(\tau+t^{\ast}\right)}{c^2} &=&
P^{ij} \frac{\partial h_{00}^{(2)}\left(\tau+t^{\ast},\ve{\xi} + c \tau \ve{\sigma}\right)}{\partial \xi^j}
\nonumber\\
\nonumber\\
&& \hspace{-2.0cm} - \sigma^i \frac{\partial h_{00}^{(2)}\left(\tau+t^{\ast},\ve{\xi} + c \tau \ve{\sigma}\right)}{\partial c \tau}
- \frac{\partial h_{0i}^{(3)}\left(\tau+t^{\ast},\ve{\xi} + c \tau \ve{\sigma}\right)}{\partial c \tau}
\nonumber\\
\nonumber\\
&& \hspace{-2.0cm} + \sigma^j P^{ik} \frac{\partial h_{0j}^{(3)}\left(\tau+t^{\ast},\ve{\xi} + c \tau \ve{\sigma}\right)}{\partial \xi^k}
+ {\cal O}\left(c^{-4}\right).
\label{transformed_geodesic_equation_B}
\end{eqnarray}

\noindent
Let us note that the first two terms are of order ${\cal O}\left(c^{-2}\right)$ and agree with Eq.~(95) in \cite{Zschocke_1PN}, while the last two terms
are of order ${\cal O}\left(c^{-3}\right)$. This fact implies that if one integrates the geodesic equation (\ref{transformed_geodesic_equation_B}) then the
first two terms in (\ref{transformed_geodesic_equation_B}) give rise to terms of the order ${\cal O}\left(c^{-2}\right)$ as well as
to terms of order ${\cal O}\left(c^{-3}\right)$, while the last two terms generate only terms of the order ${\cal O}\left(c^{-3}\right)$.
The mathematical structure of (\ref{transformed_geodesic_equation_B}) is considerably simpler than the original form in (\ref{geodesic_equation_5}),
but of more decisive importance in the integration procedure is the fact that the time-variable $\tau$ and the space-variable $\ve{\xi}$ are
independent of each other.

As final step in the transformation, the metric perturbations in (\ref{global_metric_perturbation_A}) - (\ref{global_metric_perturbation_C}) have to
be transformed in terms of these new variables $\ve{\xi}$ and $\tau$. One obtains
\begin{eqnarray} 
h^{\left(2\right)}_{00}\left(\tau+t^{\ast},\ve{\xi} + c \tau\,\ve{\sigma} \right) &=&
\sum\limits_{A=1}^{N} h^{\left(2\right) A}_{00}\left(\tau+t^{\ast},\ve{\xi} + c \tau \ve{\sigma}\right),
\nonumber\\
\label{transformed_global_metric_perturbation_A}
\end{eqnarray}

\noindent
with
\begin{widetext}
\begin{eqnarray}
h^{\left(2\right) A}_{00}\left(\tau+t^{\ast},\ve{\xi} + c \tau\,\ve{\sigma}\right) &=& + \frac{2\,G}{c^2}
\sum\limits_{l = 0}^{\infty} \frac{\left(-1\right)^l}{l!}
M_{\langle L \rangle}^A \left(\tau+t^{\ast}\right)\;\partial_{\langle L \rangle}\,\frac{1}{r^{\rm N}_A\left(\tau+t^{\ast}\right)}\,,
\label{transformed_global_metric_perturbation_B1}
\end{eqnarray}
\end{widetext}

\noindent
and
\begin{eqnarray}
h^{\left(3\right)}_{0i}\left(\tau+t^{\ast},\ve{\xi} + c \tau\,\ve{\sigma}\right) &=&
\sum\limits_{A=1}^{N} h^{\left(3\right) A}_{0i}\left(\tau+t^{\ast},\ve{\xi} + c \tau \ve{\sigma}\right),
\nonumber\\
\label{transformed_global_metric_perturbation_B2}
\end{eqnarray}

\noindent
with
\begin{widetext}
\begin{eqnarray}
h^{\left(3\right) A}_{0i}\left(\tau+t^{\ast},\ve{\xi} + c \tau\,\ve{\sigma}\right) &=&
+ \frac{4\,G}{c^3}\,\sum\limits_{l=1}^{\infty} \frac{\left(-1\right)^l}{l!}\;\dot{M}^A_{\langle i L-1 \rangle}\left(\tau+t^{\ast}\right)\,
\partial_{\langle L-1 \rangle} \frac{1}{r^{\rm N}_A\left(\tau+t^{\ast}\right)}
\nonumber\\
\nonumber\\
&& + \frac{4\,G}{c^3}\,\sum\limits_{l=1}^{\infty} \frac{\left(-1\right)^l\;l}{\left(l + 1 \right)!}\,
\epsilon_{iab}\,S^A_{\langle b L-1 \rangle}\left(\tau+t^{\ast}\right)\;\partial_{\langle a L-1 \rangle}\,
\frac{1}{r^{\rm N}_A\left(\tau+t^{\ast}\right)}
\nonumber\\
\nonumber\\
&& - \frac{4\,G}{c^3}\,v_A^i\left(\tau+t^{\ast}\right) \sum\limits_{l = 0}^{\infty} \frac{\left(-1\right)^l}{l!}
M_{\langle L \rangle}^A\left(\tau+t^{\ast}\right)\;\partial_{\langle L \rangle}\,
\frac{1}{r^{\rm N}_A\left(\tau+t^{\ast}\right)}\,,
\label{transformed_global_metric_perturbation_C}
\end{eqnarray}
\end{widetext}

\noindent
where the sum in (\ref{transformed_global_metric_perturbation_A}) and (\ref{transformed_global_metric_perturbation_B2}) runs over all massive bodies of the
Solar system. The expressions in Eqs.~(\ref{transformed_global_metric_perturbation_B1}) and (\ref{transformed_global_metric_perturbation_C}) contain the STF
spatial derivative operation $\partial_{\langle L \rangle}$,
which also has to be expressed in terms of these new variables. That issue is considered in detail in appendix \ref{Appendix_Partial_Derivative} and yields
the following expression for the STF partial derivative operation in Eqs.~(\ref{transformed_global_metric_perturbation_B1})
and (\ref{transformed_global_metric_perturbation_C}):
\begin{eqnarray}
\partial_{\langle L \rangle} &=& \underset{i_1 ... i_l}{\rm STF}\,\sum\limits_{p=0}^{l} \frac{l!}{\left(l-p\right)!\;p!}\;
\sum\limits_{q=0}^{p} \left(-1\right)^{q}\frac{p!}{\left(p-q\right)!\;q!}\;
\nonumber\\
\nonumber\\
&& \hspace{-1.2cm} \times\;\sigma^{i_1}\,...\,\sigma^{i_p}\;
P^{i_{p+1}\,j_{p+1}}\;...\;P^{i_l\,j_l}\;
\nonumber\\
&& \hspace{-1.2cm} \times\;\frac{\partial}{\partial \xi^{j_{p+1}}}\;...\;
\frac{\partial}{\partial \xi^{j_{l}}}\;
\left(\frac{\partial}{\partial c\,\tau}\right)^{p-q}
\left(\frac{\partial}{\partial c\,t^{\ast}}\right)^q\,.
\label{Transformation_Derivative_1}
\end{eqnarray}

\noindent
These expressions in (\ref{transformed_global_metric_perturbation_A}) - (\ref{Transformation_Derivative_1}) have to be inserted into the
geodesic equation (\ref{transformed_geodesic_equation_B}), which finally yields the geodesic equation for lightrays
which propagate in the gravitational field of one arbitrarily moving body $A$ in terms of these new variables $\tau$ and $\ve{\xi}$:
\begin{eqnarray}
\frac{\ddot{\ve{x}}\left(\tau+t^{\ast}\right)}{c^2} &=& \sum\limits_{A=1}^N \left[\frac{\ddot{\ve{x}}^{\cal M}_A\left(\tau+t^{\ast}\right)}{c^2}
+ \frac{\ddot{\ve{x}}^{\cal S}_A\left(\tau+t^{\ast}\right)}{c^2}\right]
\nonumber\\
\nonumber\\
&& + {\cal O}\left(c^{-4}\right)\,,
\label{transformed_geodesic_equation_C}
\end{eqnarray}

\noindent
where the indices ${\cal M}$ and ${\cal S}$ stand for mass-multipole and spin-multipole component, respectively.
That means, the linearity of geodesic equation in 1.5PN approximation allows simply to sum over all $N$ arbitrarily moving bodies just straight away.
The contributions due to the mass-multipole structure of one body $A$ is given by
\begin{widetext}
\begin{eqnarray}
\frac{\ddot{x}^{i\;{\cal M}}_A\left(\tau+t^{\ast}\right)}{c^2} &=& + \frac{2\,G}{c^2}\, P^{ij}\,\frac{\partial}{\partial \xi^j}\,
\sum\limits_{l = 0}^{\infty} \frac{\left(-1\right)^l}{l!}\;M_{\langle L \rangle}^A\left(\tau+t^{\ast}\right)
\partial_{\langle L \rangle}\;\frac{1}{r^{\rm N}_A\left(\tau+t^{\ast}\right)}
\nonumber\\
\nonumber\\
&& - \frac{2\,G}{c^2}\,\sigma^i\,\frac{\partial}{\partial c \tau}\,
\sum\limits_{l = 0}^{\infty} \frac{\left(-1\right)^l}{l!}\;M_{\langle L \rangle}^A\left(\tau+t^{\ast}\right)
\partial_{\langle L \rangle}\;\frac{1}{r^{\rm N}_A\left(\tau+t^{\ast}\right)}
\nonumber\\
\nonumber\\
&& - \frac{4\,G}{c^3}\,\frac{\partial}{\partial c \tau}\,
\sum\limits_{l=1}^{\infty} \frac{\left(-1\right)^l}{l!}\;\dot{M}^A_{\langle i L-1 \rangle}\left(\tau+t^{\ast}\right)\,
\partial_{\langle L-1 \rangle} \frac{1}{r^{\rm N}_A\left(\tau+t^{\ast}\right)}
\nonumber\\
\nonumber\\
&& + \frac{4\,G}{c^3}\,\frac{\partial}{\partial c \tau}\,v_A^i\left(\tau+t^{\ast}\right) \sum\limits_{l = 0}^{\infty} \frac{\left(-1\right)^l}{l!}
M_{\langle L \rangle}^A\left(\tau+t^{\ast}\right)\;\partial_{\langle L \rangle}\,
\frac{1}{r^{\rm N}_A\left(\tau+t^{\ast}\right)}
\nonumber\\
\nonumber\\
&& + \frac{4\,G}{c^3}\,\sigma^j P^{ik} \frac{\partial}{\partial \xi^k}\,
\sum\limits_{l=1}^{\infty} \frac{\left(-1\right)^l}{l!}\;\dot{M}^A_{\langle j L-1 \rangle}\left(\tau+t^{\ast}\right)\,
\partial_{\langle L-1 \rangle} \frac{1}{r^{\rm N}_A\left(\tau+t^{\ast}\right)}
\nonumber\\
\nonumber\\
&& - \frac{4\,G}{c^3}\,\sigma^j P^{ik} \frac{\partial}{\partial \xi^k}\,
v_A^j\left(\tau+t^{\ast}\right)\sum\limits_{l = 0}^{\infty} \frac{\left(-1\right)^l}{l!}\,
M_{\langle L \rangle}^A\left(\tau+t^{\ast}\right)\;\partial_{\langle L \rangle}\,
\frac{1}{r^{\rm N}_A\left(\tau+t^{\ast}\right)}\,,
\label{transformed_geodesic_equation_C1}
\end{eqnarray}
\end{widetext}

\noindent
and the contribution due to the spin-multipole structure of one body $A$ reads
\begin{widetext}
\begin{eqnarray}
\frac{\ddot{x}^{i\;{\cal S}}_A\left(\tau+t^{\ast}\right)}{c^2} &=&
- \frac{4\,G}{c^3}\,\frac{\partial}{\partial c \tau}\,\sum\limits_{l=1}^{\infty} \frac{\left(-1\right)^l\;l}{\left(l + 1 \right)!}\,
\epsilon_{iab}\,S^A_{\langle b L-1 \rangle}\left(\tau+t^{\ast}\right)\;\partial_{\langle a L-1 \rangle}\,
\frac{1}{r^{\rm N}_A\left(\tau+t^{\ast}\right)}
\nonumber\\
\nonumber\\
&& + \frac{4\,G}{c^3}\,\sigma^j P^{ik} \frac{\partial}{\partial \xi^k}\,
\sum\limits_{l=1}^{\infty} \frac{\left(-1\right)^l\;l}{\left(l + 1 \right)!}\,
\epsilon_{jab}\,S^A_{\langle b L-1 \rangle}\left(\tau+t^{\ast}\right)\;\partial_{\langle a L-1 \rangle}
\frac{1}{r^{\rm N}_A\left(\tau+t^{\ast}\right)}\,,
\label{transformed_geodesic_equation_C2}
\end{eqnarray}
\end{widetext}

\noindent
where the derivative operator is given by (\ref{Transformation_Derivative_1}).

By Eqs.~(\ref{transformed_geodesic_equation_C}) - (\ref{transformed_geodesic_equation_C2}) the transformation of
geodesic equation in 1.5PN approximation in terms of these new variables $\tau$ and $\ve{\xi}$ has been accomplished, which describes the propagation of
a light-signal through the field of $N$ massive bodies in arbitrary motion and having arbitrary shape and inner structure and which can also rotate
arbitrarily. Before we proceed further, three comments should be in order:

$\left({\bf i}\right)$ First, let us note that the spatial derivative operator in (\ref{Transformation_Derivative_1}) depends on time-variables $\tau$
and $t^{\ast}$, but in such a way that it does not act on time-dependent multipoles or the velocity of the body, that means:
\begin{eqnarray}
\partial_{\langle L \rangle} M^A_{\langle L \rangle}\left(\tau+t^{\ast}\right) &=& 0\,,
\label{Transformation_Derivative_20}
\\
\nonumber\\
\partial_{\langle L \rangle}\,S^A_{\langle L \rangle}\left(\tau+t^{\ast}\right) &=& 0\,,
\label{Transformation_Derivative_25}
\\
\nonumber\\
\partial_{\langle L \rangle}\,\ve{v}_A\left(\tau+t^{\ast}\right) &=& 0\,,
\label{Transformation_Derivative_26}
\end{eqnarray}

\noindent
because the construction of the derivative operator in (\ref{Transformation_Derivative_1}) is such that the derivatives with respect to variable $\tau$
cancel exactly the derivatives with respect to $t^{\ast}$ in all those functions which depend on the combination $\tau+t^{\ast}$. 
But of course $\partial_{\langle L \rangle}\,r_A^{\rm N}\left(\tau+t^{\ast}\right) \neq 0$.  

$\left({\bf ii}\right)$ Second, let us also remark that in (\ref{transformed_geodesic_equation_C1}) the STF notation for the
derivative operator has been kept. But we recall the following relation, which is a specific example of the more general relation Eq.~(A1)
in \cite{Hartmann_Soffel_Kioustelidis}:
\begin{eqnarray}
M^A_{\langle L \rangle}\,\partial_{\langle L \rangle}\,\frac{1}{r^{\rm N}_A\left(\tau+t^{\ast}\right)} &=&
M^A_{\langle L \rangle}\,\partial_{L}\,\frac{1}{r^{\rm N}_A\left(\tau+t^{\ast}\right)}\,,
\label{STF_relation_5}
\\
\nonumber\\
M^A_{\langle i L-1\rangle}\,\partial_{\langle L-1 \rangle}\,\frac{1}{r^{\rm N}_A\left(\tau+t^{\ast}\right)} &=&
M^A_{\langle i L-1 \rangle}\,\partial_{L-1}\,\frac{1}{r^{\rm N}_A\left(\tau+t^{\ast}\right)}\,.
\nonumber\\
\label{STF_relation_6}
\end{eqnarray}

\noindent
The relation in (\ref{STF_relation_5}) has allowed to replace the STF derivative operator $\partial_{\langle L \rangle}$ by $\partial_{L}$
in Eqs.~(100) - (102) in \cite{Zschocke_1PN}. Here, in view of relation (\ref{STF_relation_5}) and (\ref{STF_relation_6})
we may also replace the STF derivative operator $\partial_{\langle L \rangle}$ by $\partial_{L}$ in all terms in (\ref{transformed_geodesic_equation_C1}),
and correspondingly in the first integral in (\ref{First_Integration_9}) and (\ref{First_Integration_10}),
as well as in the second in (\ref{Second_Integration_Mass_4}) and (\ref{Second_Integration_Mass_5}).
On the other side, such replacement is not possible for the spin-multipole terms in (\ref{transformed_geodesic_equation_C2}), because of
\begin{eqnarray}
S^A_{\langle b L-1\rangle} \partial_{\langle a L-1 \rangle} \frac{1}{r^{\rm N}_A\left(\tau+t^{\ast}\right)} &\neq&
S^A_{\langle b L-1 \rangle} \partial_{a L-1} \frac{1}{r^{\rm N}_A\left(\tau+t^{\ast}\right)}\,.
\nonumber\\
\label{STF_relation_7}
\end{eqnarray}

\noindent
$\left({\bf iii}\right)$ Third, it should also be mentioned that in the limit of one massive body at rest with the origin of the coordinate-system
located at the center-of-mass and with time-independent multipoles then the geodesic equation
(\ref{transformed_geodesic_equation_C}) - (\ref{transformed_geodesic_equation_C2}) agrees with the geodesic equation given in \cite{Kopeikin1997}, notice 
the comment in \cite{Communication1}.

\section{First integration of geodesic equation}\label{First_Integration}

The coordinate velocity of the photon is determined by the first integral of geodesic equation (\ref{transformed_geodesic_equation_B}). 
In terms of the new variables we may separate the first integral of 
geodesic equation (\ref{transformed_geodesic_equation_B}) into 1PN and 1.5PN terms as follows:  
\begin{eqnarray}
\frac{\dot{\ve{x}}_{\rm 1.5PN} \left(\tau + t^{\ast}\right)}{c} &=& \ve{\sigma}  
+ \sum\limits_{A=1}^N \frac{\Delta \dot{\ve{x}}^A_{\rm 1PN}\left(\tau + t^{\ast}\right)}{c}  
\nonumber\\
\nonumber\\
&& + \sum\limits_{A=1}^N \frac{\Delta \dot{\ve{x}}^A_{\rm 1.5PN}\left(\tau + t^{\ast}\right)}{c}\,.  
\label{first_integral_geodesic_equation_A}
\end{eqnarray}

\noindent
That means, according to (\ref{transformed_geodesic_equation_C}) we may consider the light-propagation in the field of one arbitrarily moving body $A$  
and finally we have to build the sum over all massive bodies $A=1,...,N$ in order to obtain the light trajectory in the entire Solar system.  
Furthermore, according to Eq.~(\ref{transformed_geodesic_equation_C}) we split these expressions into mass-multipole terms and 
spin-multipole contributions as follows: 
\begin{eqnarray}
&&\hspace{-0.5cm} \frac{\Delta \dot{\ve{x}}_{\rm 1PN}^A \left(\tau + t^{\ast}\right)}{c} 
\!=\! \frac{\Delta \dot{\ve{x}}^{A\,{\cal M}}_{\rm 1PN}\left(\tau + t^{\ast}\right)}{c}\,,  
\label{first_integral_geodesic_equation_1PN}
\\
\nonumber\\ 
&& \hspace{-0.5cm} \frac{\Delta \dot{\ve{x}}_{\rm 1.5PN}^A \left(\tau + t^{\ast}\right)}{c} \!=\! 
\frac{\Delta \dot{\ve{x}}^{A\,{\cal M}}_{\rm 1.5PN}\left(\tau + t^{\ast}\right)}{c} 
\!+\! \frac{\Delta \dot{\ve{x}}^{A\,{\cal S}}_{\rm 1.5PN}\left(\tau + t^{\ast}\right)}{c}\,,  
\nonumber\\
\label{first_integral_geodesic_equation_15PN}
\end{eqnarray}

\noindent
where we have taken into account that in (\ref{first_integral_geodesic_equation_1PN}) there are no spin-multipoles because they are 
of the order ${\cal O}\left(c^{-3}\right)$, hence they do appear only in (\ref{first_integral_geodesic_equation_15PN}).  
We shall consider mass-multipole and spin-multipoles in the next both subsections separately.

\subsection{First integration for mass-multipoles}\label{First_Integration_Mass}  

The first integral of geodesic equation (\ref{transformed_geodesic_equation_C}) for the mass-multipole component of one massive body $A$ reads:  
\begin{eqnarray}
&& \frac{\Delta \dot{\ve{x}}^{A\,{\cal M}}_{\rm 1PN}\left(\tau + t^{\ast}\right)}{c} 
+ \frac{\Delta \dot{\ve{x}}_{\rm 1.5PN}^{A\,{\cal M}} \left(\tau + t^{\ast}\right)}{c}  
\nonumber\\
\nonumber\\
&& = \int\limits_{-\infty}^{\tau}\,d c\tau^{\prime}\,\frac{\ddot{\ve{x}}^{\cal M}_A\left(\tau^{\prime} + t^{\ast}\right)}{c^2}\,,  
\label{Additional_Equation_A}
\end{eqnarray}

\noindent
where the integrand up to the required order is given by Eq.~(\ref{transformed_geodesic_equation_C1}). Let us underline that the integration of the  
first expression on the r.h.s. in (\ref{transformed_geodesic_equation_C1}) yields terms of the order ${\cal O}\left(c^{-2}\right)$ as well as  
terms of the order ${\cal O}\left(c^{-3}\right)$. For that reason, the integral in (\ref{Additional_Equation_A}) is written as sum of 1PN and 1.5PN terms.  
In particular, for the integration of geodesic equation the following rules are important 
(cf. Eqs.~(4.9) and (4.10) in \cite{KopeikinKorobkovPolnarev2006} or Eqs.~(4.38) and (4.39) in \cite{KopeikinKorobkov2005}):  
\begin{eqnarray}
\int d c \tau^{\prime}\;\frac{\partial}{\partial c \tau^{\prime}}\,F\left(\tau^{\prime}, \ve{\xi}\right) &=& F\left(\tau^{\prime}, \ve{\xi}\right) 
+ C\left(\ve{\xi}\right), 
\label{Integration_Rule_A}
\\
\nonumber\\
\int d c \tau^{\prime}\;\frac{\partial}{\partial \xi^i}\,F\left(\tau^{\prime},\ve{\xi}\right) &=& \frac{\partial}{\partial \xi^i} \int d c \tau^{\prime}\, 
F\left(\tau^{\prime}, \ve{\xi}\right),   
\label{Integration_Rule_B}
\end{eqnarray}

\noindent
where the function $C\left(\ve{\xi}\right)$ in (\ref{Integration_Rule_A}) depends only on variable $\ve{\xi}$, thence disappears in case of definite integrals. 
The rule in (\ref{Integration_Rule_A}) and (\ref{Integration_Rule_B}) are valid if one integrates along the unperturbed light trajectory, where the derivative  
with respect to integration variable $c\,\tau^{\prime}$ acts like a total derivative; see also the first comment below Eq.~(\ref{Transformation_Derivative_2B})  
and the corresponding explanations made by Eq.~(1.19) - (1.23) in \cite{KopeikinKorobkov2005}.  
 
The integration of the first expression  
on the r.h.s. in (\ref{transformed_geodesic_equation_C1}) is shown in more detail in appendix \ref{Integral_1}, while in view of relation (\ref{Integration_Rule_A}) 
the integrals of the second, third, and fourth expression in (\ref{transformed_geodesic_equation_C1}) are straightforward.  
The fifth term in (\ref{transformed_geodesic_equation_C1}) can be integrated by parts  
using relation (\ref{Relation_1}) and is shown in more detail in appendix \ref{Integral_2}, while the integration of the sixth term goes very similar. 
Altogether, for the 1PN terms one obtains:  
\begin{widetext}
\begin{eqnarray}
\frac{\Delta \dot{x}^{i\;{\cal M}}_{A\;{\rm 1PN}} \left(\tau+t^{\ast}\right)}{c} &=&
- \frac{2\,G}{c^2}\,
\sum\limits_{l = 0}^{\infty} \frac{\left(-1\right)^l}{l!}\;M_{\langle L \rangle}^A\left(\tau+t^{\ast}\right)
\partial_{\langle L \rangle}\;
\frac{d_A^i\left(\tau+t^{\ast}\right)}{r^{\rm N}_A\left(\tau+t^{\ast}\right)-\ve{\sigma}\cdot\ve{r}^{\rm N}_A\left(\tau+t^{\ast}\right)}
\,\frac{1}{r^{\rm N}_A\left(\tau+t^{\ast}\right)} 
\nonumber\\
\nonumber\\
&& - \frac{2\,G}{c^2}\,\sigma^i
\sum\limits_{l = 0}^{\infty} \frac{\left(-1\right)^l}{l!}\;M_{\langle L \rangle}^A\left(\tau+t^{\ast}\right)
\partial_{\langle L \rangle}\;\frac{1}{r^{\rm N}_A\left(\tau+t^{\ast}\right)} 
\nonumber\\
\nonumber\\
&& + {\cal O}\left(\frac{v_A}{c}\,\dot{M}_L^A\right) + {\cal O}\left(\ddot{M}_L^A\right) + {\cal O}\left(\frac{v_A^2}{c^2} M_L^A\right).  
\label{First_Integration_9}
\end{eqnarray}
\end{widetext}

\noindent 
In the first term on the r.h.s. in (\ref{First_Integration_9}) we have used relation (\ref{Relation_2}).  
For the 1.5PN terms one obtains:  
\begin{widetext}
\begin{eqnarray}
\frac{\Delta \dot{x}^{i\;{\cal M}}_{A\;{\rm 1.5PN}} \left(\tau+t^{\ast}\right)}{c} &=&
+ \frac{2\,G}{c^3}\, 
\sum\limits_{l = 1}^{\infty} \frac{\left(-1\right)^l}{l!}\;\dot{M}_{\langle L \rangle}^A\left(\tau+t^{\ast}\right)
\partial_{\langle L \rangle}\;
\frac{d_A^i\left(\tau+t^{\ast}\right)}{r^{\rm N}_A\left(\tau+t^{\ast}\right)-\ve{\sigma}\cdot\ve{r}^{\rm N}_A\left(\tau+t^{\ast}\right)}
\nonumber\\
\nonumber\\
&& \hspace{-3.0cm} + \frac{2\,G}{c^3}\, 
\ve{\sigma} \cdot \ve{v}_A\left(\tau+t^{\ast}\right)  
\sum\limits_{l = 0}^{\infty} \frac{\left(-1\right)^l}{l!}\;M_{\langle L \rangle}^A\left(\tau+t^{\ast}\right)
\partial_{\langle L \rangle}\;
\frac{d_A^i\left(\tau+t^{\ast}\right)}{r^{\rm N}_A\left(\tau+t^{\ast}\right)-\ve{\sigma}\cdot\ve{r}^{\rm N}_A\left(\tau+t^{\ast}\right)}
\,\frac{1}{r^{\rm N}_A\left(\tau+t^{\ast}\right)} 
\nonumber\\
\nonumber\\
&& \hspace{-3.0cm} - \frac{2\,G}{c^3}\, P^{ij}\,\frac{\partial}{\partial \xi^j}\,
\sum\limits_{l = 0}^{\infty} \frac{\left(-1\right)^l}{l!}\;M_{\langle L \rangle}^A\left(\tau+t^{\ast}\right)
\partial_{\langle L \rangle}\;
\frac{\ve{v}_A\left(\tau+t^{\ast}\right) \cdot \ve{d}_A\left(\tau+t^{\ast}\right)}
{r^{\rm N}_A\left(\tau+t^{\ast}\right) - \ve{\sigma} \cdot \ve{r}^{\rm N}_A\left(\tau+t^{\ast}\right)} 
\nonumber\\
\nonumber\\
&& \hspace{-3.0cm} - \frac{4\,G}{c^3}
\sum\limits_{l=1}^{\infty} \frac{\left(-1\right)^l}{l!}\;\dot{M}^A_{\langle i L-1 \rangle}\left(\tau+t^{\ast}\right)\,
\partial_{\langle L-1 \rangle} \frac{1}{r^{\rm N}_A\left(\tau+t^{\ast}\right)}
+ \frac{4\,G}{c^3}\,v_A^i\left(\tau+t^{\ast}\right) \sum\limits_{l = 0}^{\infty} \frac{\left(-1\right)^l}{l!}
M_{\langle L \rangle}^A\left(\tau+t^{\ast}\right)\;\partial_{\langle L \rangle}\,
\frac{1}{r^{\rm N}_A\left(\tau+t^{\ast}\right)}
\nonumber\\
\nonumber\\
&& \hspace{-3.0cm} - \frac{4\,G}{c^3}\,\sigma^j 
\sum\limits_{l=1}^{\infty} \frac{\left(-1\right)^l}{l!}\;\dot{M}^A_{\langle j L-1 \rangle}\left(\tau+t^{\ast}\right)\,
\partial_{\langle L-1 \rangle}\,
\frac{d_A^i\left(\tau+t^{\ast}\right)}{r^{\rm N}_A\left(\tau+t^{\ast}\right)-\ve{\sigma}\cdot\ve{r}^{\rm N}_A\left(\tau+t^{\ast}\right)} 
\,\frac{1}{r^{\rm N}_A\left(\tau+t^{\ast}\right)} 
\nonumber\\
\nonumber\\
&& \hspace{-3.0cm} + {\cal O}\left(\frac{v_A}{c}\,\dot{M}_L^A\right) + {\cal O}\left(\ddot{M}_L^A\right) + {\cal O}\left(\frac{v_A^2}{c^2} M_L^A\right),   
\label{First_Integration_10}
\end{eqnarray}
\end{widetext}

\noindent
where we recall $\dot{M}_A = 0$.  
In the second and sixth term on the r.h.s. in (\ref{First_Integration_10}) we have used (\ref{Relation_2}), while in the first term 
on the r.h.s. in (\ref{First_Integration_10}) we have used relation (\ref{Relation_3}).   
For the third term in (\ref{First_Integration_10}) one might want to use relation (\ref{Relation_4}), but 
actually it does not simplify that expression significantly. 
The derivative operator $\partial_{\langle L \rangle}$ in (\ref{First_Integration_9}) and (\ref{First_Integration_10}) in terms of the 
new variables $\ve{\xi}$, $\tau$, $t^{\ast}$ is given by (\ref{Transformation_Derivative_1}). 

Let us recall, that in 1PN approximation the derivative operator in Eq.~(\ref{Transformation_Derivative_1}) can be replaced by the expression 
in (\ref{Simplified_Differentialoperator}), because the derivatives with respect to variable $t^{\ast}$ produce terms of the order  
${\cal O}\left(c^{-3}\right)$; see also text below Eq.~(\ref{Appendix_Partial_Derivative_10}). Then, keeping in mind relation (\ref{STF_relation_5}),  
one may easily show that the 1PN expression in Eq.~(\ref{First_Integration_9}) agrees with Eq.~(111) in \cite{Zschocke_1PN}. 
In \cite{Zschocke_1PN} it has been demonstrated that in case of bodies at rest ($\ve{v}_A=0$) having time-independent  
mass-multipoles and located at the origin of coordinate system ($\ve{x}_A=0$) our result in (\ref{First_Integration_9}) 
agrees with the time-derivative of Eqs.~(33) and (36) in \cite{Kopeikin1997}. It should also be noticed that the derivative of (\ref{First_Integration_10})  
with respect to variable $c\tau$ yields the expression in (\ref{transformed_geodesic_equation_C1}).  

\subsection{First integration for spin-multipoles}\label{First_Integration_Spin}  

The first integral of geodesic equation (\ref{transformed_geodesic_equation_C}) for the spin-multipole component of one massive body $A$ reads: 
\begin{eqnarray}
\frac{\Delta \dot{\ve{x}}_{\rm 1.5PN}^{A\;{\cal S}} \left(\tau + t^{\ast}\right)}{c} &=&  
\int\limits_{-\infty}^{\tau}\,d c\tau^{\prime}\,\frac{\ddot{\ve{x}}^{\cal S}_A\left(\tau^{\prime} + t^{\ast}\right)}{c^2}\,,  
\label{Additional_Equation_B}
\end{eqnarray}

\noindent
where the integrand up to the required order is given by Eq.~(\ref{transformed_geodesic_equation_C2}). The integration in (\ref{Additional_Equation_B}) can be 
performed straightforward and one obtains:  
\begin{widetext}
\begin{eqnarray}
\frac{\Delta \dot{x}^{i\;{\cal S}}_{A\;{\rm 1.5PN}} \left(\tau+t^{\ast}\right)}{c} &=&
- \frac{4\,G}{c^3}\,\sum\limits_{l=1}^{\infty} \frac{\left(-1\right)^l\;l}{\left(l + 1 \right)!}\,
\epsilon_{iab}\,S^A_{\langle b L-1 \rangle}\left(\tau+t^{\ast}\right)\;\partial_{\langle a L-1 \rangle}\,
\frac{1}{r^{\rm N}_A\left(\tau+t^{\ast}\right)}
\nonumber\\
\nonumber\\
&& \hspace{-2.0cm} - \frac{4\,G}{c^3}\,\sigma^j 
\sum\limits_{l=1}^{\infty} \frac{\left(-1\right)^l\;l}{\left(l + 1 \right)!}\,
\epsilon_{jab}\,S^A_{\langle b L-1 \rangle}\left(\tau+t^{\ast}\right)\;\partial_{\langle a L-1 \rangle}\,
\frac{d_A^i\left(\tau+t^{\ast}\right)}{r^{\rm N}_A\left(\tau+t^{\ast}\right)-\ve{\sigma}\cdot\ve{r}^{\rm N}_A\left(\tau+t^{\ast}\right)}
\,\frac{1}{r^{\rm N}_A\left(\tau+t^{\ast}\right)} 
\nonumber\\
\nonumber\\
&& \hspace{-2.0cm} + {\cal O}\left(\dot{S}_L^A\right) + {\cal O}\left(\frac{v_A}{c}\,S_L^A\right).  
\label{First_Integration_15}
\end{eqnarray}
\end{widetext}

\noindent
Let us remark that the second term in (\ref{First_Integration_15}) is obtained by integration by parts, using (\ref{Relation_1})  
and afterwards making use of relation (\ref{Relation_2}).  
Note that the derivative operator $\partial_{\langle L \rangle}$ in terms of the new variables $\ve{\xi}$, $\tau$, $t^{\ast}$ 
is given by (\ref{Transformation_Derivative_1}). In the appendix \ref{Appendix_Spin} it is shown that 
in the limit of bodies at rest and stationary spin-multipoles our result in (\ref{First_Integration_15})  
agrees with Eqs.~(32) and (37) in \cite{Kopeikin1997}, up to an overall sign which has been clarified \cite{Communication1}. We also note that  
the derivative of (\ref{First_Integration_15}) with respect to variable $c\tau$ yields the expression in (\ref{transformed_geodesic_equation_C2}).  

Let us remark that neglecting terms of the order ${\cal O}\left(\ddot{M}_L^A\right)$,  
$\displaystyle {\cal O}\left(\frac{v_A}{c}\,\dot{M}_L^A\right)$ and $\displaystyle {\cal O}\left(\frac{v_A^2}{c^2}\,M_L^A\right)$ 
in Eqs.~(\ref{First_Integration_9}) and (\ref{First_Integration_10}), 
and neglecting terms of the order ${\cal O}\left(\dot{S}_L^A\right)$ and $\displaystyle {\cal O}\left(\frac{v_A}{c}\,S_L^A\right)$ 
in Eq.~(\ref{First_Integration_15})  
is consistent with the fact that the DSX metric in Eqs.~(\ref{global_metric_perturbation_A}) - (\ref{global_metric_perturbation_C})
does also not contain such terms because they are beyond 1.5PN approximation.

\section{Second integration of geodesic equation}\label{Second_Integration}

The light trajectory of the photon is determined by the second integration of geodesic equation (\ref{transformed_geodesic_equation_B}), 
and can be written as follows:  
\begin{eqnarray}
\ve{x}_{\rm 1.5PN} \left(\tau + t^{\ast}\right) &=& \ve{\xi} + c \tau \ve{\sigma} 
+ \sum\limits_{A=1}^N \Delta \ve{x}^A_{\rm 1PN}\left(\tau+t^{\ast},\tau_0+t^{\ast}\right)
\nonumber\\
&& + \sum\limits_{A=1}^N \Delta \ve{x}^A_{\rm 1.5PN}\left(\tau+t^{\ast}, \tau_0+t^{\ast}\right),  
\label{second_integral_geodesic_equation}
\end{eqnarray}

\noindent
where the sum runs over all massive bodies $A=1,...,N$ of the Solar system. 
Like in the case of first integration, we split these expressions into mass-multipole terms and spin-multipole contributions as follows:
\begin{eqnarray}
\Delta \ve{x}_{\rm 1PN}^A \left(\tau + t^{\ast},\tau_0+t^{\ast}\right)
&=& \Delta \ve{x}^{A\,{\cal M}}_{\rm 1PN}\left(\tau + t^{\ast},\tau_0+t^{\ast}\right),
\nonumber\\
\label{second_integral_geodesic_equation_1PN}
\\
\nonumber\\
\Delta \ve{x}_{\rm 1.5PN}^A \left(\tau + t^{\ast},\tau_0+t^{\ast}\right) &=& 
\Delta \ve{x}^{A\,{\cal M}}_{\rm 1.5PN}\left(\tau + t^{\ast},\tau_0+t^{\ast}\right)
\nonumber\\
\nonumber\\
&& \hspace{-1.0cm} + \Delta \ve{x}^{A\,{\cal S}}_{\rm 1.5PN}\left(\tau + t^{\ast},\tau_0+t^{\ast}\right), 
\label{second_integral_geodesic_equation_15PN}
\end{eqnarray}

\noindent
where in (\ref{second_integral_geodesic_equation_1PN}) there are no spin-multipoles because they are terms  
of the order ${\cal O}\left(c^{-3}\right)$ and consequently they do appear only in (\ref{second_integral_geodesic_equation_15PN}).
We will consider the mass-multipole and the spin-multipole components separately.   

\subsection{Second integration for mass-multipoles}\label{Second_Integration_Mass}  

The mass-multipole terms in (\ref{second_integral_geodesic_equation_1PN}) and (\ref{second_integral_geodesic_equation_15PN}) read  
\begin{eqnarray}
&& \Delta \ve{x}^{A\,{\cal M}}_{\rm 1PN}\left(\tau + t^{\ast},\tau_0+t^{\ast}\right) 
+ \Delta \ve{x}^{A\,{\cal M}}_{\rm 1.5PN}\left(\tau + t^{\ast},\tau_0+t^{\ast}\right) 
\nonumber\\ 
\nonumber\\ 
&&\!=\! \int\limits_{\tau_0}^{\tau} d c\tau^{\prime} \left[\frac{\Delta \dot{\ve{x}}^{A\,{\cal M}}_{{\rm 1PN}}\left(\tau^{\prime}+t^{\ast}\right)}{c} 
\!+\! \frac{\Delta \dot{\ve{x}}^{{A\,\cal M}}_{{\rm 1.5PN}}\left(\tau^{\prime}+t^{\ast}\right)}{c}\right],  
\nonumber\\
\label{Second_Integration_10}
\end{eqnarray}

\noindent
where the first and second integrand on the r.h.s. in (\ref{Second_Integration_10}) is given by Eq.~(\ref{First_Integration_9}) and (\ref{First_Integration_10}), 
respectively. Let us underline, that the integration of the first integrand yields terms of the order ${\cal O}\left(c^{-2}\right)$ as well as of the order  
${\cal O}\left(c^{-3}\right)$. Therefore, the integral in (\ref{Second_Integration_10}) is written as sum of 1PN and 1.5PN terms, while after the integration 
one may separate the 1PN and 1.5PN terms. Inserting (\ref{First_Integration_9}) and (\ref{First_Integration_10}) into (\ref{Second_Integration_10}) yields  
all in all $8$ integrals $I_3\,...\,I_{10}$.  
In favor of clear arrangement, each of these integrals is considered separately in the appendix \ref{Appendix_Integrals_Mass_Multipoles}, 
and their solutions are given by Eqs.~(\ref{Integral_A_15}), (\ref{Integral_B_10}), (\ref{Integral_C_15}), (\ref{Integral_D_10}), 
(\ref{Integral_E_10}), (\ref{Integral_F_10}), (\ref{Integral_G_10}), (\ref{Integral_H_10}).  
Altogether, for the mass-multipole terms to order ${\cal O}\left(c^{-2}\right)$ we obtain: 
\begin{widetext}
\begin{eqnarray}
\Delta \ve{x}^{{A\,\cal M}}_{{\rm 1PN}}\left(\tau+t^{\ast},\tau_0+t^{\ast}\right) &=& \Delta \ve{x}^{{A\,\cal M}}_{{\rm 1PN}}\left(\tau+t^{\ast}\right)
- \Delta \ve{x}^{{A\,\cal M}}_{{\rm 1PN}}\left(\tau_0+t^{\ast}\right),
\nonumber\\
\nonumber\\
\nonumber\\
\Delta x^{i\,{\cal M}}_{A\;{\rm 1PN}}\left(\tau+t^{\ast}\right) &=&
- \frac{2\,G}{c^2}\,
\sum\limits_{l = 0}^{\infty} \frac{\left(-1\right)^l}{l!}\;M_{\langle L \rangle}^A\left(\tau+t^{\ast}\right)
\partial_{\langle L \rangle}\;
\frac{d^i_A\left(\tau+t^{\ast}\right)}{r^{\rm N}_A\left(\tau+t^{\ast}\right)-\ve{\sigma}\cdot\ve{r}^{\rm N}_A\left(\tau+t^{\ast}\right)} 
\nonumber\\
\nonumber\\
&& + \frac{2\,G}{c^2}\,\sigma^i\,
\sum\limits_{l = 0}^{\infty} \frac{\left(-1\right)^l}{l!}\;
M_{\langle L \rangle}^A\left(\tau+t^{\ast}\right)
\partial_{\langle L \rangle}\;
\ln \left[r^{\rm N}_A\left(\tau+t^{\ast}\right) - \ve{\sigma}\cdot\ve{r}^{\rm N}_A\left(\tau+t^{\ast}\right)\right] 
\nonumber\\
\nonumber\\
&& + {\cal O}\left(\frac{v_A}{c}\,\dot{M}_L^A\right) + {\cal O}\left(\ddot{M}_L^A\right) + {\cal O}\left(\frac{v_A^2}{c^2} M_L^A\right). 
\label{Second_Integration_Mass_4}
\end{eqnarray}
\end{widetext}

\noindent
For the mass-multipole terms to order ${\cal O}\left(c^{-3}\right)$ one obtains:  
\begin{widetext}
\begin{eqnarray}
\Delta \ve{x}^{{A\,\cal M}}_{{\rm 1.5PN}}\left(\tau+t^{\ast},\tau_0+t^{\ast}\right) &=& \Delta \ve{x}^{{A\,\cal M}}_{{\rm 1.5PN}}\left(\tau+t^{\ast}\right)
- \Delta \ve{x}^{{A\,\cal M}}_{{\rm 1.5PN}}\left(\tau_0+t^{\ast}\right),
\nonumber\\
\nonumber\\
\nonumber\\
\Delta x^{i\,{\cal M}}_{A\;{\rm 1.5PN}}\left(\tau+t^{\ast}\right) &=& + \frac{2\,G}{c^3}\,
\sum\limits_{l = 1}^{\infty} \frac{\left(-1\right)^l}{l!}\;\dot{M}_{\langle L \rangle}^A\left(\tau+t^{\ast}\right)
\partial_{\langle L \rangle}\;
d_A^i\left(\tau+t^{\ast}\right)\,
\frac{\ve{\sigma}\cdot\ve{r}^{\rm N}_A\left(\tau+t^{\ast}\right)}{r^{\rm N}_A\left(\tau+t^{\ast}\right)-\ve{\sigma}\cdot\ve{r}^{\rm N}_A\left(\tau+t^{\ast}\right)}
\nonumber\\
\nonumber\\
&& \hspace{-3.0cm} - \frac{2\,G}{c^3}\,
\sum\limits_{l = 1}^{\infty} \frac{\left(-1\right)^l}{l!}\;\dot{M}_{\langle L \rangle}^A\left(\tau+t^{\ast}\right)
\partial_{\langle L \rangle}\;
d_A^i\left(\tau+t^{\ast}\right)\,
\ln \left[r^{\rm N}_A\left(\tau+t^{\ast}\right)-\ve{\sigma}\cdot\ve{r}^{\rm N}_A\left(\tau+t^{\ast}\right)\right]
\nonumber\\
\nonumber\\
&& \hspace{-3.0cm} - \frac{2\,G}{c^3}\,P^{ij}\,\frac{\partial}{\partial \xi^j}\,
\sum\limits_{l = 0}^{\infty} \frac{\left(-1\right)^l}{l!}\;M_{\langle L \rangle}^A\left(\tau+t^{\ast}\right)
\partial_{\langle L \rangle}\;
\ve{v}_A\left(\tau+t^{\ast}\right) \cdot \ve{d}_A\left(\tau+t^{\ast}\right)\,
\frac{\ve{\sigma}\cdot\ve{r}^{\rm N}_A\left(\tau+t^{\ast}\right)}{r^{\rm N}_A\left(\tau+t^{\ast}\right)-\ve{\sigma} \cdot \ve{r}^{\rm N}_A\left(\tau+t^{\ast}\right)}
\nonumber\\
\nonumber\\
&& \hspace{-3.0cm} + \frac{2\,G}{c^3}\,P^{ij}\,\frac{\partial}{\partial \xi^j}\,
\sum\limits_{l = 0}^{\infty} \frac{\left(-1\right)^l}{l!}\;M_{\langle L \rangle}^A\left(\tau+t^{\ast}\right)
\partial_{\langle L \rangle}\;
\ve{v}_A\left(\tau+t^{\ast}\right) \cdot \ve{d}_A\left(\tau+t^{\ast}\right)\,
\ln \left[r^{\rm N}_A\left(\tau+t^{\ast}\right)-\ve{\sigma} \cdot \ve{r}^{\rm N}_A\left(\tau+t^{\ast}\right)\right]
\nonumber\\
\nonumber\\
&& \hspace{-3.0cm} - \frac{2\,G}{c^3}\,\sigma^i 
\sum\limits_{l = 1}^{\infty} \frac{\left(-1\right)^l}{l!}\, 
\dot{M}_{\langle L \rangle}^A\left(\tau+t^{\ast}\right)
\partial_{\langle L \rangle}  
\bigg[r^{\rm N}_A\left(\tau+t^{\ast}\right) + \ve{\sigma}\cdot \ve{r}^{\rm N}_A\left(\tau+t^{\ast}\right)
\ln \left[r^{\rm N}_A\left(\tau+t^{\ast}\right) - \ve{\sigma}\cdot\ve{r}^{\rm N}_A\left(\tau+t^{\ast}\right)\right]\bigg]
\nonumber\\
\nonumber\\
&& \hspace{-3.0cm} + \frac{2\,G}{c^3}\,\sigma^i\,\ve{\sigma}\cdot\ve{v}_A\left(\tau+t^{\ast}\right)\,
\sum\limits_{l = 0}^{\infty} \frac{\left(-1\right)^l}{l!}\;
M_{\langle L \rangle}^A\left(\tau+t^{\ast}\right)
\partial_{\langle L \rangle}\;
\ln \left[r^{\rm N}_A\left(\tau+t^{\ast}\right) - \ve{\sigma}\cdot\ve{r}^{\rm N}_A\left(\tau+t^{\ast}\right)\right]
\nonumber\\
\nonumber\\
&& \hspace{-3.0cm} + \frac{2\,G}{c^3}\,\sigma^i\,
\sum\limits_{l = 0}^{\infty} \frac{\left(-1\right)^l}{l!}\;
M_{\langle L \rangle}^A\left(\tau+t^{\ast}\right)
\partial_{\langle L \rangle}\,
\frac{\ve{v}_A\left(\tau+t^{\ast}\right)\cdot\ve{d}_A\left(\tau+t^{\ast}\right)}
{r^{\rm N}_A\left(\tau+t^{\ast}\right) - \ve{\sigma}\cdot \ve{r}^{\rm N}_A\left(\tau+t^{\ast}\right)}
\nonumber\\
\nonumber\\
&& \hspace{-3.0cm} + \frac{4\,G}{c^3}
\sum\limits_{l=1}^{\infty} \frac{\left(-1\right)^l}{l!}\;\dot{M}^A_{\langle i L-1 \rangle}\left(\tau+t^{\ast}\right)\,
\partial_{\langle L-1 \rangle}
\ln \left[r^{\rm N}_A\left(\tau + t^{\ast}\right) - \ve{\sigma}\cdot \ve{r}^{\rm N}_A\left(\tau+t^{\ast}\right)\right]
\nonumber\\
\nonumber\\
&& \hspace{-3.0cm} - \frac{4\,G}{c^3}\,v_A^i\left(\tau+t^{\ast}\right)\sum\limits_{l = 0}^{\infty} \frac{\left(-1\right)^l}{l!}
M_{\langle L \rangle}^A\left(\tau+t^{\ast}\right)\;\partial_{\langle L \rangle}\,
\ln \left[r^{\rm N}_A\left(\tau + t^{\ast}\right) - \ve{\sigma}\cdot \ve{r}^{\rm N}_A\left(\tau+t^{\ast}\right)\right]
\nonumber\\
\nonumber\\
&& \hspace{-3.0cm} - \frac{4\,G}{c^3}\,\sigma^j\,
\sum\limits_{l=1}^{\infty} \frac{\left(-1\right)^l}{l!}\;\dot{M}^A_{\langle j L-1 \rangle}\left(\tau+t^{\ast}\right)\,
\partial_{\langle L-1 \rangle}
\frac{d_A^i\left(\tau+t^{\ast}\right)}{r_A^{\rm N}\left(\tau+t^{\ast}\right) - \ve{\sigma}\cdot \ve{r}_A^{\rm N}\left(\tau+t^{\ast}\right)}
\nonumber\\
\nonumber\\
&& \hspace{-3.0cm} + {\cal O}\left(\frac{v_A}{c}\,\dot{M}_L^A\right) + {\cal O}\left(\ddot{M}_L^A\right) + {\cal O} \left(\frac{v_A^2}{c^2}\,M_L^A\right).  
\label{Second_Integration_Mass_5}
\end{eqnarray}
\end{widetext}

\noindent
Notice, that the derivative operator $\partial_{\langle L \rangle}$ in (\ref{Second_Integration_Mass_4}) and (\ref{Second_Integration_Mass_5}) in terms of the 
new variables $\ve{\xi}$, $\tau$, $t^{\ast}$ is given by (\ref{Transformation_Derivative_1}).  
One may demonstrate, that (\ref{Second_Integration_Mass_4}) and (\ref{Second_Integration_Mass_5}) are  
consistent with (\ref{First_Integration_9}) and (\ref{First_Integration_10}). That means,  
the derivative of (\ref{Second_Integration_Mass_4}) and (\ref{Second_Integration_Mass_5}) with respect to variable $\tau$ coincides with the expressions 
in (\ref{First_Integration_9}) and (\ref{First_Integration_10}) up to terms of the order ${\cal O}\left(c^{-4}\right)$.  
For such a proof one has to use the relations (\ref{Relation_1}) and (\ref{Integral_A_20}) and one must take into account (\ref{time_derivative_impact_vector})  
and (\ref{time_derivative_mass_multipoles}). 

The 1PN solution in Eq.~(\ref{Second_Integration_Mass_4}) coincides with Eq.~(137) in \cite{Zschocke_1PN}. Recall that in 1PN approximation the  
derivative operator in Eq.~(\ref{Transformation_Derivative_1}) can be replaced by the simplified expression in Eq.~(\ref{Simplified_Differentialoperator}) 
(cf. Eq.~(101) in \cite{Zschocke_1PN}), because derivatives with respect to variable $t^{\ast}$ generate terms of the order ${\cal O}\left(c^{-3}\right)$;  
see also comments below Eq.~(\ref{Appendix_Partial_Derivative_10}). Furthermore, in \cite{Zschocke_1PN} it has already been shown that in case of bodies at rest  
and located at the origin of coordinate system our result in (\ref{Second_Integration_Mass_4}) agrees with Eqs.~(33) and (36) in \cite{Kopeikin1997}.

\subsection{Second integration for spin-multipoles}\label{Second_Integration_Spin}  

The spin-multipole terms in (\ref{second_integral_geodesic_equation_15PN}) read  
\begin{eqnarray}
\Delta\ve{x}^{{A\,\cal S}}_{{\rm 1.5PN}}\left(\tau+t^{\ast},\tau_0+t^{\ast}\right) &=& 
\int\limits_{\tau_0}^{\tau}\,d c\tau^{\prime}\,\frac{\Delta \dot{\ve{x}}^{\cal S}_A\left(\tau^{\prime}+t^{\ast}\right)}{c}\,, 
\nonumber\\
\label{Second_Integration_15}
\end{eqnarray}

\noindent
where the integrand in (\ref{Second_Integration_15}) is given by the expressions in Eq.~(\ref{First_Integration_15}).   
The second expression on the r.h.s. in Eq.~(\ref{First_Integration_15}) is rewritten by means of relation (\ref{Relation_2}) and then, by means  
of relations (\ref{Appendix_Integral_I1_25}) and (\ref{Integral_A_6}), we may integrate by parts. We obtain the following solution:  
\begin{widetext}
\begin{eqnarray}
\Delta\ve{x}^{{A\,\cal S}}_{{\rm 1.5PN}}\left(\tau+t^{\ast},\tau_0+t^{\ast}\right) &=& 
\Delta\ve{x}^{{A\,\cal S}}_{{\rm 1.5PN}}\left(\tau+t^{\ast}\right) - \Delta\ve{x}^{{A\,\cal S}}_{{\rm 1.5PN}}\left(\tau_0+t^{\ast}\right),
\nonumber\\
\nonumber\\
\Delta x^{i\,{\cal S}}_{A\,{\rm 1.5PN}}\left(\tau+t^{\ast}\right) &=&
+ \frac{4\,G}{c^3}\,\sum\limits_{l=1}^{\infty} \frac{\left(-1\right)^l\;l}{\left(l + 1 \right)!}\,
\epsilon_{iab}\,S^A_{\langle b L-1 \rangle}\left(\tau+t^{\ast}\right)\;\partial_{\langle a L-1 \rangle}\,
\ln \left[r^{\rm N}_A\left(\tau+t^{\ast}\right) - \ve{\sigma}\cdot \ve{r}^{\rm N}_A\left(\tau+t^{\ast}\right)\right] 
\nonumber\\
\nonumber\\
&& - \frac{4\,G}{c^3}\,\sigma^j  
\sum\limits_{l=1}^{\infty} \frac{\left(-1\right)^l\;l}{\left(l + 1 \right)!}\,
\epsilon_{jab}\,S^A_{\langle b L-1 \rangle}\left(\tau+t^{\ast}\right)\;\partial_{\langle a L-1 \rangle}\,
\frac{d_A^i\left(\tau+t^{\ast}\right)}{r_A^{\rm N}\left(\tau+t^{\ast}\right) - \ve{\sigma}\cdot \ve{r}_A^{\rm N}\left(\tau+t^{\ast}\right)}  
\nonumber\\
\nonumber\\
&& + {\cal O}\left(\dot{S}_L^A\right) + {\cal O}\left(\frac{v_A}{c}\,S_L^A\right),  
\label{Second_Integration_Spin_5}
\end{eqnarray}
\end{widetext}

\noindent
where for the second expression we also have used relation (\ref{Relation_3}).   
The derivative operator $\partial_{\langle L \rangle}$ in terms of the new variables $\ve{\xi}$, $\tau$, $t^{\ast}$ 
is given by (\ref{Transformation_Derivative_1}).
One may easily check, that (\ref{Second_Integration_Spin_5}) is consistent with (\ref{First_Integration_15}), in the sense that  
the derivative of (\ref{Second_Integration_Spin_5}) with respect to variable $\tau$ just yields the expression in (\ref{First_Integration_15}) 
up to terms of the order ${\cal O}\left(c^{-4}\right)$. For that proof  
simply apply the relations (\ref{Relation_1}) and (\ref{Integral_A_20}) and take into account (\ref{time_derivative_impact_vector}).  
Furthermore, in appendix \ref{Appendix_Spin} it is shown that  
in the limit of bodies at rest and time-independent spin-multipoles our result in (\ref{Second_Integration_Spin_5})
agrees with Eqs.~(33) and (38) in \cite{Kopeikin1997}, up to an overall sign which has been clarified \cite{Communication1}.  

We underline again that neglecting terms of the order ${\cal O}\left(\ddot{M}_L^A\right)$,  
$\displaystyle {\cal O}\left(\frac{v_A}{c}\,\dot{M}_L^A\right)$ and $\displaystyle {\cal O}\left(\frac{v_A^2}{c^2}\,M_L^A\right)$  
in Eqs.~(\ref{Second_Integration_Mass_4}) and (\ref{Second_Integration_Mass_5}),
and the neglecting terms of the order ${\cal O}\left(\dot{S}_L^A\right)$ and $\displaystyle {\cal O}\left(\frac{v_A}{c}\,S_L^A\right)$ 
in Eq.~(\ref{Second_Integration_Spin_5}) is in coincidence with the DSX metric in Eqs.~(\ref{global_metric_perturbation_A}) - (\ref{global_metric_perturbation_C})  
where such terms do not occur because they are beyond 1.5PN approximation.

\section{Light trajectory in the field of spin-dipoles}\label{Section5}  

In our previous investigation \cite{Zschocke_1PN} the light trajectory in the field of $N$ arbitrarily moving mass-monopoles, mass-dipoles, and 
mass-quadrupoles has been considered as specific examples of the general solution, see Eqs.~(139), (140), and (143) - (148) in \cite{Zschocke_1PN}, respectively.  
Here we will consider the light trajectory in the field of $N$ arbitrarily moving spin-dipoles as specific example of the general solution. 
It may also serve as a further instructive example about how the presented approach runs.

\subsection{Light trajectory in the field of $N$ arbitrarily moving spin-dipoles} 

The rotational motion of a real body like the Sun, Earth, or Jupiter, is a highly complicated physical subject, because these bodies are not rigid monopoles  
and the rotational motion can therefore not be described by a simple spin-dipole, but must   
be expressed by the full set of time-dependent spin-multipoles $S_L^A\left(t\right)$ with $l=1,2,3,...$. On the other side, the main impact among all  
spin-multipoles on light deflection is of course given by the first summand in (\ref{Second_Integration_Spin_5}) which is proportional to the  
intrinsic spin vector $\ve{S}_A\left(t\right)$ of body $A$ and which is called spin-dipole. It is also well-known that for sub-micro-arcsecond astrometry   
the light trajectory in the field of a spin-dipole is of specific importance, because the  
light deflection caused by the spin-dipole of a body at rest amounts to be $0.7\,\mu{\rm as}$ for grazing rays at the Sun,
$0.2\,\mu{\rm as}$ for grazing rays at Jupiter, and $0.04\,\mu{\rm as}$ for grazing rays at Saturn \cite{Klioner2003a,Klioner1991}.
Therefore, we will consider the light trajectory in the field of one arbitrarily moving body with time-dependent spin-dipole in more detail in this section. 

According to Eq.~(\ref{second_integral_geodesic_equation}) with (\ref{Second_Integration_15}) and (\ref{Second_Integration_Spin_5}),  
the light trajectory in the field of $N$ arbitrarily moving spin-dipoles reads:  
\begin{eqnarray}
\ve{x}_{\rm S} \left(\tau + t^{\ast}\right) &=& \ve{\xi} + c\,\tau\,\ve{\sigma}  
\nonumber\\
\nonumber\\
&& \hspace{-2.0cm} + \sum\limits_{A=1}^N \bigg(\Delta \ve{x}_A^{\rm S}\left(\tau+t^{\ast}\right)- \Delta \ve{x}_A^{\rm S}\left(\tau_0+t^{\ast}\right)\bigg), 
\label{arbitrarily_moving_spin_dipole_5}
\end{eqnarray}

\noindent
where 
\begin{widetext}
\begin{eqnarray}
\Delta x_A^{i\;{\rm S}}\left(\tau+t^{\ast}\right) &=&
- \frac{2\,G}{c^3}\,\epsilon_{iab}\,S^A_b\left(\tau+t^{\ast}\right)\;\partial_a\,
\ln \left[r^{\rm N}_A\left(\tau+t^{\ast}\right) - \ve{\sigma}\cdot \ve{r}^{\rm N}_A\left(\tau+t^{\ast}\right)\right]
\nonumber\\
\nonumber\\
&& + \frac{2\,G}{c^3}\,\sigma^j\,\epsilon_{jab}\,S^A_b\left(\tau+t^{\ast}\right)\;\partial_a\,  
\frac{d_A^i\left(\tau+t^{\ast}\right)}{r_A^{\rm N}\left(\tau+t^{\ast}\right) - \ve{\sigma}\cdot \ve{r}_A^{\rm N}\left(\tau+t^{\ast}\right)}\,. 
\label{arbitrarily_moving_spin_dipole_10}
\end{eqnarray}
\end{widetext}

\noindent
The derivative operator in terms of the variables $\ve{\xi},\tau,t^{\ast}$ is given by (\ref{Transformation_Derivative_1}), which
for one index reads:
\begin{eqnarray}
\partial_a &=& P^{ak}\,\frac{\partial}{\partial \xi^k} + \sigma^a\,\frac{\partial}{\partial c \tau} - \sigma^a\,\frac{\partial}{\partial c t^{\ast}}\,.
\label{arbitrarily_moving_spin_dipole_15}
\end{eqnarray}

\noindent
By inserting (\ref{arbitrarily_moving_spin_dipole_15}) into (\ref{arbitrarily_moving_spin_dipole_10}), we encounter the following individual terms: 
\begin{eqnarray}
&& P^{ak}\,\frac{\partial}{\partial \xi^k}\,\ln \left[r^{\rm N}_A\left(\tau+t^{\ast}\right) - \ve{\sigma}\cdot \ve{r}^{\rm N}_A\left(\tau+t^{\ast}\right)\right]
\nonumber\\ 
\nonumber\\ 
&& = \frac{d_A^a\left(\tau+t^{\ast}\right)}{r^{\rm N}_A\left(\tau+t^{\ast}\right)}
\,\frac{1}{r^{\rm N}_A\left(\tau+t^{\ast}\right) - \ve{\sigma}\cdot \ve{r}^{\rm N}_A\left(\tau+t^{\ast}\right)}\,,
\label{arbitrarily_moving_spin_dipole_20}
\end{eqnarray}

\noindent
and 
\begin{eqnarray}
&& \sigma^a\left(\frac{\partial}{\partial c \tau} - \frac{\partial}{\partial c t^{\ast}}\right) 
\ln \left[r^{\rm N}_A\left(\tau+t^{\ast}\right) - \ve{\sigma}\cdot \ve{r}^{\rm N}_A\left(\tau+t^{\ast}\right)\right] 
\nonumber\\ 
\nonumber\\ 
&& = - \frac{\sigma^a}{r^{\rm N}_A\left(\tau+t^{\ast}\right)}\,, 
\label{arbitrarily_moving_spin_dipole_25}
\end{eqnarray}

\noindent
and
\begin{eqnarray}
&& P^{ak}\,\frac{\partial}{\partial \xi^k}\,
\frac{d_A^i\left(\tau+t^{\ast}\right)}{r_A^{\rm N}\left(\tau+t^{\ast}\right) - \ve{\sigma}\cdot \ve{r}_A^{\rm N}\left(\tau+t^{\ast}\right)} 
\nonumber\\
\nonumber\\
&& = - \frac{d_A^a\left(\tau+t^{\ast}\right)\;d_A^i\left(\tau+t^{\ast}\right)}
{\left(r_{\rm A}^{\rm N}\left(\tau+t^{\ast}\right) - \ve{\sigma}\cdot\ve{r}_{\rm A}^{\rm N}\left(\tau+t^{\ast}\right)\right)^2}\,
\frac{1}{r_{\rm A}^{\rm N}\left(\tau+t^{\ast}\right)}
\nonumber\\
\nonumber\\
&& \;\;\;\; + \frac{P^{ai}}{r_{\rm A}^{\rm N}\left(\tau+t^{\ast}\right) - \ve{\sigma}\cdot \ve{r}_{\rm A}^{\rm N}\left(\tau+t^{\ast}\right)}\,, 
\label{arbitrarily_moving_spin_dipole_30}
\end{eqnarray}

\noindent
and we recall $\epsilon_{jab}\,\sigma^a\,\sigma^j =0$. Inserting (\ref{arbitrarily_moving_spin_dipole_20}) - (\ref{arbitrarily_moving_spin_dipole_30}) 
into (\ref{arbitrarily_moving_spin_dipole_10}) yields 
\begin{widetext}
\begin{eqnarray}
\Delta\ve{x}_A^{\rm S}\left(\tau+t^{\ast}\right) &=&
+ \frac{2\,G}{c^3}\, 
\,\frac{\ve{S}_A\left(\tau+t^{\ast}\right) \times \ve{d}_A\left(\tau+t^{\ast}\right)}
{r^{\rm N}_A\left(\tau+t^{\ast}\right) - \ve{\sigma}\cdot \ve{r}^{\rm N}_A\left(\tau+t^{\ast}\right)}\;
\frac{1}{r^{\rm N}_A\left(\tau+t^{\ast}\right)}
+ \frac{2\,G}{c^3}\,\frac{\ve{\sigma} \times \ve{S}_A\left(\tau+t^{\ast}\right)}{r^{\rm N}_A\left(\tau+t^{\ast}\right)} 
\nonumber\\
\nonumber\\
\nonumber\\
&& - \frac{2\,G}{c^3}\,
\frac{\ve{\sigma} \cdot \left(\ve{d}_A\left(\tau+t^{\ast}\right) \times \ve{S}_A\left(\tau+t^{\ast}\right)\right)}
{\left(r_{\rm A}^{\rm N}\left(\tau+t^{\ast}\right) - \ve{\sigma}\cdot\ve{r}_{\rm A}^{\rm N}\left(\tau+t^{\ast}\right)\right)^2}\,
\frac{\ve{d}_A\left(\tau+t^{\ast}\right)}{r_{\rm A}^{\rm N}\left(\tau+t^{\ast}\right)}
- \frac{2\,G}{c^3}\,
\frac{\ve{\sigma} \times \ve{S}_A\left(\tau+t^{\ast}\right)}
{r_{\rm A}^{\rm N}\left(\tau+t^{\ast}\right) - \ve{\sigma}\cdot \ve{r}_{\rm A}^{\rm N}\left(\tau+t^{\ast}\right)}\,,  
\label{arbitrarily_moving_spin_dipole_35}
\end{eqnarray}
\end{widetext}

\noindent
where the notation $\epsilon_{ijk}\,a_j\,b_k = \left(\ve{a} \times \ve{b}\right)^i$ has been used. The complete expression for the light trajectory  
in 1.5PN approximation in the field of $N$ arbitrarily moving and time-dependent intrinsic spin-dipoles is finally obtained by
inserting (\ref{arbitrarily_moving_spin_dipole_35}) into (\ref{arbitrarily_moving_spin_dipole_5}).

As mentioned in the introductory section, 
in \cite{KopeikinMashhoon2002} the light trajectory in post-Minkowskian approximation in the field of $N$ arbitrarily moving pointlike spin-dipoles  
has been determined. That means, the pointlike objects in \cite{KopeikinMashhoon2002} may even be in ultra-relativistic motion, while our 1.5PN solution  
in (\ref{arbitrarily_moving_spin_dipole_35}) is valid for extended bodies with spin-dipole but in slow-motion along arbitrary worldlines.  
In appendix \ref{Solution_Kopeikin_Mashhoon} it is shown that our result in (\ref{arbitrarily_moving_spin_dipole_35}) agrees with the results in 
\cite{KopeikinMashhoon2002} for the light trajectory up to terms of the order ${\cal O}\left(c^{-4}\right)$.  
One may also verify that in the limit of time-independent spin-dipoles, $\ve{S}_A = {\rm const}$, and in the limit of uniform motion,  
$\ve{v}_A = {\rm const}$, our result in (\ref{arbitrarily_moving_spin_dipole_35}) agrees with Eq.~(26) in \cite{Deng_2015} in GR,  
noticing that constant terms cancel each other according to Eq.~(\ref{arbitrarily_moving_spin_dipole_5}).

\subsection{Light trajectory in the field of $N$ bodies at rest with spin-dipole} 

In this section we will consider the case of light propagation in the field of $N$ spin-dipoles at rest and compare with results in the literature. 
For time-independent spin-dipole $\ve{S}_A = {\rm const.}$ and for one body at rest located at $\ve{x}_A = {\rm const.}$ in the global reference system  
we have $\ve{r}_{\rm A}^{\rm N}\left(\tau+t^{\ast}\right) \rightarrow \ve{r}_{\rm A}^{\rm N} = \ve{\xi} + c\,\tau\,\ve{\sigma} - \ve{x}_A$ and  
$\ve{d}_A\left(\tau+t^{\ast}\right) \rightarrow \ve{d}_A = \ve{r}_A^{\rm N} - \ve{\sigma}\,\left(\ve{\sigma}\cdot\ve{r}_A^{\rm N}\right)$ 
where $d_A$ is the time-independent impact vector defined by Eq.~(\ref{notation_2}).  
From (\ref{arbitrarily_moving_spin_dipole_5}) we obtain the light trajectory in the field of $N$ bodies at rest with time-independent spin-dipoles: 
\begin{eqnarray}
\ve{x}_{\rm S} \left(\tau + t^{\ast}\right) &=& \ve{\xi} + c\,\tau\,\ve{\sigma}
+ \sum\limits_{A=1}^N \bigg(\Delta \ve{x}_A^{\rm S}\left(\tau\right)- \Delta \ve{x}_A^{\rm S}\left(\tau_0\right)\bigg),
\nonumber\\
\label{Light_Trajectory_Spin_At_Rest_5}
\end{eqnarray}

\noindent 
where from (\ref{arbitrarily_moving_spin_dipole_35}) we obtain the following expression for the correction-term: 
\begin{eqnarray}
\Delta\ve{x}_A^{\rm S}\left(\tau\right) &=& \frac{2\,G}{c^3}\,\frac{\ve{S}_A \times \ve{d}_A}{d_A^2}\,\frac{\ve{\sigma}\cdot\ve{r}_A^{\rm N}}{r_A^{\rm N}}  
+ \frac{2\,G}{c^3}\,\frac{\ve{\sigma} \times \ve{S}_A}{r_A^{\rm N}}  
\nonumber\\
\nonumber\\
&& - \frac{2\,G}{c^3}\,\ve{\sigma} \cdot \left(\ve{d}_A \times \ve{S}_A\right) \,\frac{\ve{d}_A}{d_A^4}\,
\frac{\left(r_A^{\rm N} + \ve{\sigma}\cdot\ve{r}_A^{\rm N}\right)^2}{r_A^{\rm N}}  
\nonumber\\
\nonumber\\
&& - \frac{2\,G}{c^3}\,\ve{\sigma} \times \ve{S}_A\;\frac{r_A^{\rm N} + \ve{\sigma}\cdot\ve{r}_A^{\rm N}}{d_A^2}\,, 
\label{Light_Trajectory_Spin_At_Rest_10}
\end{eqnarray}
 
\noindent
where a time-independent term $\displaystyle \frac{2\,G}{c^3}\,\frac{\ve{S}_A \times \ve{d}_A}{d_A^2} = {\rm const.}$ has been omitted 
because this term will be cancelled in view of (\ref{Light_Trajectory_Spin_At_Rest_5}). The time-dependence of (\ref{Light_Trajectory_Spin_At_Rest_5}) 
and (\ref{Light_Trajectory_Spin_At_Rest_10}) is solely caused by the time-dependence of the unperturbed lightray in (\ref{variable_5}).  
In order to obtain the form of the expression in (\ref{Light_Trajectory_Spin_At_Rest_10}) we have also used  
$d_A^2 = \left(r_A^{\rm N} - \ve{\sigma}\cdot\ve{r}_A^{\rm N}\right)\,\left(r_A^{\rm N} + \ve{\sigma}\cdot\ve{r}_A^{\rm N}\right)$.  
The expression in (\ref{Light_Trajectory_Spin_At_Rest_5}) - (\ref{Light_Trajectory_Spin_At_Rest_10}) agrees with the solution in Eq.~(56) 
in \cite{Klioner1991}, where the trajectory of a photon as function of time has been determined in the field of 
$N$ bodies at rest in post-Newtonian approximation for the lightrays. It is  
straightforward to show that the time-derivative $\partial_{c\,\tau}\Delta\ve{x}_A^{\rm S}\left(\tau\right)$ coincides with Eq.~(59) in \cite{Klioner1991}.

\section{Time-delay}\label{Observable_Effects_Time_Delay}

In the previous sections we have determined the light trajectory of a light-signal which propagates through the metric field 
of the Solar system, that means through the gravitational field of $N$ arbitrarily moving massive bodies. However, the 
light trajectory is not an observable at all. In real astrometric measurements one of the most important observable quantity   
concerns the time delay of some light-signal propagating in the Solar system.  
The considerations here are similar to what has been discussed in \cite{Zschocke_1PN} about observable effects, but with the 
extension to 1.5PN approximation.  
Especially, we will assume that the light source is located at $\ve{x}_0 = \ve{x}\left(t_0\right)$ where $t_0$ is the moment of 
emission of the light-signal, and the observer is located at $\ve{x}_1 = \ve{x}\left(t_1\right)$ where $t_1$ is the moment of reception of the 
light-signal by the observer. Furthermore, both the light source and the observer are assumed to be at rest with respect  
to the global reference system.  

In the pioneering work \cite{Shapiro1}, {\it Shapiro} has considered the general-relativistic effect of time delay of a light-signal which propagates  
through the gravitational field of a static and spherically symmetric massive body. Especially, {\it Shapiro} has drawn the attention to the fact about  
the measurability of that additional test of relativity by radar technology. In fact, the Shapiro time delay was discovered soon afterwards \cite{Shapiro2}.  
It might be useful to realize that the reason for the time delay is not only laying upon the fact that the light-trajectory is curved but also because the  
speed of a photon is decelerated in the gravitational field of a monopole at rest. While the classical Shapiro effect is originally related to a time delay of a 
light-signal in the monopole-field, it became a matter of common knowledge to call the time delay of a light-signal in any gravitational field just Shapiro effect. 

For describing the Shapiro-effect, we introduce a vector pointing from the light source at the moment of emission toward the observer at the moment of reception, 
which in terms of the new variables reads 
\begin{eqnarray}
\ve{R} &=& \ve{x}\left(\tau_1 + t^{\ast}\right) - \ve{x}\left(\tau_0 + t^{\ast}\right)\,,
\label{Shapiro_5}
\\
\nonumber\\
\ve{k} &=& \frac{\ve{R}}{R}\,, 
\label{Shapiro_10}
\end{eqnarray}

\noindent
where $\ve{k}$ is just the corresponding unit vector with $ R = \left|\ve{R}\right|$ being the absolute value of $\ve{R}$. Using very similar steps  
as in \cite{KopeikinSchaefer1999_Gwinn_Eubanks}, we obtain from Eq.~(\ref{second_integral_geodesic_equation}) the following expression for the time delay  
in the gravitational field of $N$ arbitrarily moving massive bodies in 1.5PN approximation that means up to terms of the order ${\cal O}\left(c^{-4}\right)$: 
\begin{eqnarray}
c \left(\tau_1 - \tau_0 \right) &=& R + \Delta c \tau_{\rm 1PN} + \Delta c \tau_{\rm 1.5PN}\,,  
\label{Shapiro_15}
\\
\nonumber\\
\Delta c \tau_{\rm 1PN} &=& - \sum\limits_{A=1}^N \ve{k}\cdot\left[\Delta\ve{x}^A_{{\rm 1PN}}\left(\tau_1 + t^{\ast}, \tau_0 + t^{\ast}\right)\right],  
\nonumber\\
\label{Shapiro_15_A}
\\
\nonumber\\
\Delta c \tau_{\rm 1.5PN} &=& - \sum\limits_{A=1}^N \ve{k} \cdot \left[\Delta \ve{x}^A_{{\rm 1.5PN}}\left(\tau_1 + t^{\ast}, \tau_0 + t^{\ast}\right)\right],  
\nonumber\\
\label{Shapiro_15_B}
\end{eqnarray}

\noindent
where the sum runs over all massive bodies and the expressions for $\Delta \ve{x}^A_{{\rm 1PN}}$ and $\Delta \ve{x}^A_{{\rm 1.5PN}}$ are given by  
Eqs.~(\ref{second_integral_geodesic_equation_1PN}) and (\ref{second_integral_geodesic_equation_15PN}) with (\ref{Second_Integration_Mass_4}), 
(\ref{Second_Integration_Mass_5}) and (\ref{Second_Integration_Spin_5}), respectively.  
The 1.5PN relation (\ref{Shapiro_15}) generalizes the 1PN relation (154) in \cite{Zschocke_1PN}.  

We will consider the time delay in (\ref{Shapiro_15}) of a light-signal caused by $N$ arbitrarily moving bodies in some more detail, but will restrict ourselves 
on the case of $N$ moving bodies with monopole-structure (M) , quadrupole-structure ($J_2$), and spin-dipole-structure (S).    
Higher multipoles are so tiny that they are negligible in the time delay effect. These first terms in the general formula (\ref{Shapiro_15}) read  
\begin{eqnarray}
c \left(t_1 - t_0 \right) &=& R + \Delta\,c t^{\rm M}_{\rm 1PN} + \Delta\,c t^{J_2}_{\rm 1PN}  
\nonumber\\
\nonumber\\
&& \hspace{0.35cm} + \Delta\,c t^{\rm M}_{\rm 1.5PN} + \Delta\,c t^{\rm S}_{\rm 1.5PN}\,,  
\label{Shapiro_Delay} 
\end{eqnarray}

\noindent
which are instructive examples and do allow for a cross-check with known results in the literature. 

Furthermore, as mentioned in the introductory section, there are several proposals to ESA for future space-based missions,  
like ASTROD \cite{Astrod1,Astrod2}, LATOR \cite{Lator1,Lator2}, ODYSSEY \cite{Odyssey}, SAGAS \cite{Sagas}, TIPO \cite{TIPO},  
which aim at time-transfer accuracies of two separated clocks within the Solar system of up to $10\,{\rm ps}$. The question arises about the ability of  
such extremely-precise astrometry missions, especially designed for tests of relativity in the Solar system, to detect some 1.5PN terms in the 
Shapiro effect which will be discussed in this section.  

In general, the light-signal will be assumed to be emitted at a space-time point with  
BCRS coordinates $\ve{x}_0,t_0$ and received by an observer at a space-time point with BCRS coordinates $\ve{x}_1,t_1$. 
We also introduce the following notations: $\ve{r}^{\,0}_A = \ve{x}_0 - \ve{x}_A\left(t_0\right)$, $\ve{r}^{\,1}_A = \ve{x}_1 - \ve{x}_A\left(t_1\right)$, 
$R = \left|\ve{x}_0 - \ve{x}_1\right|$, $\ve{v}^{\,0}_A = \ve{v}_A\left(t_0\right)$, $\ve{v}^{\,1}_A = \ve{v}_A\left(t_1\right)$, 
$\ve{d}_A^{\,0} = \ve{d}_A\left(t_0\right)$,  
$\ve{d}_A^{\,1} = \ve{d}_A\left(t_1\right)$. Furthermore, we notice that  
$\ve{\sigma} = \ve{k} + {\cal O}\left(c^{-2}\right)$ according to Eq.~(\ref{Light_Deflection_15}) given below, that means we may replace the vector 
$\ve{\sigma}$ in favor of vector $\ve{k}$ whenever it is reasonable.

\subsection{Moving mass-monopole}\label{Shapiro_Monopoles} 

We will consider the time delay in (\ref{Shapiro_15}) of a light-signal caused by an arbitrarily moving monopole.  

\subsubsection{In terms of coordinate time} 

From Eqs.~(\ref{Second_Integration_Mass_4}) and (\ref{Second_Integration_Mass_5}) we obtain in the field of arbitrary-moving monopoles ($l=0$) the   
expressions $\Delta \ve{x}_{\rm 1PN}^{A\,{\rm M}}$ and $\Delta \ve{x}_{\rm 1.5PN}^{A\,{\rm M}}$, respectively. According to  
Eqs.~(\ref{Shapiro_15}) - (\ref{Shapiro_15_B}) and using $t_0 = \tau_0 + t^{\ast}$ and $t_1 = \tau_1 + t^{\ast}$ 
we obtain up to terms of the order ${\cal O}\left(c^{-4}\right)$:
\begin{eqnarray}
\Delta\,c t^{\rm M} &=& \Delta\,c t^{\rm M}_{\rm 1PN} + \Delta\,c t^{\rm M}_{\rm 1.5PN}\,, 
\label{Shapiro_Monopoles_1_A}
\\
\nonumber\\
\nonumber\\
\Delta\,c t^{\rm M}_{\rm 1PN} &=& - \sum\limits_{A=1}^N \ve{k} \cdot \Delta \ve{x}_{\rm 1PN}^{A\,{\rm M}}\left(t_1,t_0\right)  
\nonumber\\
\nonumber\\
&=& - \sum\limits_{A=1}^N \frac{2\,G\,M_A}{c^2}\,
\ln \frac{r_A^{\,1} - \ve{\sigma}\cdot\ve{r}_A^{\,1}}{r_A^{\,0} - \ve{\sigma}\cdot\ve{r}_A^{\,0}}\,, 
\label{Shapiro_Monopoles_1_B}
\\
\nonumber\\
\nonumber\\
\Delta\,c t^{\rm M}_{\rm 1.5PN} &=& - \sum\limits_{A=1}^N \ve{k} \cdot \Delta \ve{x}_{\rm 1.5PN}^{A\,{\rm M}}\left(t_1,t_0\right)  
\nonumber\\
\nonumber\\
&=& + \sum\limits_{A=1}^N \frac{2\,G\,M_A}{c^3} 
\left(\ve{\sigma}\cdot\ve{v}^{\,1}_A\right) \ln \left(r_A^{\,1} - \ve{\sigma}\cdot\ve{r}_A^{\,1}\right)  
\nonumber\\
\nonumber\\
&& - \sum\limits_{A=1}^N \frac{2\,G\,M_A}{c^3}  
\left(\ve{\sigma}\cdot\ve{v}^{\,0}_A\right) \ln \left(r_A^{\,0} - \ve{\sigma}\cdot\ve{r}_A^{\,0}\right) 
\nonumber\\
\nonumber\\
&& \hspace{-1.7cm} - \sum\limits_{A=1}^N \frac{2\,G\,M_A}{c^3}\,\left(\frac{\ve{v}_A^{\,1} \cdot \ve{d}_A^{\,1}}{r_A^{\,1} - \ve{\sigma}\cdot\ve{r}_A^{\,1}} 
- \frac{\ve{v}_A^{\,0} \cdot \ve{d}_A^{\,0}}{r_A^{\,0} - \ve{\sigma}\cdot\ve{r}_A^{\,0}}\right).   
\label{Shapiro_Monopoles_1_C}
\end{eqnarray}

\noindent
In the limit of monopoles at rest only the term in Eq.~(\ref{Shapiro_Monopoles_1_B}) remains which then represents the well-known classical Shapiro effect 
\cite{MTW,Brumberg1991,Poisson_Will,Kopeikin_Efroimsky_Kaplan} which is growing logarithmically with $R$, while   
in our result (\ref{Shapiro_Monopoles_1_C}) the argument of the logarithm depends on the worldline of the arbitrary-moving body $\ve{x}_A\left(t\right)$.  

One may verify that our result for the Shapiro delay for arbitrarily moving monopoles in Eq.~(\ref{Shapiro_Monopoles_1_B}) - (\ref{Shapiro_Monopoles_1_C}),  
agrees in the limit of uniform motion with Eq.~(20) in \cite{Comparison_Shapiro_1}, with Eq.~(45) in \cite{Hees_Bertone_Poncin_Lafitte_2014a},  
and with Eq.~(33) in \cite{Soffel_Han} up to terms of the order ${\cal O}\left(c^{-4}\right)$. 
In this respect we recall that the term in the last line in (\ref{Shapiro_Monopoles_1_C}) can be written as follows:  
\begin{eqnarray}
&& \frac{\ve{v}_A^{\,1} \cdot \ve{d}_A^{\,1}}{r_A^{\,1} - \ve{\sigma}\cdot\ve{r}_A^{\,1}}
- \frac{\ve{v}_A^{\,0} \cdot \ve{d}_A^{\,0}}{r_A^{\,0} - \ve{\sigma}\cdot\ve{r}_A^{\,0}}
\nonumber\\
\nonumber\\
&& = \frac{\ve{v}_A^{\,1} \cdot \ve{r}_A^{\,1} - r_A^{\,1}\left(\ve{\sigma}\cdot\ve{v}_A^{\,1}\right)}{r_A^{\,1} - \ve{\sigma}\cdot\ve{r}_A^{\,1}}
- \frac{\ve{v}_A^{\,0} \cdot \ve{r}_A^{\,0} - r_A^{\,0}\left(\ve{\sigma}\cdot\ve{v}_A^{\,0}\right)}{r_A^{\,0} - \ve{\sigma}\cdot\ve{r}_A^{\,0}}
\nonumber\\
\nonumber\\
&& \;\; + \,\ve{\sigma}\cdot \left(\ve{v}_A^{\,1} - \ve{v}_A^{\,0}\right)\,, 
\label{Shapiro_Monopoles_2}
\end{eqnarray}

\noindent
where the term in the last line is proportional to the acceleration of the massive body $A$ and vanishes in case of uniform motion. 
The neglect of this term, as suggested in \cite{KopeikinSchaefer1999}, is well-justified because a simple estimate reveals that such terms are extremely small
and far out of detectability even for future astrometry missions.
An estimate of the absolute value of the 1PN time delay formula in Eq.~(\ref{Shapiro_Monopoles_1_B}) for one body $A$ and
assuming an astrometric configuration with $\ve{\sigma}\cdot\ve{r}_A^{\,0} \simeq - r_A^{\,0}$ and $\ve{\sigma}\cdot\ve{r}_A^{\,1} \simeq r_A^{\,1}$,
is given by \cite{Poisson_Will}:
\begin{eqnarray}
\left|\Delta\,t^{\rm M}_{\rm 1PN}\right| &\le& \frac{2 G M_A}{c^3}\,
\ln \frac{4\,r_A^{\,1}\,r_A^{\,0}}{\left(d_A^1\right)^2}\,.
\label{Shapiro_Monopoles_1_D}
\end{eqnarray}

\noindent
A very similar estimate of the absolute value of the 1.5PN correction in Eq.~(\ref{Shapiro_Monopoles_1_C}) for one body $A$ and
same configuration yields
\begin{eqnarray}
\left|\Delta\,t^{\rm M}_{\rm 1.5PN}\right| &\le&
\frac{v_A}{c}\,\left|\Delta\,t^{\rm M}_{\rm 1PN}\right| + \frac{4 G M_A}{c^3}\,\frac{v_A}{c}\,\frac{r_A^{\,1}}{d^1_A}\,.
\label{Shapiro_Monopoles_1_E}
\end{eqnarray}

\noindent
The second term in (\ref{Shapiro_Monopoles_1_E}) is proportional to $\sim r_A^{\,1}/d^1_A$, which for grazing rays becomes a large quantity.  
For instance, for Jupiter we would get $r_A^{\,1}/d^1_A \sim 10^4$ which spoils the effect of the tiny factor $v_A/c \sim 10^{-5}$ which is typical for 
1.5PN corrections. This large term is solely caused by the term in the last line in Eq.~(\ref{Shapiro_Monopoles_1_C}).  
Below, we will consider the expressions for light deflection where we will encounter this large term again, cf. text below Eq.~(34) in \cite{Klioner2003a}.  
As we will show in the next subsection, this large factor $r_A^{\,1}/d^1_A$ is related to the retardation of gravitational action.  

\subsubsection{In terms of retarded time} 

Gravitational action travels with the finite speed of light and this effect cannot be ignored in high-precision astrometry, 
as it has been outlined long time ago \cite{Hellings1986,KopeikinSchaefer1999,KopeikinMashhoon2002,Klioner2003a,KlionerPeip2003}. 
In order to take account for that effect we follow  
the arguments of the investigations in \cite{KopeikinSchaefer1999,KopeikinMashhoon2002,Klioner2003a,KlionerPeip2003,KopeikinMakarov2007}, 
which have shown that the position of the massive body must not be taken at the time of observation, $\ve{x}_A\left(t_1\right)$,  
but at the retarded time-moment, $\ve{x}_A\left(t_1^{\rm ret}\right)$. In general, the retarded time is defined by an implicit relation,  
\begin{eqnarray}
t^{\rm ret} &=& t - \frac{\left|\ve{x}\left(t\right) - \ve{x}_A\left(t^{\rm ret}\right)\right|}{c}\,,  
\label{Retarded_Time}
\end{eqnarray}
 
\noindent
where $t$ is the coordinate time. For the special case where $t$ is the time of emission $t_0$ or the time of reception $t_1$ see Eq.~(\ref{Retardation_t0}).   
Actually, the retarded time is a function of the position of body under consideration and, therefore, an index $A$ should also be attached at $t^{\rm ret}$ 
but for simpler notation such label is omitted. According to Eqs.~(47) - (48) in \cite{Zschocke_Soffel}, the retarded position can be  
series-expanded and leads to the following relations for any instant of time:  
\begin{eqnarray}
\ve{r}_A\left(t^{\rm ret}\right) &=& \ve{r}_A\left(t\right) + r_A\left(t\right)\,\frac{\ve{v}_A\left(t\right)}{c} + {\cal O}\left(c^{-2}\right), 
\label{Retardation_A}
\\
\nonumber\\
r_A\left(t^{\rm ret}\right) &=& r_A\left(t\right) + \frac{\ve{r}_A\left(t\right) \cdot \ve{v}_A\left(t\right)}{c} + {\cal O}\left(c^{-2}\right).  
\label{Retardation_B}
\end{eqnarray}

\noindent
These relations allow one to rewrite identically the expressions in (\ref{Shapiro_Monopoles_1_A}) - (\ref{Shapiro_Monopoles_1_C}) into the following form 
up to terms of the order ${\cal O}\left(c^{-4}\right)$: 
\begin{eqnarray}
\Delta\,c t^{\rm M} &=& \Delta\,c t^{\rm M}_{\rm 1PN} + \Delta\,c t^{\rm M}_{\rm 1.5PN}\,,
\label{Retardation_C}
\\
\nonumber\\
\nonumber\\
\Delta\,c t^{\rm M}_{\rm 1PN} &=& - \sum\limits_{A=1}^N \frac{2\,G\,M_A}{c^2}\,
\ln \frac{r_A\left(t^{\rm ret}_1\right) - \ve{\sigma}\cdot\ve{r}_A\left(t^{\rm ret}_1\right)}
{r_A\left(t^{\rm ret}_0\right) - \ve{\sigma}\cdot\ve{r}_A\left(t^{\rm ret}_0\right)}\,,
\nonumber\\
\label{Retardation_D}
\\
\nonumber\\
\nonumber\\
\Delta\,c t^{\rm M}_{\rm 1.5PN} &=& + \sum\limits_{A=1}^N \frac{2\,G\,M_A}{c^3}\,  
\nonumber\\
\nonumber\\
&& \hspace{-0.5cm} \times \left(\ve{\sigma}\cdot\ve{v}_A\left(t^{\rm ret}_1\right)\right) 
\ln \left(r_A\left(t^{\rm ret}_1\right) - \ve{\sigma}\cdot\ve{r}_A\left(t^{\rm ret}_1\right)\right)  
\nonumber\\
\nonumber\\
&& - \sum\limits_{A=1}^N \frac{2\,G\,M_A}{c^3}\,
\nonumber\\
\nonumber\\
&& \hspace{-0.5cm} \times \left(\ve{\sigma}\cdot\ve{v}_A\left(t^{\rm ret}_0\right)\right)  
\ln \left(r_A\left(t^{\rm ret}_0\right) - \ve{\sigma}\cdot\ve{r}_A\left(t^{\rm ret}_0\right)\right),  
\nonumber\\
\label{Retardation_E}
\end{eqnarray}

\noindent
where (cf. Eq.~(\ref{Retarded_Time})):  
\begin{eqnarray}
t_n^{\rm ret} &=& t_n - \frac{\left|\ve{r}_A\left(t_n^{\rm ret}\right)\right|}{c}, \;\; n=0,1\,.  
\label{Retardation_t0}
\end{eqnarray}

\noindent
The solution for the time delay in (\ref{Retardation_C}) - (\ref{Retardation_E}) agrees with Eq.~(51) in \cite{KopeikinSchaefer1999}.  
Especially, we notice that the term in the last line of Eq.~(\ref{Shapiro_Monopoles_1_C}) has been absorbed in (\ref{Retardation_D}).   
Consequently, if one uses the expression for the time delay in terms of retarded time, Eqs.~(\ref{Retardation_C}) - (\ref{Retardation_E}), then  
one obtains the following correct estimate for the time delay in 1.5PN approximation:
\begin{eqnarray}
\left|\Delta\,t^{\rm M}_{\rm 1PN}\right| &\le& \frac{2 G M_A}{c^3}\,
\ln \frac{4\,r_A\left(t^{\rm ret}_0\right)\,r_A\left(t^{\rm ret}_1\right)}{d_A^2\left(t^{\rm ret}_1\right)}\,, 
\label{Shapiro_Monopoles_1_F1}
\\
\nonumber\\
\left|\Delta\,t^{\rm M}_{\rm 1.5PN}\right| &\le& \frac{v_A}{c}\,\left|\Delta\,t^{\rm M}_{\rm 1PN}\right|.  
\label{Shapiro_Monopoles_1_F2}
\end{eqnarray}
 
\noindent
For numerical values of the upper bound in Eq.~(\ref{Shapiro_Monopoles_1_F1}) and Eq.~(\ref{Shapiro_Monopoles_1_F2}) see Table \ref{Table2}. 
\begin{table}[h!]
\begin{tabular}{| c | c | c | c |}
\hline
\hline
&&&\\[-7pt]
Parameter  
&\hbox to 20mm{\hfill Sun \hfill}
&\hbox to 20mm{\hfill Jupiter \hfill}
&\hbox to 20mm{\hfill Saturn \hfill}\\[3pt]
\hline
&&&\\[-7pt]
$GM_A/c^2\,[{\rm m}]$ & $1476$ & $1.4$ & $0.4$ \\[3pt]
$P_A\,[{\rm m}]$ & $696 \times 10^6$ & $71.5 \times 10^6$ & $60.3 \times 10^6$ \\[3pt]
$J_2^A$ & $2 \times 10^{-7}$ & $14.696 \times 10^{-3}$ & $16.291 \times 10^{-3}$ \\[3pt]
$J_4^A$ & $ - $ & $ - 0.587 \times 10^{-3}$ & $ - 0.936 \times 10^{-3}$ \\[3pt]
$J_6^A$ & $ - $ & $0.034 \times 10^{-3}$ & $0.086 \times 10^{-3}$ \\[3pt]
$J_8^A$ & $ - $ & $ - 2.5 \times 10^{-6}$ & $ - 10.0 \times 10^{-6}$ \\[3pt]
$J_{10}^A$ & $ - $ & $0.21 \times 10^{-6}$ & $2.0 \times 10^{-6}$ \\[3pt]
$S_A\,[{\rm kg}\,{\rm m}^2/\,{\rm s}]$ & $1.64 \times 10^{41}$ & $4.15 \times 10^{38}$ & $7.13 \times 10^{37}$ \\[3pt]
$r_{A}^1\,[{\rm m}]$ & $0.147 \times 10^{12}$ & $0.59 \times 10^{12}$   & $1.20 \times 10^{12}$ \\[3pt]
$v_A/c$ & $4 \times 10^{-8}$ & $4.4 \times 10^{-5}$ & $3.2 \times 10^{-5}$ \\[3pt]
\hline
\hline
\end{tabular}
\caption{Numerical parameters for mass $M_A$, radius $P_A$, actual coefficients of zonal harmonics $J_n^A$, distance between observer and body $r_{A}^1$,   
orbital velocity $v_A$ of Sun, Jupiter and Saturn \cite{JPL}. 
The value for $J^A_2$ for the Sun is taken from \cite{J_2_Sun}, while $J_n^A$ with $n=2,4,6$ for Jupiter and Saturn are taken from \cite{Book_Zonal_Harmonics}, 
while $J_n^A$ with $n=8,10$ for Jupiter and Saturn are taken from \cite{Zonal_Harmonics_Jupiter} and \cite{Zonal_Harmonics_Saturn}, respectively. 
The spin angular momenta $S_A$ are determined from the moment of inertia $I_A$ with the ratio $\displaystyle \frac{I_A}{M_A P_A^2} = 0.059, 0.254, 0.210$ for 
Sun, Jupiter, Saturn, respectively from NASA planetary fact sheets.  
For the distance between light-source and body we assume $r_{A}^0 = 10^{13}\,{\rm m}$  
so that the light-source is within the near-zone of the Solar system, while $r_A^1$ is computed under assumption that the observer (spacecraft)
is located at Lagrange point $L_2$, i.e. $1.5 \times 10^9\,{\rm m}$ from the Earth's orbit.}
\label{Table1}
\end{table}

\subsection{Moving spin-dipole}\label{Shapiro_Spin_Dipoles} 

Now let us consider the time delay in (\ref{Shapiro_15}) of a light-signal caused by $N$ arbitrarily moving spin-dipoles. 

\subsubsection{In terms of coordinate time} 

From (\ref{Second_Integration_Spin_5}) we obtain in the field of arbitrary-moving spin-dipoles ($l=1$) the expression for $\Delta \ve{x}_{\rm 1.5PN}^{A\,{\rm S}}$, 
as given by Eq.~(\ref{arbitrarily_moving_spin_dipole_35}). According to Eq.~(\ref{Shapiro_15_B}) we obtain for the Shapiro-delay the following expression 
up to terms of the order ${\cal O}\left(c^{-4}\right)$:  
\begin{eqnarray}
\Delta\,c t^{\rm S}_{\rm 1.5PN} &=& - \sum\limits_{A=1}^N \ve{k} \cdot \Delta \ve{x}_{\rm 1.5PN}^{A\,{\rm S}}\left(t_1,t_0\right)  
\nonumber\\ 
\nonumber\\ 
&=&  
- \frac{2 G}{c^3} 
\nonumber\\ 
\nonumber\\ 
&& \hspace{-2.0cm} \times \sum\limits_{A=1}^N \left[\frac{\ve{\sigma} \cdot \left(\ve{S}^{\,1}_A \times \ve{d}^{\,1}_A\right)}{\left(d^{\,1}_A\right)^2}  
\frac{\ve{\sigma}\cdot\ve{r}^{\,1}_A}{r^{\,1}_A}  
- \frac{\ve{\sigma} \cdot \left(\ve{S}^{\,0}_A \times \ve{d}^{\,0}_A\right)}{\left(d^{\,0}_A\right)^2}  
\frac{\ve{\sigma}\cdot\ve{r}^{\,0}_A}{r^{\,0}_A} \right]\!,
\nonumber\\
\label{Shapiro_Spin_2}
\end{eqnarray}

\noindent
where $\ve{S}^{\,1}_A = \ve{S}_A\left(t_1\right)$ and $\ve{S}^{\,0}_A = \ve{S}_A\left(t_0\right)$ are the spin-dipoles of body $A$ 
at time observation-time $t_1$ and at emission-time $t_0$ respectively. It can be checked that in the limit of bodies at rest our result  
in (\ref{Shapiro_Spin_2}) agrees with Eq.~(72) in \cite{Klioner1991}. Furthermore, by very similar steps as used in appendix \ref{Solution_Kopeikin_Mashhoon} 
one may verify an agreement of our solution in Eq.~(\ref{Shapiro_Spin_2}) with Eqs.~(48) - (50) in \cite{KopeikinMashhoon2002} in case of slow motion;  
note that the global spin-tensor in \cite{KopeikinMashhoon2002} has to be reexpressed in terms of intrinsic spin-dipole, for instance by means 
of the relations Eqs.~(B.8) and (C.10) in \cite{Zschocke_Soffel} and the retarded time has to be series-expanded in terms of global coordinate-time.  
An estimate of the upper bound of Eq.~(\ref{Shapiro_Spin_2}) yields  
\begin{eqnarray}
\left| \Delta\,t^{\rm S}_{\rm 1.5PN}\right| &\le& \frac{4 G}{c^4}\,\frac{S_A^1}{d^1_A}\,,  
\label{Shapiro_Spin_3}
\end{eqnarray}

\noindent
which agrees with the estimate in Eq.~(75) in \cite{Klioner1991} for grazing rays and spin-dipoles at rest. 

\subsubsection{In terms of retarded time} 

In view of relations (\ref{Retardation_A}) - (\ref{Retardation_B}) and up to terms of the order ${\cal O}\left(c^{-4}\right)$ 
one may perform the following replacements in Eq.~(\ref{Shapiro_Spin_2}): 
\begin{eqnarray}
\ve{r}_A^n &\rightarrow& \ve{r}_A\left(t^{\rm ret}_n\right), \;\; n=0,1\,,  
\label{Shapiro_Spin_4}
\\
\nonumber\\
\ve{S}_A^n &\rightarrow& \ve{S}_A\left(t^{\rm ret}_n\right), \;\; n=0,1\,,  
\label{Shapiro_Spin_5}
\\
\nonumber\\
\ve{d}_A^n &\rightarrow& \ve{d}_A\left(t^{\rm ret}_n\right), \;\; n=0,1\,.  
\label{Shapiro_Spin_6}
\end{eqnarray}
 
\noindent
The upper bound is then given by    
\begin{eqnarray}
\left| \Delta\,t^{\rm S}_{\rm 1.5PN}\right| &\le& \frac{4 G}{c^4}\,\frac{S_A\left(t^{\rm ret}_1\right)}{d_A\left(t^{\rm ret}_1\right)}\,.   
\label{Shapiro_Spin_7}
\end{eqnarray}

\noindent
For numerical values of the upper bound in Eq.~(\ref{Shapiro_Spin_7}) see Table \ref{Table2}.   
\begin{table}[h!]
\begin{tabular}{| c | c | c | c |}
\hline
\hline
&&&\\[-7pt]
Term
&\hbox to 20mm{\hfill Sun \hfill}
&\hbox to 20mm{\hfill Jupiter \hfill}
&\hbox to 20mm{\hfill Saturn \hfill}\\[3pt]
\hline
&&&\\[-7pt]
$\Delta t_{\rm 1PN}^{\rm M}$ & $160\,\mu{\rm s}$ & $0.2\,\mu{\rm s}$ & $0.06\,\mu{\rm s}$ \\[3pt]
$\Delta t_{\rm 1PN}^{J_2}$ & $3.3 \times 10^{-3}\,{\rm ns}$ & $0.2\,{\rm ns}$ & $0.07\,{\rm ns}$ \\[3pt]
\hline
$\Delta t_{\rm 1.5PN}^{\rm M}$ & $6 \times 10^{-3}\,{\rm ns}$ & $9 \times 10^{-3}\,{\rm ns}$ & $2 \times 10^{-3}\,{\rm ns}$ \\[3pt]
$\Delta t_{\rm 1.5PN}^{\rm S}$ & $8 \times 10^{-3}\,{\rm ns}$ & $2 \times 10^{-4}\,{\rm ns}$ & $4 \times 10^{-5}\,{\rm ns}$ \\[3pt]
\hline
\hline
\end{tabular}
\caption{The numerical magnitude for time delay in the field of one Solar system body (either Sun, Jupiter or Saturn) according to
the upper limits given by Eqs.~(\ref{Shapiro_Monopoles_1_F1}), (\ref{Shapiro_Monopoles_1_F2}), (\ref{Shapiro_Spin_7}), and (\ref{Estimate_Q3}). 
The parameters for Sun and giant planets Jupiter and Saturn are summarized in Table~\ref{Table1}.
The given numerical values are determined for grazing lightrays, that means the impact parameter equals the radius of the massive body: $d_A = P_A$.
The given magnitude for time delay should be compared with the aimed accuracies of future astrometry missions proposed to ESA like
ASTROD \cite{Astrod1,Astrod2}, LATOR \cite{Lator1,Lator2}, ODYSSEY \cite{Odyssey}, SAGAS \cite{Sagas}, or TIPO \cite{TIPO}, which aim at an  
accuracy in the determination of time delay for a light-signal better than $\Delta t \sim 0.1\,{\rm ns}$. Accordingly, 1.5PN effects
in time delay will surely not be detectable even within the very next generation of high-precision space-based astrometry missions.}
\label{Table2}
\end{table}

\subsection{Time-delay for moving mass-quadrupole}\label{Shapiro_Quadrupoles} 

\subsubsection{In terms of coordinate time} 

From Eqs.~(\ref{Second_Integration_Mass_4}) and (\ref{Second_Integration_Mass_5}) we obtain in the field of arbitrary-moving quadrupoles ($l=2$) the
expressions $\Delta \ve{x}_{\rm 1PN}^{A\,J_2}$ and $\Delta \ve{x}_{\rm 1.5PN}^{A\,J_2}$, respectively. Then, according to
Eqs.~(\ref{Shapiro_15}) - (\ref{Shapiro_15_B}) we obtain for the time delay: 
\begin{eqnarray}
\Delta\,c t^{J_2} &=& \Delta\,c t^{J_2}_{\rm 1PN} + \Delta\,c t^{J_2}_{\rm 1.5PN}\,,
\label{Time_Delay_Quadrupole_1}
\\
\nonumber\\
\nonumber\\
\Delta\,c t^{J_2}_{\rm 1PN} &=& - \sum\limits_{A=1}^N \ve{k} \cdot \Delta \ve{x}_{\rm 1PN}^{A\,J_2}\left(t_1,t_0\right), 
\label{Time_Delay_Quadrupole_2}
\\
\nonumber\\
\Delta\,c t^{J_2}_{\rm 1.5PN} &=& - \sum\limits_{A=1}^N \ve{k} \cdot \Delta \ve{x}_{\rm 1.5PN}^{A\,J_2}\left(t_1,t_0\right). 
\label{Time_Delay_Quadrupole_3}
\end{eqnarray}

\noindent 
Actually, the expression $\Delta \ve{x}_{\rm 1PN}^{J_2}$ has already been presented in its explicit form by Eq.~(144) in \cite{Zschocke_1PN}. In view of their 
involved structure, $\Delta \ve{x}_{\rm 1PN}^{J_2}$ as well as $\Delta \ve{x}_{\rm 1.5PN}^{J_2}$ will not be given here.  
The estimate of (\ref{Time_Delay_Quadrupole_2})  
and (\ref{Time_Delay_Quadrupole_3}) proceeds very similar to what has been done in detail in \cite{Zschocke_Klioner}. For an axisymmetric body 
one obtains after some amount of algebra:  
\begin{eqnarray} 
\left| \Delta\,t^{J_2}_{\rm 1PN}\right| &\le& 3\,\left|J^A_2\right|\,\frac{G M_A}{c^3}\,, 
\label{Estimate_Q1} 
\\
\nonumber\\
\left| \Delta\,t^{J_2}_{\rm 1.5PN}\right| &\le& \frac{v_A}{c}\,\left| \Delta\,t^{J_2}_{\rm 1PN}\right|  
+ 6\,\left|J^A_2\right|\,\frac{G M_A}{c^3}\,\frac{v_A}{c}\,\frac{r_A^1}{d_A^1}\,,   
\nonumber\\
\label{Estimate_Q2}
\end{eqnarray} 

\noindent 
where $J_2$ is the actual coefficient of second zonal harmonics. 
The estimate in (\ref{Estimate_Q1}) agrees with the estimate for quadrupoles at rest, cf. Eq.~(26) in \cite{Zschocke_Klioner}.  
Like in (\ref{Shapiro_Monopoles_1_E}), we encounter in (\ref{Estimate_Q2}) we encounter a large term which is proportional to $\sim r_A^1/d_A^1$.

\subsubsection{In terms of retarded time} 
 
With the aid of relations (\ref{Retardation_A}) - (\ref{Retardation_B}) one rewrites $\Delta \ve{x}_{\rm 1PN}^{J_2}\left(t\right)$ and 
$\Delta \ve{x}_{\rm 1.5PN}^{J_2}\left(t\right)$ in terms of retarded time. Formally, one may also replace $M_{a b}^A\left(t_n\right)$ by 
$M_{a b}^A\left(t^{\rm ret}_n\right),\;n=0,1$, but the impact of such replacement on time delay is negligible.   
Then, after considerable amount of algebra, one obtains the correct estimates in 1.5PN correction, which are given by: 
\begin{eqnarray}
\left| \Delta\,t^{J_2}_{\rm 1PN}\right| &\le& 3\,\left|J^A_2\right|\,\frac{G M_A}{c^3}\,,
\label{Estimate_Q3}
\\
\nonumber\\
\left| \Delta\,t^{J_2}_{\rm 1.5PN}\right| &\le& \frac{v_A\left(t^{\rm ret}_1\right)}{c}\,\left| \Delta\,t^{J_2}_{\rm 1PN}\right|. 
\label{Estimate_Q4}
\end{eqnarray}
 
\noindent
The numerical magnitude of the 1PN correction in (\ref{Estimate_Q3}) is given in Table~\ref{Table2}, while  
the 1.5PN correction in (\ref{Estimate_Q4}) is by far much below the detectability of future astrometry missions and will not be given in Table \ref{Table2}.  

In view of the tininess of $\Delta\,t^{J_2}_{\rm 1PN}$ it becomes obvious that higher multipole terms are negligible in the time delay 
and, therefore, will not be considered here.

\section{Light-deflection}\label{Observable_Effects_Light_Deflection} 

The light deflection is of fundamental importance in astrometric measurements. 
Like in the previous section, we assume the light source to be located at $\ve{x}_0 = \ve{x}\left(t_0\right)$ where $t_0$ is the moment of
emission of the light-signal, and the observer is located at $\ve{x}_1 = \ve{x}\left(t_1\right)$ where $t_1$ is the moment of reception of the
light-signal by the observer. Both the light source and the observer are assumed to be at rest with respect
to the global reference system.

The light deflection is defined by the angle $\varphi$ between unit vector $\ve{k}$ and   
the unit tangent vector $\ve{n}$ of the lightray at the observers position: $\varphi = \arcsin \left|\ve{k} \times \ve{n}\right|$ \cite{Klioner2003a}. 
In 1.5PN approximation the unit tangent vector at the observer is given by  
\begin{eqnarray}
\ve{n}_{\rm 1.5PN}\left(\tau_1+t^{\ast}\right) &=& \frac{\dot{\ve{x}}_{\rm 1.5PN}\left(\tau_1+t^{\ast}\right)}
{\left|\dot{\ve{x}}_{\rm 1.5PN}\left(\tau_1+t^{\ast}\right)\right|}\,.
\label{Light_Deflection_5}
\end{eqnarray}

\noindent
By inserting Eq.~(\ref{first_integral_geodesic_equation_A}) into (\ref{Light_Deflection_5}), we obtain  
\begin{eqnarray}
\ve{n}_{\rm 1.5PN}\left(\tau_1+t^{\ast}\right) \!&=&\! \ve{\sigma} 
+ \sum\limits_{A=1}^N\ve{\sigma} \times \left(\frac{\Delta \dot{\ve{x}}^A_{\rm 1PN}\left(\tau_1+t^{\ast}\right)}{c} \times \ve{\sigma}\right)  
\nonumber\\
\nonumber\\
&& \hspace{-1.0cm} + \sum\limits_{A=1}^N \ve{\sigma} \times
\left(\frac{\Delta \dot{\ve{x}}^A_{\rm 1.5PN}\left(\tau_1+t^{\ast}\right)}{c} \times \ve{\sigma}\right), 
\label{Light_Deflection_10}
\end{eqnarray}

\noindent
where $\Delta \dot{\ve{x}}^A_{\rm 1PN}$ and $\Delta \dot{\ve{x}}^A_{\rm 1.5PN}$ are given 
by (\ref{first_integral_geodesic_equation_1PN}) and (\ref{first_integral_geodesic_equation_15PN}), respectively, with the expressions  
in Eqs.~(\ref{First_Integration_9}), (\ref{First_Integration_10}) and (\ref{First_Integration_15}).  
The 1.5PN relation (\ref{Light_Deflection_10}) generalizes the 1PN relation (156) in \cite{Zschocke_1PN}.  

The expression in (\ref{Light_Deflection_10}) for the unit tangent vector along the light trajectory at observers position is valid 
in case of stars, which means in case of light sources which are at far distances from the observer. 
For astrometry within the Solar system we need to obtain an expression which is valid for light sources at finite distances from the observer.  
In order to obtain such an expression we use the following relation among the vectors $\ve{k}$ and $\ve{\sigma}$,   
\begin{eqnarray}
\ve{\sigma} &=& \ve{k} - \frac{1}{R} \sum\limits_{A=1}^N 
\left[\ve{k}\times \bigg(\Delta\ve{x}^A_{\rm 1PN}\left(\tau_1 + t^{\ast}, \tau_0 + t^{\ast}\right)\times\ve{k}\bigg)\right]  
\nonumber\\
\nonumber\\
&& \hspace{-0.7cm} - \frac{1}{R} \sum\limits_{A=1}^N 
\left[\ve{k}\times \bigg(\Delta\ve{x}^A_{\rm 1.5PN}\left(\tau_1 + t^{\ast}, \tau_0+t^{\ast}\right)\times\ve{k}\bigg)\right],  
\label{Light_Deflection_15}
\end{eqnarray}

\noindent
where $\Delta \ve{x}^A_{\rm 1PN}$ and $\Delta \ve{x}^A_{\rm 1.5PN}$ are given by Eqs.~(\ref{second_integral_geodesic_equation_1PN}) and  
(\ref{second_integral_geodesic_equation_15PN}), respectively, with the expressions in  
Eqs.~(\ref{Second_Integration_Mass_4}), (\ref{Second_Integration_Mass_5}) and (\ref{Second_Integration_Spin_5}).

The relation follows from the definitions (\ref{Shapiro_5}) and (\ref{Shapiro_10}) and with the aid of the expression for the  
light trajectory in (\ref{second_integral_geodesic_equation}) and for the Shapiro effect in (\ref{Shapiro_15}).  
The 1.5PN expression in (\ref{Light_Deflection_15}) generalizes the 1PN relation (157) in \cite{Zschocke_1PN}.  
We also notice that the first line in (\ref{Light_Deflection_15}) agrees with Eq.~(66) in \cite{Klioner2003a}. 
By inserting (\ref{Light_Deflection_15}) into (\ref{Light_Deflection_10}) we finally arrive at the following expression for the unit 
tangent vector at the observers position:  
\begin{eqnarray}
\ve{n}_{\rm 1.5PN}\left(\tau_1+t^{\ast}\right) &=& \ve{k}
\nonumber\\
\nonumber\\
&& \hspace{-2.5cm} - \frac{1}{R} \sum\limits_{A=1}^N 
\left[\ve{k} \times \bigg( \Delta \ve{x}^A_{\rm 1PN} \left(\tau_1 + t^{\ast}, \tau_0 + t^{\ast}\right) \times \ve{k}\bigg)\right]  
\nonumber\\
\nonumber\\
&& \hspace{-2.5cm} + \sum\limits_{A=1}^N \ve{k} \times \left(\frac{\Delta \dot{\ve{x}}^A_{\rm 1PN}\left(\tau_1+t^{\ast}\right)}{c} \times \ve{k}\right)
\nonumber\\
\nonumber\\
&& \hspace{-2.5cm} - \frac{1}{R} \sum\limits_{A=1}^N 
\left[\ve{k} \times \bigg(\Delta \ve{x}^A_{\rm 1.5PN} \left(\tau_1 + t^{\ast}, \tau_0 + t^{\ast}\right) \times \ve{k}\bigg)\right]  
\nonumber\\
\nonumber\\
&& \hspace{-2.5cm} + \sum\limits_{A=1}^N \ve{k} \times \left(\frac{\Delta \dot{\ve{x}}^A_{\rm 1.5PN}\left(\tau_1+t^{\ast}\right)}{c} \times \ve{k}\right).  
\label{Light_Deflection_20}
\end{eqnarray}

\noindent
The 1.5PN relation in (\ref{Light_Deflection_20}) generalizes the 1PN relation (158) in \cite{Zschocke_1PN}.  
The formula (\ref{Light_Deflection_20}) is valid for light sources at finite distance. In the limit of infinite spatial distances, $R \rightarrow \infty$, 
the relation (\ref{Light_Deflection_20}) changes into the expression in (\ref{Light_Deflection_10}).  
 
In summary of this section, the expression for the time delay in (\ref{Shapiro_15}) and for the unit tangent vector in (\ref{Light_Deflection_20}) are valid  
for a light-signal which has been emitted by a source located at finite spatial distances, and which propagates through the Solar system, that means 
through the gravitational field of $N$ arbitrarily moving bodies and having arbitrary  
shape and inner structure and which can be in arbitrary rotational motion.  

If the light-source is located at infinity, i.e. $R \rightarrow \infty$ and in a good approximation realized by stars or quasars, 
then the light deflection angle of a light-signal in the field of $N$ arbitrarily moving bodies in 1.5PN approximation is determined by  
\begin{eqnarray}
\varphi &=& \left|\ve{\sigma} \times \ve{n}_{\rm 1.5PN}\right|,
\label{Deflection1}
\end{eqnarray}

\noindent
where $\ve{n}_{\rm 1.5PN}$ is given by Eq.~(\ref{Light_Deflection_10}).
If the light-source is located at finite distance, i.e. $R$ is finite and in a good approximation realized by Solar system objects,  
then the light deflection angle of a light-signal in the field of $N$ arbitrarily moving bodies in 1.5PN approximation is defined by
\begin{eqnarray}
\varphi &=& \left|\ve{k} \times \ve{n}_{\rm 1.5PN}\right|,  
\label{Deflection2}
\end{eqnarray}

\noindent
where $\ve{n}_{\rm 1.5PN}$ is given by Eq.~(\ref{Light_Deflection_20}). The relation (\ref{Deflection1}) is of simpler structure than (\ref{Deflection2}), but 
which equation can be utilized depends on how far the light-source is. For our preliminary considerations here it will be sufficient to consider light-source 
at infinity, that means to apply just relation (\ref{Deflection1}). Like in case of Shapiro delay, we will consider the light deflection caused by $N$  
arbitrarily moving bodies in some more detail, by considering bodies with mass-multipole structure and spin-dipole-structure. 
An estimate is also given for spin-octupole. The terms which are of relevance for nas-accuracy read  
\begin{eqnarray}
\varphi &=& \varphi^{\rm M}_{\rm 1PN} + \sum\limits_{n = 2}^{10} \varphi^{\rm J_n}_{\rm 1PN} 
\nonumber\\
\nonumber\\
&& + \varphi^{\rm M}_{\rm 1.5PN} + \varphi^{\rm J_2}_{\rm 1.5PN} + \varphi^{\rm S}_{\rm 1.5PN} + \varphi^{\rm SO}_{\rm 1.5PN}\,.  
\label{Light_Deflection_M_S}
\end{eqnarray}

\noindent
In what follows we will consider these terms in some detail and give some estimates of their magnitude.

\subsection{Light deflection for moving mass-monopole}\label{Deflection_Monopole}

\subsubsection{In terms of coordinate time} 

From (\ref{First_Integration_9}) and (\ref{First_Integration_10}) we obtain for the coordinate velocity of the photon in the field of $N$ arbitrarily 
moving monopoles:  
\begin{eqnarray}
\frac{\Delta \dot{\ve{x}}^{\rm M}_{\rm 1PN}\left(t_1\right)}{c} &=& - \frac{2 G}{c^2} \sum\limits_{A=1}^{N}\frac{M_A}{r^{\,1}_A}  
\left[\frac{\ve{d}^{\,1}_A}{r^{\,1}_A - \ve{\sigma}\cdot\ve{r}^{\,1}_A} + \ve{\sigma}\right],  
\nonumber\\
\label{Light_Deflection_M_1}
\end{eqnarray}

\noindent
\begin{eqnarray}
\frac{\Delta \dot{\ve{x}}^{\rm M}_{\rm 1.5PN}\left(t_1\right)}{c} &=& + \frac{2 G}{c^2} \sum\limits_{A=1}^{N}\frac{M_A}{r^{\,1}_A}
\frac{\ve{\sigma}\cdot\ve{v}_A^{\,1}}{c}\,\frac{\ve{d}^{\,1}_A}{r^{\,1}_A - \ve{\sigma}\cdot\ve{r}^{\,1}_A}  
\nonumber\\
\nonumber\\
&& \hspace{-2.5cm} + \frac{4 G}{c^2} \sum\limits_{A=1}^{N}\frac{M_A}{r^{\,1}_A}\,\frac{\ve{v}_A^{\,1}}{c}  
- \frac{2 G}{c^2} \sum\limits_{A=1}^{N} 
\frac{M_A}{r^{\,1}_A - \ve{\sigma}\cdot\ve{r}^{\,1}_A}\,\frac{\ve{\sigma}\times\left(\ve{v}_A^{\,1} \times \ve{\sigma}\right)}{c}  
\nonumber\\
\nonumber\\
&& \hspace{-2.5cm} + \frac{2 G}{c^2} \sum\limits_{A=1}^{N}\frac{M_A}{\left(r^{\,1}_A - \ve{\sigma}\cdot\ve{r}^{\,1}_A\right)^2}\,
\frac{\ve{d}_A^{\,1}\cdot\ve{v}_A^{\,1}}{c}\,\frac{\ve{d}_A^{\,1}}{r^{\,1}_A}\,.  
\label{Light_Deflection_M_2}
\end{eqnarray}

\noindent
In the limit of uniformly moving bodies our result in (\ref{Light_Deflection_M_1}) - (\ref{Light_Deflection_M_2}) agrees with Eq.~(6.3) and (6.5) 
in \cite{KlionerKopeikin1992} and with Eq.~(20) in \cite{Deng_2015} up to terms of the order ${\cal O}\left(v_A^2/c^2\right)$. 
By inserting (\ref{Light_Deflection_M_1}) - (\ref{Light_Deflection_M_2}) into (\ref{Light_Deflection_5}) and then into (\ref{Deflection1}) we obtain 
the light deflection angle, which for one massive body $A$ can be estimated as follows:  
\begin{eqnarray}
\varphi^{\rm M}_{\rm 1PN} &=& \left|\ve{\sigma} \times \frac{\Delta \dot{\ve{x}}^{\rm M}_{\rm 1PN}\left(t_1\right)}{c}\right| 
\le \frac{4 G M_A}{c^2 d_A^1}\,, 
\label{Light_Deflection_M_3}
\\
\nonumber\\
\nonumber\\
\varphi^{\rm M}_{\rm 1.5PN} &=& \left|\ve{\sigma} \times \frac{\Delta \dot{\ve{x}}^{\rm M}_{\rm 1.5PN}\left(t_1\right)}{c}\right| 
\nonumber\\ 
\nonumber\\ 
&\le& \varphi^{\rm M}_{\rm 1PN}\,\frac{v_A}{c} + \frac{8 G M_A}{c^2 d_A}\,\frac{v_A}{c}\,\frac{r_A^1}{d_A^1}\,.
\label{Light_Deflection_M_4}
\end{eqnarray}

\noindent
Like in Eqs.~(\ref{Shapiro_Monopoles_1_E}) and (\ref{Estimate_Q2}), we encounter again the typical large term in (\ref{Light_Deflection_M_4}) which 
is proportional to $\sim r_A^1/d_A^1$ and originates from the last two terms in (\ref{Light_Deflection_M_2}). This large term is solely caused by the  
retardation of gravitational action. That means, the use of the time-moment of reception at the body's position, $\ve{x}_A\left(t_1\right)$  
in Eq.~(\ref{Light_Deflection_M_1}) causes a significant error in the determination of light deflection for moving bodies. This peculiarity has been 
recognized long time ago, for instance see text below Eq.~(34) in \cite{Klioner2003a}. 
Especially, this issue has thoroughly and comprehensively been solved for moving pointlike bodies in the investigations
\cite{KopeikinSchaefer1999,KopeikinMashhoon2002,Klioner2003a,KlionerPeip2003}. 
In the next subsection we will further elucidate this fact.

\subsubsection{In terms of retarded time} 

From the physical point of view, it is obvious that instead of $t_1$ one has to use the retarded time-moment for the  
position of the massive body in (\ref{Light_Deflection_M_1}). That means, with the aid of relations (\ref{Retardation_A}) - (\ref{Retardation_B})  
one may show that Eqs.~(\ref{Light_Deflection_M_1}) - (\ref{Light_Deflection_M_2}) can be rewritten as follows:  
\begin{eqnarray}
\frac{\Delta \dot{\ve{x}}^{\rm M}_{\rm 1PN}\left(t_1\right)}{c} &=& - \frac{2 G}{c^2} \sum\limits_{A=1}^{N}\frac{M_A}{r_A\left(t^{\rm ret}_1\right)}
\nonumber\\
\nonumber\\
&&\hspace{-0.5cm} \times 
\left[\frac{\ve{d}_A\left(t^{\rm ret}_1\right)}{r_A\left(t^{\rm ret}_1\right) - \ve{\sigma}\cdot\ve{r}_A\left(t^{\rm ret}_1\right)} + \ve{\sigma}\right],
\label{Light_Deflection_M_7}
\\
\nonumber\\
\nonumber\\
\frac{\Delta \dot{\ve{x}}^{\rm M}_{\rm 1.5PN}\left(t_1\right)}{c} &=& + \frac{2 G}{c^3} \sum\limits_{A=1}^{N}\frac{M_A}{r_A\left(t^{\rm ret}_1\right)}
\ve{\sigma}\cdot\ve{v}_A\left(t^{\rm ret}_1\right)
\nonumber\\
\nonumber\\
&& \times \frac{\ve{d}_A\left(t^{\rm ret}_1\right)}{r_A\left(t^{\rm ret}_1\right) - \ve{\sigma}\cdot\ve{r}_A\left(t^{\rm ret}_1\right)}
\nonumber\\
\nonumber\\
&& \hspace{-2.5cm} + \frac{4 G}{c^3} \sum\limits_{A=1}^{N}\frac{M_A}{r_A\left(t^{\rm ret}_1\right)}\,\ve{v}_A\left(t^{\rm ret}_1\right)
\nonumber\\
\nonumber\\
&& \hspace{-2.5cm} + \frac{2 G}{c^3} \sum\limits_{A=1}^{N} \frac{M_A}{r_A\left(t^{\rm ret}_1\right)}\, 
\frac{\ve{d}_A\left(t^{\rm ret}_1\right)}{r_A\left(t^{\rm ret}_1\right)}\,
\ve{\sigma}\cdot\ve{v}_A\left(t^{\rm ret}_1\right)  
\nonumber\\
\nonumber\\
&& \hspace{-2.5cm} - \frac{2 G}{c^3} \ve{\sigma} \sum\limits_{A=1}^{N} \frac{M_A}{r_A\left(t^{\rm ret}_1\right)}\,
\frac{\ve{r}_A\left(t^{\rm ret}_1\right) \cdot \ve{v}_A\left(t^{\rm ret}_1\right)}{r_A\left(t^{\rm ret}_1\right)}
\nonumber\\
\nonumber\\
&& \hspace{-2.5cm} - \frac{2 G}{c^3} \sum\limits_{A=1}^{N} \frac{M_A}{r_A\left(t^{\rm ret}_1\right)}\,
\frac{\ve{d}_A\left(t^{\rm ret}_1\right)}{r_A\left(t^{\rm ret}_1\right) - \ve{\sigma}\cdot\ve{r}_A\left(t^{\rm ret}_1\right)} 
\nonumber\\
\nonumber\\
&& \times \frac{\ve{v}_A\left(t^{\rm ret}_1\right) \cdot \ve{d}_A\left(t^{\rm ret}_1\right)}{r_A\left(t^{\rm ret}_1\right)}\,. 
\label{Light_Deflection_M_8}
\end{eqnarray}

\noindent
The last two terms in (\ref{Light_Deflection_M_2}) do not explicitly appear in (\ref{Light_Deflection_M_8}), because they are absorbed 
in (\ref{Light_Deflection_M_7}). Accordingly, instead of (\ref{Light_Deflection_M_4}) we obtain the following correct estimates  
for the 1PN and 1.5PN corrections in (\ref{Light_Deflection_M_7}) and (\ref{Light_Deflection_M_8}), respectively:  
\begin{eqnarray}
\varphi^{\rm M}_{\rm 1PN} &\le& \frac{4 G M_A}{c^2 d_A\left(t_1^{\rm ret}\right)}\,,  
\label{Light_Deflection_M_5}
\\
\nonumber\\
\nonumber\\
\varphi^{\rm M}_{\rm 1.5PN} &\le& \varphi^{\rm M}_{\rm 1PN}\,\frac{v_A\left(t_1^{\rm ret}\right)}{c}\,.  
\label{Light_Deflection_M_6}
\end{eqnarray}

\noindent 
The given upper limit in (\ref{Light_Deflection_M_6}) agrees with Eq.~(42) and (46) in \cite{KopeikinMakarov2007}, and with the results in \cite{Klioner2003a}.  
For numerical values of the upper bound in Eq.~(\ref{Light_Deflection_M_5}) and (\ref{Light_Deflection_M_6}) see Table \ref{Table3}.

\subsection{Light deflection for moving spin-dipole}\label{Deflection_Spin}

\subsubsection{In terms of coordinate time} 

The coordinate velocity of a light-signal propagating in the field of arbitrarily moving spin-dipoles can either be obtained from (\ref{First_Integration_15})  
using (\ref{arbitrarily_moving_spin_dipole_15}), or simply by time-differentiation of Eq.~(\ref{arbitrarily_moving_spin_dipole_35}), and reads: 
\begin{eqnarray}
\frac{\Delta \dot{\ve{x}}^{\rm S}_{\rm 1.5PN}\left(t_1\right)}{c} &=& - \frac{2 G}{c^3} \sum\limits_{A=1}^{N}
\frac{\ve{\sigma}\times \ve{S}_A^1}{\left(r_A^1\right)^3}\,\left(\ve{\sigma}\cdot\ve{r}_A^1\right)  
\nonumber\\
\nonumber\\
&& \hspace{-2.5cm} + \frac{2 G}{c^3} \sum\limits_{A=1}^{N} 
\frac{\ve{S}^1_A \times \ve{d}^1_A}{\left(r_A^1\right)^2}\,\frac{1}{r_A^1 - \ve{\sigma}\cdot\ve{r}_A^1}\left(1 - \frac{\ve{\sigma}\cdot\ve{r}_A^1}{r_A^1}\right)  
\nonumber\\
\nonumber\\
&& \hspace{-2.5cm} + \frac{2 G}{c^3} \sum\limits_{A=1}^{N}\ve{d}^1_A\,\left[\ve{\sigma}\cdot\left(\ve{d}^1_A\times\ve{S}_A^1\right)\right]  
\frac{\ve{\sigma}\cdot\ve{r}_A^1}{\left(r_A^1\right)^3}\,\frac{1}{\left(r_A^1-\ve{\sigma}\cdot\ve{r}_A^1\right)^2}  
\nonumber\\
\nonumber\\
&& \hspace{-2.5cm} - \frac{4 G}{c^3} \sum\limits_{A=1}^{N}\ve{d}^1_A\,\left[\ve{\sigma}\cdot\left(\ve{d}^1_A\times\ve{S}_A^1\right)\right]
\frac{1}{\left(r_A^1\right)^2}\,\frac{1}{\left(r_A^1-\ve{\sigma}\cdot\ve{r}_A^1\right)^2}
\nonumber\\
\nonumber\\
&& \hspace{-2.5cm} - \frac{2 G}{c^3} \sum\limits_{A=1}^{N} \frac{\ve{\sigma}\times\ve{S}_A^1}{r_A^1}\,\frac{1}{r_A^1-\ve{\sigma}\cdot\ve{r}_A^1}\,.   
\label{Light_Deflection_S_1}
\end{eqnarray}

\noindent
One may verify that in the limit of bodies at rest our result agrees with Eq.~(59) in \cite{Klioner1991}.  
An upper bound for the magnitude of the light deflection is given by 
\begin{eqnarray}
\varphi^{\rm S}_{\rm 1.5PN} &=& \left|\ve{\sigma} \times \frac{\Delta \dot{\ve{x}}^{\rm S}_{\rm 1.5PN}\left(t_1\right)}{c}\right| 
\le \frac{4 G S^1_A}{c^3 \left(d_A^1\right)^2}\,,  
\label{Light_Deflection_S_2}
\end{eqnarray}

\noindent
in agreement with the estimate given by Eq.~(65) in \cite{Klioner1991} for lightrays which propagate in the equatorial plane of a rotating body at rest.

\subsubsection{In terms of retarded time} 

Following the same arguments as in the above considerations, we may replace all expression in (\ref{Light_Deflection_S_1})
by their retarded expressions according to Eqs.~(\ref{Shapiro_Spin_4}) - (\ref{Shapiro_Spin_6}) for $n=1$. Then the estimate of light deflections yields: 
\begin{eqnarray}
\varphi^{\rm S}_{\rm 1.5PN} \!\! &\le& \frac{4 G S_A\left(t^{\rm ret}_1\right)}{c^3 \left(d_A\left(t^{\rm ret}_1\right)\right)^2}\,,  
\label{Light_Deflection_S_3}
\end{eqnarray}

\noindent
which formally agrees with the estimate in Eq.~(\ref{Light_Deflection_S_2}).  
For numerical values of the upper bound in Eq.~(\ref{Light_Deflection_S_3}) see Table \ref{Table3}.  
\begin{table}[h!]
\begin{tabular}{| c | c | c | c |}
\hline
\hline
&&&\\[-7pt]
Term
&\hbox to 20mm{\hfill Sun [\muas]\hfill} 
&\hbox to 20mm{\hfill Jupiter [\muas] \hfill}
&\hbox to 20mm{\hfill Saturn [\muas]\hfill}\\[3pt] 
\hline
&&&\\[-7pt]
$\varphi_{\rm 1PN}^{\rm M}$ & $1.75 \times 10^6$ & $16.3 \times 10^3$ & $5.8 \times 10^3$ \\[3pt]
$\varphi_{\rm 1PN}^{J_2}$ & $1$ & $240$ & $95$ \\[3pt]
$\varphi_{\rm 1PN}^{J_4}$ & $ - $ & $9.6$ & $5.46$ \\[3pt]
$\varphi_{\rm 1PN}^{J_6}$ & $ - $ & $0.56$ & $0.50$ \\[3pt]
$\varphi_{\rm 1PN}^{J_8}$ & $ - $ & $0.04$ & $0.06$ \\[3pt]
$\varphi_{\rm 1PN}^{J_{10}}$ & $ - $ & $0.003$ & $0.01$ \\[3pt]
\hline
$\varphi_{\rm 1.5PN}^{\rm M}$ & $0.1$ & $0.8$ & $0.2$ \\[3pt]
$\varphi_{\rm 1.5PN}^{J_2}$ & $ - $ & $ 0.011$ & $0.003$ \\[3pt]
$\varphi_{\rm 1.5PN}^{\rm S}$ & $0.7$ & $0.2$ & $0.04$ \\[3pt]
$\varphi_{\rm 1.5PN}^{\rm SO}$ & $ - $ & $0.015$ & $0.006$ \\[3pt]
\hline
\hline
\end{tabular}
\caption{The numerical magnitude for light deflection in the field of one Solar system body (either Sun, Jupiter or Saturn) according to 
the upper limits given by Eqs.~(\ref{Light_Deflection_M_5}), (\ref{Light_Deflection_M_6}), (\ref{Light_Deflection_S_3}),  
(\ref{Light_Deflection_Q_3}), (\ref{Light_Deflection_Q_4}) and (\ref{Light_Deflection_Higher_Multipoles_1}).  
The parameters for Sun and giant planets Jupiter and Saturn are summarized in Table~\ref{Table1}. The given numerical values are determined for grazing lightrays,  
that means the impact parameter equals the radius of the massive body: $d_A = P_A$. For the light deflection in the field of spin-octupole, 
$\varphi_{\rm 1.5PN}^{\rm SO}$, we take the results of Ref.~\cite{Jan-Meichsner_Diploma_Thesis} where the light deflection in the field of one rotating body 
at rest and having constant mass density has been determined. Blank entries indicate that the effect is smaller than $1\,{\rm nas}$.  
In view of the fact that astrometry on sub-\muas-level implies an accuracy for $\varphi$ at least better than $0.1\,\muas$, the 1.5PN effects 
in light deflection become detectable within the very next generation of high-precision space-based astrometry missions.}
\label{Table3}
\end{table}

\subsection{Light deflection for moving mass-quadrupole}\label{Deflection_Quadrupole}

\subsubsection{In terms of coordinate time} 

The 1PN correction to the coordinate velocity of the lightray in the field $N$ arbitrarily-moving bodies with quadrupole structure,  
$\Delta \dot{\ve{x}}_{\rm 1PN}^{J_2}\left(t\right)$, has already been given Eq.~(117) in \cite{Zschocke_1PN} and can also be deduced from 
Eq.~(\ref{First_Integration_9}), while the 1.5PN correction $\Delta \dot{\ve{x}}_{\rm 1.5PN}^{J_2}\left(t\right)$ from Eq.~(\ref{First_Integration_10}).  
In view of the complexity of these terms, we will not present these expressions in their explicit form. We just mention that  
the estimation of these terms proceeds similar to the procedure performed in \cite{Zschocke_Klioner}. After some considerable amount of algebra one obtains:  
\begin{eqnarray}
\varphi^{J_2}_{\rm 1PN} &=& \left|\ve{\sigma} \times \frac{\Delta \dot{\ve{x}}^{J_2}_{\rm 1PN}\left(t_1\right)}{c}\right|
\le \frac{4 G M_A}{c^2}\left|J_2^A\right|\frac{\left(P_A\right)^2}{\left(d_A^1\right)^3}\,,
\nonumber\\
\label{Light_Deflection_Q_1}
\\
\nonumber\\
\nonumber\\
\varphi^{J_2}_{\rm 1.5PN} &=& \left|\ve{\sigma} \times \frac{\Delta \dot{\ve{x}}^{J_2}_{\rm 1.5PN}\left(t_1\right)}{c}\right|
\nonumber\\
\nonumber\\
&\le& \varphi^{J_2}_{\rm 1PN}\,\frac{v_A}{c}
+ \frac{8 G M_A}{c^2}\left|J_2^A\right|\,\frac{v_A}{c}\,\frac{\left(P_A\right)^2}{\left(d_A^1\right)^3}\,\frac{r_A^1}{d^1_A}\,.
\nonumber\\
\label{Light_Deflection_Q_2}
\end{eqnarray}

\noindent
The estimate of the 1PN quadrupole term in (\ref{Light_Deflection_Q_1}) is equal to the much simpler case of quadrupoles at rest, 
cf. Eq.~(41) in \cite{Klioner1991} and Eq.~(13) in \cite{Zschocke_Klioner}.  
The second term in (\ref{Light_Deflection_Q_2}) is proportional to $\sim r_A^1/d^1_A$ which for grazing rays becomes large.   
Like in Eqs.~(\ref{Shapiro_Monopoles_1_E}), (\ref{Estimate_Q2}), and (\ref{Light_Deflection_M_4}), this term is caused by the 
finite speed of gravitational action.

\subsubsection{In terms of retarded time}  

One may rewrite the expression for $\Delta \dot{\ve{x}}_{\rm 1PN}^{J_2}\left(t\right)$ and $\Delta \dot{\ve{x}}_{\rm 1.5PN}^{J_2}\left(t\right)$ in terms of 
retarded time by means of Eqs.~(\ref{Retardation_A}) - (\ref{Retardation_B}), and formally one may also replace  
$M_{a b}\left(t_1\right) \rightarrow M_{a b}\left(t^{\rm ret}_1\right)$. Then the estimation of the 1PN and 1.5PN correction terms  
in the quadrupole light deflection becomes
\begin{eqnarray}
\varphi^{J_2}_{\rm 1PN} &\le& \frac{4 G M_A}{c^2}\left|J_2^A\right|\frac{\left(P_A\right)^2}{\left(d_A\left(t^{\rm ret}_1\right)\right)^3}\,,  
\label{Light_Deflection_Q_3}
\\
\nonumber\\
\varphi^{J_2}_{\rm 1.5PN} &\le& \varphi^{J_2}_{\rm 1PN}\,\frac{v_A\left(t^{\rm ret}_1\right)}{c}\,,
\label{Light_Deflection_Q_4}
\end{eqnarray}

\noindent
which agrees with Eqs.~(44) and (46) in \cite{KopeikinMakarov2007}. 
The numerical magnitude of these upper bounds (\ref{Light_Deflection_Q_3}) and (\ref{Light_Deflection_Q_4}) can be found in Table \ref{Table3}.

\subsection{Light deflection for higher mass-multipoles}\label{Deflection_Multipoles}

The 1PN solution (\ref{First_Integration_9}) and the 1.5PN solution (\ref{First_Integration_10}) for moving bodies with full mass-multipole 
structure allow to determine the light deflection in the field of moving mass-multipoles to any order in $l$. However, the expressions for 
$\Delta \dot{\ve{x}}_{\rm 1PN}^{J_n}$ and $\Delta \dot{\ve{x}}_{\rm 1.5PN}^{J_n}$ ($J_n$ are the actual zonal harmonic coefficients 
of the massive body) become more and more involved the higher the  
order of the mass-multipoles are and imply a considerable amount of algebra. The investigation of these terms will be postponed for awhile. 
In meanwhile let us consider an educated guess that the light deflection in the field of higher mass-multipoles is determined by the following relation:  
\begin{eqnarray}
\varphi^{J_n}_{\rm 1PN} &=& \left|\ve{\sigma} \times \frac{\Delta \dot{\ve{x}}^{\rm J_n}_{\rm 1PN}\left(t_1\right)}{c}\right|
\le \frac{4 G M_A}{c^2}\frac{\left|J_n^A\right|\,\left(P_A\right)^n}{\left(d_A\left(t^{\rm ret}_1\right)\right)^{n+1}}\,,
\nonumber\\
\label{Light_Deflection_Higher_Multipoles_1}
\\
\nonumber\\
\varphi^{J_n}_{\rm 1.5PN} &=& \left|\ve{\sigma} \times \frac{\Delta \dot{\ve{x}}^{\rm J_n}_{\rm 1.5PN}\left(t_1\right)}{c}\right|
\le \varphi^{J_n}_{\rm 1PN}\,\frac{v_A\left(t^{\rm ret}_1\right)}{c}\,,
\label{Light_Deflection_Higher_Multipoles_2}
\end{eqnarray}

\noindent
which in case of $n=2$ agrees with Eqs.~(\ref{Light_Deflection_Q_3}) - (\ref{Light_Deflection_Q_4}). The suggestion in  
Eqs.~(\ref{Light_Deflection_Higher_Multipoles_1}) - (\ref{Light_Deflection_Higher_Multipoles_2}) is based on the considerations above 
and triggered by the fact that in the limit of bodies at rest formula (\ref{Light_Deflection_Higher_Multipoles_1}) agrees with the results 
in \cite{Poncin_Lafitte_Teyssandier_2008}.  
Numerical values for (\ref{Light_Deflection_Higher_Multipoles_1}) are presented in Table~\ref{Table3}, while  
(\ref{Light_Deflection_Higher_Multipoles_2}) yields values below $1\,{\rm nas}$ for $n \ge 3$. A detailed proof of (\ref{Light_Deflection_Higher_Multipoles_1})  
and (\ref{Light_Deflection_Higher_Multipoles_2}) and a comparison of formula (\ref{Light_Deflection_Higher_Multipoles_2}) 
with \cite{Hees_Bertone_Poncin_Lafitte_2014a} will be presented in a subsequent investigation.

\section{Summary and Outlook}\label{Summary_Outlook}

During the last 25 years, astrometric measurements have made an impressive advancement from milli-arcsecond level of accuracy by the ESA  
astrometry mission {\it Hipparchos} \cite{Hipparcos1,Hipparcos2} toward micro-arcsecond level of accuracy by the ESA astrometry mission {\it Gaia} \cite{GAIA}.  
Ever since, applied relativity has evolved into one of the basic components of modern astrometry, the branch of science which includes the whole machinery of
advanced astrometric measurements, especially: (1) theory of reference systems, (2) precise description of light trajectory from the celestial light source 
toward the observer, (3) relativistic modeling of real observations, (4) determination of the metric of the Solar system in post-Newtonian approximation  
(weak-field slow-motion approximation) or post-Minkowskian approximation (weak-field approximation) and beyond,  
(5) multipole expansion of metric tensor of Solar system, (6) relativistic data reduction of astrometric measurements, and  
(7) determination of ephemeris of the Solar system bodies and of the observer accurate enough for a given accuracy.  

But for all that stunning progress, the step from micro-arcsecond toward nano-arsecond astrometry will be a long-term ambition, which implies many challenges  
on theoretical as well as technological side. While a few of these issues have been mentioned in the introductory section, most of these  
challenges and especially their elaborated details cannot be foreseen at present. But for any actual ambitions about sub-micro-arcsecond astrometry  
two of these problems are of decisive importance: first to establish a set of accurate reference systems and reference frames for exact data reduction,  
and second to provide an accurate modeling of light trajectory from the celestial light source through the Solar system toward the observer.  
As it has been mentioned in the introductory section, especially these two highly important issues have also been emphasized by the 
ESA-Senior-Survey-Committee (SSC) in response of the selection of science themes for future space-based astrometry missions \cite{SSC}. The presented  
investigation is mainly devoted to these two specific subjects. Especially, in order to arrive at a precise modeling of light-propagation through the 
Solar system, two difficult aspects have carefully to be treated:  

({\bf 1}) First, in compliance with the requirements of the IAU recommendations \cite{IAU_Resolution1,IAU_Resolution2}, one has to introduce  
one global reference system (BCRS) and $N$ local reference systems (GCRS-like), one for each massive body, which  
allow to describe the global metric of the Solar system in terms of intrinsic mass-multipoles and intrinsic spin-multipoles  
the massive bodies, that means for the metric perturbations $h_{\alpha\beta}\left(M^A_L,S^A_L\right)$, as mentioned by Eq.~(\ref{Global_Metric_B}).  

({\bf 2}) Second, for sub-micro-arcsecond or even nano-arcsecond-astrometry one has to describe the light trajectory in the field 
of arbitrarily moving massive bodies, that means as a function of their worldlines $\ve{x}_A\left(t\right)$, because a series expansion like in  
Eq.~(\ref{worldline_introduction}) is unsuitable for several reasons discussed in the introductory section. The worldlines can be concretized by 
Solar system ephemeris \cite{JPL} at any stage of the calculations.  

In a previous investigation \cite{Zschocke_1PN} we have obtained a solution in 1PN approximation for the light trajectory through the Solar system in 
full agreement with these both requirements ({\bf 1}) and ({\bf 2}). As outlined in more detail in \cite{Zschocke_1PN} and also mentioned in the 
introductory section, for high-precision astrometry on sub-\muas-level or nas-level of accuracy the 1PN approximation is not sufficient at all.  
Instead, it is inevitable to determine the light trajectory through the Solar system in 1.5PN approximation and to reconcile the entire approach 
with the important requirements ({\bf 1}) and ({\bf 2}). Such an approach has been developed here in the presented investigation. Accordingly, the main results  
of our investigation are given by the first integration of geodesic equation in Eq.~(\ref{first_integral_geodesic_equation_A}) and 
by the second integration of geodesic equation in Eq.~(\ref{second_integral_geodesic_equation}):  
\begin{eqnarray}
\dot{\ve{x}}_{\rm 1.5PN} &=& c\,\ve{\sigma} + \Delta \dot{\ve{x}}_{\rm 1PN} + \Delta \dot{\ve{x}}_{\rm 1.5PN}\,,  
\label{Summary_1}
\\
\nonumber\\
\ve{x}_{\rm 1.5PN} &=& \ve{\xi} + c \tau \ve{\sigma} + \Delta \ve{x}_{\rm 1PN} + \Delta \ve{x}_{\rm 1.5PN}\,,
\label{Summary_2}
\end{eqnarray}

\noindent
where the time-argument $\tau + t^{\ast}$ have been omitted here for simpler notation. The terms in (\ref{Summary_1}) for one body $A$  
are given by Eqs.~(\ref{First_Integration_9}), (\ref{First_Integration_10}), and (\ref{First_Integration_15}), respectively, and the terms in (\ref{Summary_2})  
for one body $A$ are given by Eqs.~(\ref{Second_Integration_Mass_4}), (\ref{Second_Integration_Mass_5}), and (\ref{Second_Integration_Spin_5}), respectively.  
 
In view of the complexity of the solution in (\ref{Summary_1}) and (\ref{Summary_2}), several cross-checks have been performed:  
\begin{enumerate}
\item[$\bullet$] time-derivative of (\ref{first_integral_geodesic_equation_A}) yields (\ref{transformed_geodesic_equation_C}).  
\item[$\bullet$] time-derivative of (\ref{second_integral_geodesic_equation}) yields (\ref{first_integral_geodesic_equation_A}).  
\item[$\bullet$] our results agree with \cite{Kopeikin1997} for bodies at rest and time-independent mass-multipoles in 1PN approximation.  
\item[$\bullet$] our results agree with \cite{Kopeikin1997} for bodies at rest and time-independent spin-multipoles in 1.5PN approximation.  
\item[$\bullet$] our results agree with \cite{KopeikinMashhoon2002} for arbitrarily moving bodies with spin-dipole in 1.5PN approximation.  
\item[$\bullet$] our results agree with \cite{Klioner1991} for bodies at rest with spin-dipole.  
\end{enumerate}

Further cross-checks in 1PN approximation have already been done in \cite{Zschocke_1PN} for the case of light-propagation in the field of bodies with 
mass-monopole, mass-dipole, mass-quadrupole structure and bodies at rest with full mass-multipole structure.  

The numerical magnitude about the impact of mass-multipoles and spin-multipoles on light deflection, presented in Table~\ref{Table3}, reveal that  
the first mass-multipoles up to order $l=10$ and the first spin-multipoles up to order $l=3$ have to be taken into account for astrometry on nano-arcsecond 
level of accuracy.  
This fact is important in view of the complexity of the 1.5PN solution for the light trajectory, because it allows to simplify that solution considerably.  
However, more detailed investigations are very necessary in order to simplify the massive computations in astrometric data reduction as much as possible.  

The approach presented has further to be developed into several directions before the conditions are complied for a complete modeling of light-propagation  
through the Solar system on sub-\muas $\;$ or nas-level of accuracy. In particular, the following issues may serve as minimal supplement to the list of aspects 
which have already been mentioned in the introductory section:  
 
${\bf A.}$ The model for the light trajectory has to implement some terms in 2PN approximation, which can formally be written as follows: 
\begin{eqnarray}
\dot{\ve{x}}_{\rm 2PN} &=& c\,\ve{\sigma} + \Delta \dot{\ve{x}}_{\rm 1PN} + \Delta \dot{\ve{x}}_{\rm 1.5PN} + \Delta \dot{\ve{x}}_{\rm 2PN}\,,
\label{Summary_3}
\\
\nonumber\\ 
\ve{x}_{\rm 2PN} &=& \ve{\xi} + c\,\tau \ve{\sigma} + \Delta \ve{x}_{\rm 1PN} + \Delta \ve{x}_{\rm 1.5PN} + \Delta \ve{x}_{\rm 2PN}\,, 
\nonumber\\ 
\label{Summary_4}
\end{eqnarray}
 
\noindent  
where (\ref{Summary_3}) and (\ref{Summary_4}) represents the coordinate velocity and the trajectory of the light-signal, respectively. 
The 2PN corrections have been determined for the case of monopoles at rest \cite{Brumberg1987,Brumberg1991} and later recalculated in 
progressing investigations in \cite{KlionerKopeikin1992,Klioner_Zschocke,Deng_Xie} and also within this work, see appendix \ref{Appendix_2PN}. 
It is clear that for a comprehensive theory of light propagation 
aiming at sub-\muas-level of accuracy it needs carefully to be scrutinized which 2PN corrections beyond the monopole part are of relevance for such 
extremely-precise astrometry.  

${\bf B.}$ A fundamental prerequisite in order to gain further progress in the theory of light propagation in 2PN approximation, one necessarily needs  
to determine the space-space part of the BCRS as well as of the GCRS metric tensor including all terms of the order ${\cal O}\left(c^{-4}\right)$.  
However, an extension of these global and local reference systems to the post-post-Newtonian order is a highly involved assignment of a task and 
is presently an active field of research \cite{Chinese_Xu_Wu,Xu_Gong_Wu_Soffel_Klioner,MinazzolliChauvineau2009,KS} and far from being completed.  

${\bf C.}$ In the first instance, the post-Newtonian approach of the DSX formalism allows for astrometry in the near-zone of the Solar system. However, astrometric  
measurements of stars or extragalactic celestial objects are subject to far-zone astrometry, which requires a matching procedure of two asymptotic solutions:  
the near-zone solution and the far-zone solution for the light trajectory \cite{Kopeikin_Efroimsky_Kaplan}. Such matching approach has been 
proposed in \cite{KlionerKopeikin1992,Will_2003}, which has to be further developed in such a way to be in line with the requirements of nas-astrometry. 

${\bf D.}$ The unique interpretation of observational data implies a hierarchy of several reference systems \cite{Brumberg1991,Kopeikin_Efroimsky_Kaplan}: 
\begin{enumerate} 
\item[a)] BCRS $\left(x^0, x^1, x^2, x^3\right)$ for description of the light trajectory in the Solar system,  
\item[b)] GCRS-like $\left(X_A^0, X_A^1, X_A^2, X_A^3\right)$, one for each body $A=1,...,N$ of the Solar system in order to define the intrinsic multipoles,   
\item[c)] CoMRS $\left({\cal X}^0, {\cal X}^1, {\cal X}^2, {\cal X}^3\right)$ which is co-moving with the observer,  
\item[d)] ToRS $\left(z_a^0, z_a^1, z_a^2, z_a^3\right)$, one for each ground-station $a=1,...,n$ on Earth which are involved in data reduction,  
\end{enumerate} 

\noindent
where CoMRS stands for {\it co-moving reference system} and ToRS denotes {\it topocentric reference system}.  
The light trajectory in our investigation is given in the BCRS, but that is of course not sufficient for a comprehensive astrometric model of light propagation.  
In particular, the presented solution has to be transformed into the reference system which is co-moving with a free-falling observer (CoMRS) \cite{COMRS}.  
This transformation takes account for aberrational effects. Especially, it has to be clarified whether or not the CoMRS in \cite{COMRS},  
which was primarily intended for the Gaia mission, is also sufficient for the requirements on nas-level of accuracy.

${\bf E.}$ The basic assumption of post-Newtonian expansion is that all retardations of the gravitational actions are small. In the model presented  
the effect of retardation has been implemented in a more or less heuristic manner, in order to provide a proper estimation for the upper limit of time delay  
and light deflection. This procedure needs to be scrutinized in considerably more detail. Especially, it has to be clarified 
how the retardation of gravitational action has to be implemented based on clear theoretical foundation in the entire approach.  
The solution of this problem is related to the far-zone astrometry about how the presented solution in the near-zone 
can be matched with the solution for the lightray in the far-zone of the Solar system \cite{KlionerKopeikin1992,Kopeikin_Efroimsky_Kaplan}.  

In summary, a precise determination of light trajectory up to a given accuracy is of fundamental importance in the theory of any astrometric measurements.  
Besides considerable effort which has still to be done in near future, we come to the conclusion that a complete modeling of light trajectory from  
celestial light sources through the Solar system toward the observer is accomplishable also for extremely high-precision astrometry on 
sub-\muas $\;$ and even on nano-arcsecond level of accuracy.

\section{Acknowledgment}

This work was supported by the Deutsche Forschungsgemeinschaft (DFG).

\appendix

\section{Notations}\label{Notation}

Throughout the article the following notations are in use:

\begin{itemize}

\item $G$ is the Newtonian constant of gravitation.

\item $c$ is the vacuum speed of light in flat Minkowski space.

\item Lower case Latin indices $a$, $b$, \dots, $i$, $j$, \dots take values 1,2,3.

\item Lower case Greek indices $\alpha$, $\beta$, \dots, $\mu$, $\nu$, \dots take values 0,1,2,3.

\item $\delta_{ij} = \delta^{ij} = {\rm diag} \left(+1,+1,+1\right)$ is Kronecker delta.

\item The three-dimensional coordinate quantities (''three-vectors'') referred to
the spatial axes of the corresponding reference system are set in
boldface: $\ve{a}$.

\item The contravariant components of ''three-vectors'' are $a^{i} = \left(a^1,a^2,a^3\right)$.

\item The contravariant components of ''four-vectors'' are $a^{\mu} = \left(a^0,a^1,a^2,a^3\right)$.

\item Repeated indices imply the Einstein's summation irrespective of
their positions (e.g. $a^i\,b^i=a^1\,b^1+a^2\,b^2+a^3\,b^3$ and
$a^\alpha\,b^\alpha=a^0\,b^0+a^1\,b^1+a^2\,b^2+a^3\,b^3$).

\item The absolute value (Euclidean norm) of a ''three-vector'' $\ve{a}$ is
denoted as $|\ve{a}|$ or, simply, $a$ and can be computed as
$a=|\ve{a}|=(a^1\,a^1+a^2\,a^2+a^3\,a^3)^{1/2}$.

\item The scalar product of any two ''three-vectors'' $\ve{a}$ and $\ve{b}$
with respect to the Euclidean metric $\delta_{ij}$ is denoted by
$\ve{a}\,\cdot\,\ve{b}$ and can be computed as
$\ve{a}\,\cdot\,\ve{b}=\delta_{ij}\,a^i\,b^j=a^i\,b^i$.

\item The vector product of any two ''three-vectors'' $\ve{a}$ and $\ve{b}$
is designated by $\ve{a}\times\ve{b}$ and can be computed as
$\left(\ve{a}\times\ve{b}\right)^i=\varepsilon_{ijk}\,a^j\,b^k$, where
$\varepsilon_{ijk}=(i-j)(j-k)(k-i)/2$ is the fully antisymmetric
Levi-Civita symbol.

\item The global coordinate system is denoted by lower-case letters: $\left(c t, \ve{x}\right)$.

\item The local coordinate system of a massive body A is denoted by upper-case letters: $\left(c T_A, \ve{X}_A\right)$.

\item The photon trajectory is denoted by $\ve{x}\left(t\right)$.
In order to distinguish the photon's spatial coordinate $\ve{x}\left(t\right)$
from the spatial coordinate $\ve{x}$ of the global system,
the time-dependence of photon's spatial coordinate will
everywhere be shown explicitly throughout the article.

\item The worldline of massive body A is denoted by $\ve{x}_A\left(t\right)$
or $\ve{x}_A\left(T_A\right)$.

\item Partial derivatives in the global coordinate system:
$\displaystyle \partial_{\mu} = \frac{\partial}{\partial x^{\mu}}$ or
$\displaystyle \partial_{i} = \frac{\partial}{\partial x^{i}}$.

\item Partial derivatives in the local coordinate system of body A:
$\displaystyle {\cal D}^A_{\alpha} = \frac{\partial}{\partial X_A^{\alpha}}$ or
$\displaystyle {\cal D}^A_{a} = \frac{\partial}{\partial X_A^{a}}$.

\item $n! = n \left(n-1\right)\left(n-2\right)\cdot\cdot\cdot 2 \cdot 1$ is the faculty for positive integer;
$0! = 1$.

\item $L=i_1 i_2 ...i_l$ is a Cartesian multi-index of a given tensor $T$, that means
$T_L \equiv T_{i_1 i_2 \,.\,.\,.\,i_l}$, and each index $i_1,i_2,...,i_l$ runs from $1$ to $3$
(i.e. over the Cartesian coordinate label).

\item Two identical multi-indices imply summation, e.g.:
$\partial_L\,T_L \equiv \sum\limits_{i_1\,.\,.\,.\,i_l}\,\partial_{i_1\,.\,.\,.\,i_l}\,T_{i_1\,.\,.\,.\,i_l}$.

\item The symmetric tracefree (STF) part of a tensor $T_L$ is defined by Eq.~(A2) in \cite{Zschocke_1PN} and denoted by 
$T_{\langle L\rangle}$.  

\end{itemize}

\section{Notation of impact vectors:}\label{Impact_Vectors}

Before we distinguish between the case of massive bodies at rest and massive bodies in motion, we consider the  
unperturbed lightray in flat Minkowskian space-time, which in Cartesian coordinates is given by the expression in (\ref{Introduction_1}), 
\begin{eqnarray}
\ve{x}_{\rm N}\left(t\right) &=& \ve{x}_0 + c\left(t - t_0\right)\ve{\sigma}\,,
\label{Unperturbed_light_ray_10}
\end{eqnarray}

\noindent
which describes a straight line and where the subscript N stands for Newtonian limit. By Eq.~(\ref{variable_2}) we have introduced the following impact vector:  
\begin{eqnarray}
\ve{\xi} &=& \ve{\sigma} \times \left(\ve{x}_{\rm N}\left(t\right) \times \ve{\sigma}\right)
= \ve{\sigma} \times \left(\ve{x}_0 \times \ve{\sigma}\right), 
\label{notation_2}
\\
\nonumber\\
d &=& \left|\ve{\xi}\right|.
\label{notation_2a}
\end{eqnarray}

\noindent
The impact vector in (\ref{notation_2}) points from the origin of the global system (BCRS) toward the point of closest approach of the
unperturbed lightray to that origin. The impact vector in (\ref{notation_2}) is time-independent, both in case of massive bodies at rest as well
as in case of massive bodies in motion.

\subsection{Massive bodies at rest:}

Massive bodies at rest means their positions remain constant with respect to the global reference system: $\ve{x}_A = {\rm const}$.
We will make use of the following notation for the vector from the massive body at rest toward the photon propagating along the exact light trajectory:
\begin{eqnarray}
\ve{r}_A &=& \ve{x}\left(t\right) - \ve{x}_A \,,
\label{notation_3a}
\end{eqnarray}

\noindent
with the absolute value $r_A=\left|\ve{r}_A\right|$.
The vector from the massive body at rest toward the photon along the unperturbed light trajectory reads:
\begin{eqnarray}
\ve{r}^{\rm N}_A &=& \ve{x}_{\rm N}\left(t\right) - \ve{x}_A
\nonumber\\
&=& \ve{x}_0 + c\left(t - t_0\right)\ve{\sigma} - \ve{x}_A\,,
\label{notation_3b}
\end{eqnarray}

\noindent
with the absolute value $r_A^{\rm N}=\left|\ve{r}^{\rm N}_A \right|$, and obviously $\ve{r}_A = \ve{r}^{\rm N}_A + {\cal O}\left(c^{-2}\right)$.
We also need the vector from the massive body at rest toward the photon at the moment of signal-emission:
\begin{eqnarray}
\ve{r}^0_A &=& \ve{x}_0 - \ve{x}_A\,,
\label{notation_3c}
\end{eqnarray}

\noindent
with the absolute value $r_A^0=\left|\ve{r}^0_A \right|$.
Note that in case of massive bodies at rest there will be no time-argument in $\ve{r}_A$ and $\ve{r}^{\rm N}_A$,
irrespective of the fact that the distance between the photon and the body actually depends on time due to the propagation of the photon.
In case of massive bodies at rest we introduce the following impact-vector:
\begin{eqnarray}
\ve{d}_A &=& \ve{\sigma} \times \left(\ve{r}^{\rm N}_A \times \ve{\sigma}\right), \quad d_A = \left|\ve{d}_A\right|\,.
\label{notation_4}
\end{eqnarray}

\noindent
The impact-vector in (\ref{notation_4}) is time-independent, $\ve{\dot{d}}_A = 0$, and points from the origin of local coordinate system
of massive body $A$ toward the unperturbed lightray at the time of closest approach to that origin,
defined by 
\begin{eqnarray}
{\rm t}_A^{\ast} &=& t_0 - \frac{\ve{\sigma} \cdot \left(\ve{x}_0 - \ve{x}_A\right)}{c}
+ {\cal O}\left(c^{-2}\right)\,,
\label{time_of_closest_approach_t_0}
\\
&=& t_1 - \frac{\ve{\sigma} \cdot \left(\ve{x}_1 - \ve{x}_A\right)}{c}
+ {\cal O}\left(c^{-2}\right)\,. 
\label{time_of_closest_approach_t_1}
\end{eqnarray}

\noindent 
Notice that the term {\it weak gravitational field} implies $\displaystyle d_A \gg \frac{G\,M_A}{c^2}$.

\subsection{Massive bodies in motion:}

In case of massive bodies in motion, their positions become time-dependent: $\ve{x}_A \left(t\right)$.
Then we will make use of the following notation for the vector from the massive body toward the photon propagating along the
exact light trajectory:
\begin{eqnarray}
\ve{r}_A\left(t\right) &=& \ve{x}\left(t\right) - \ve{x}_A\left(t\right),
\label{notation_5a}
\end{eqnarray}

\noindent
with the absolute value $r_A\left(t\right)=\left|\ve{r}_A\left(t\right)\right|$.
The vector from the massive body in motion toward the photon along the unperturbed light trajectory reads:
\begin{eqnarray}
\ve{r}^{\rm N}_A\left(t\right) &=& \ve{x}_{\rm N}\left(t\right) - \ve{x}_A\left(t\right)
\nonumber\\
&=& \ve{x}_0 + \,c \left(t - t_0\right)\ve{\sigma} - \ve{x}_A\left(t\right)\,,
\label{notation_5b}
\end{eqnarray}

\noindent
with the absolute value $r_A^{\rm N}\left(t\right)=\left|\ve{r}^{\rm N}_A\left(t\right)\right|$
and obviously $\ve{r}_A\left(t\right) = \ve{r}^{\rm N}_A\left(t\right) + {\cal O}\left(c^{-2}\right)$.
We also will need the vector from the massive body toward the photon at the time-moment of
emission of the light-signal, given by
\begin{eqnarray}
\ve{r}^{\rm N}_A\left(t_0\right) &=& \ve{x}_0 - \ve{x}_A\left(t_0\right)\,,
\label{notation_5c}
\end{eqnarray}

\noindent
with the absolute value $r^{\rm N}_A\left(t_0\right) = \left|\ve{r}^{\rm N}_A\left(t_0\right)\right|$.
In case of massive bodies in motion we introduce the following impact vector:
\begin{eqnarray}
\ve{d}_A\left(t\right) &=& \ve{\sigma} \times \left(\ve{r}^{\rm N}_A \left(t\right) \times \ve{\sigma}\right),   
\label{notation_6}
\end{eqnarray}

\noindent
with the absolute value $d_A\left(t\right) = \left|\ve{d}_A\left(t\right)\right|$. 
The impact-vector in (\ref{notation_6}) is time-dependent, $\ve{\dot{d}}_A \neq 0$, and points from the origin of local
coordinate system of massive body $A$ toward the unperturbed lightray at the time of closest approach to that origin.
The time-dependence of the impact-vector in (\ref{notation_6}) is solely caused by the motion of the massive body, that
means a time-derivative of (\ref{notation_6}) is proportional to the orbital velocity of this body,
$\ve{\dot{d}}_A\left(t\right) = \ve{\sigma} \times \left(\ve{\sigma} \times \ve{v}_A\left(t\right)\right)$.
The term {\it weak gravitational field} implies $\displaystyle d_A\left(t_A^{\ast}\right) \gg \frac{G\,M_A}{c^2}$
for the time of closest approach of the lightray to the massive body, which are given by 
\begin{eqnarray}
t_A^{\ast} &=& t_0 - \frac{\ve{\sigma} \cdot \left(\ve{x}_0 - \ve{x}_A\left(t_A^{\ast}\right)\right)}{c}
+ {\cal O}\left(c^{-2}\right)\,,
\label{time_of_closest_approach_t_0_moving_body}
\\
&=& t_1 - \frac{\ve{\sigma} \cdot \left(\ve{x}_1 - \ve{x}_A\left(t_A^{\ast}\right)\right)}{c}
+ {\cal O}\left(c^{-2}\right)\,,  
\label{time_of_closest_approach_t_1_moving_body}
\end{eqnarray}

\noindent
and which slightly differ from the expressions (\ref{time_of_closest_approach_t_0}) and (\ref{time_of_closest_approach_t_1}) 
by the time-argument of the spatial coordinates of the massive body.

\section{Partial derivative operator}\label{Appendix_Partial_Derivative} 

The spatial derivative in terms of the new variables $\tau$ and $\ve{\xi}$ has been given by  
relation (\ref{Transformation_Derivative_2A}) which is valid for any smooth function $F\left(t,\ve{x}\right)$, that means   
\begin{eqnarray}
&& \hspace{-0.5cm} \frac{\partial F\left(t,\ve{x}\right)}{\partial x^i}
\Bigg|_{\ve{x}=\ve{x}_{\rm N}\mbox{\normalsize $\left(t\right)$}}
\nonumber\\
&& \hspace{-0.3cm} = \left(P^{ij} \frac{\partial}{\partial \xi^j} + \sigma^i\,\frac{\partial}{\partial c\,\tau}
- \sigma^i \frac{\partial}{\partial c t^{\ast}}\right)
F\left(t^{\ast} + \tau,\ve{\xi} + c \tau\,\ve{\sigma}\right). 
\nonumber\\
\label{Appendix_Partial_Derivative_5}
\end{eqnarray}

\noindent
According to the metric perturbations in (\ref{global_metric_perturbation_A}) - (\ref{global_metric_perturbation_C})
we have to consider the STF partial derivative operation in Eq.~(\ref{spatial_derivative}), which reads  
\begin{eqnarray}
\partial_{\langle L \rangle} &=& \underset{i_1 ... i_l}{\rm STF}\;\frac{\partial}{\partial x^{i_1}}...\frac{\partial}{\partial x^{i_l}}\,.
\label{STF_derivative_operation}
\end{eqnarray}

\noindent
In order to express the spatial derivative operation in (\ref{STF_derivative_operation}) in terms of these new variables, we apply the binomial theorem:  
\begin{eqnarray}
\left(a + b + c\right)^l &=&\sum\limits_{p=0}^{l}\, 
\left( \begin{array}[c]{l}
l \\
\displaystyle
p \end{array}\right) \,a^{l-p}\,\sum\limits_{q=0}^{p}\, 
\left( \begin{array}[c]{l}
p \\
\displaystyle
q \end{array}\right) \,b^{p-q}\,c^{q}\,, 
\nonumber\\
\label{Binomial_Theorem}
\end{eqnarray}

\noindent
where the binomial coefficients are defined by
\begin{eqnarray}
\left( \begin{array}[c]{l}
l \\
\displaystyle
p \end{array}\right) = \frac{l!}{\left(l-p\right)!\,p!}\,,\quad
\left( \begin{array}[c]{l}
p \\
\displaystyle
q \end{array}\right) = \frac{p!}{\left(p-q\right)!\,q!}\,.
\label{Binomial_Coefficient}
\end{eqnarray}

\noindent  
In virtue of the binomial theorem in (\ref{Binomial_Theorem}), we obtain for the STF partial derivative  
operator in (\ref{STF_derivative_operation}) in terms of the new variables $\tau$ and $\ve{\xi}$ the following expression: 
\begin{eqnarray}
\partial_{\langle L \rangle} &=& \underset{i_1 ... i_l}{\rm STF}\,\sum\limits_{p=0}^{l} \frac{l!}{\left(l-p\right)!\;p!}\;
\sum\limits_{q=0}^{p} \left(-1\right)^{q}\frac{p!}{\left(p-q\right)!\;q!}\;
\nonumber\\
\nonumber\\
&& \hspace{-1.1cm} \times \sigma^{i_1}\,...\,\sigma^{i_p}\;
P^{i_{p+1}\,j_{p+1}}\;...\;P^{i_l\,j_l}\;
\frac{\partial}{\partial \xi^{j_{p+1}}}\;...\;
\frac{\partial}{\partial \xi^{j_{l}}}\;
\nonumber\\
\nonumber\\
&& \hspace{-1.1cm} \times \left(\frac{\partial}{\partial c\,\tau}\right)^{p-q} 
\left(\frac{\partial}{\partial c\,t^{\ast}}\right)^q\,.
\label{Appendix_Partial_Derivative_10}
\end{eqnarray}

\noindent
The same expression for $\partial_{\langle L \rangle}$ has been used in \cite{KopeikinKorobkov2005} (cf. Eqs.~(4.42) - (4.43) 
ibid.); note the symmetry $p-q \leftrightarrow q$ of the expression in (\ref{Appendix_Partial_Derivative_10}).  
The derivatives with respect to variable $c\,t^{\ast}$ act only on $M_L^A\left(\tau+t^{\ast}\right)$ and $\ve{x}_A\left(\tau+t^{\ast}\right)$,  
hence the partial derivatives $\displaystyle \left(\frac{\partial}{\partial c\,t^{\ast}}\right)^q$ produce terms of the order
${\cal O}\left(c^{-q}\right)$. For that reason it was possible to neglect all derivatives with respect to variable $c\,t^{\ast}$  
in 1PN approximation which has been investigated in \cite{Zschocke_1PN}.    
If one neglects such derivatives (i.e. take only the terms with $q=0$ in Eq.~(\ref{Appendix_Partial_Derivative_10})), then 
we would obtain the simpler derivative operator:  
\begin{eqnarray}
\partial^{q=0}_{\langle L\rangle} &=& \underset{i_1 ... i_l}{\rm STF}\, \sum\limits_{p=0}^{l} \frac{l!}{\left(l-p\right)!\;p!}\;
\sigma^{i_1}\,...\,\sigma^{i_p}\;
P^{i_{p+1}\,j_{p+1}}\;...\;P^{i_l\,j_l}\;
\nonumber\\
&& \times \frac{\partial}{\partial \xi^{j_{p+1}}}\;...\;
\frac{\partial}{\partial \xi^{j_{l}}}\;\left(\frac{\partial}{\partial c\tau}\right)^p\,, 
\label{Simplified_Differentialoperator}
\end{eqnarray}

\noindent
which coincides with the expression as given by Eq.~(24) in \cite{Kopeikin1997} or by Eq.~(101) in \cite{Zschocke_1PN} where it was 
allowed to omit the STF operation because of relation (\ref{STF_relation_5}).  

\section{Derivatives} 

In this appendix we will summarize some useful spatial-derivatives and time-derivatives. 
Throughout this appendix all time arguments are omitted in order to simplify the notations, that means  
\begin{eqnarray}
r_A^{\rm N} \equiv r_A^{\rm N}\left(\tau + t^{\ast}\right),
\label{argument_1}
\\
\nonumber\\
\ve{r}_A^{\rm N} \equiv \ve{r}_A^{\rm N}\left(\tau + t^{\ast}\right),
\label{argument_2}
\\
\nonumber\\
\ve{d}_A \equiv \ve{d}_A\left(\tau + t^{\ast}\right),
\label{argument_3}
\\
\nonumber\\
\ve{v}_A \equiv \ve{v}_A\left(\tau + t^{\ast}\right).
\label{argument_4}
\end{eqnarray}

\subsection{Spatial-derivatives} 

In this appendix some relevant relations for spatial-derivatives are summarized. 
The vector $\ve{r}_A^{\rm N}$ depends only on the variables $\ve{\xi}$, $\tau$ and $\ve{x}_A\left(\tau+t^{\ast}\right)$.  
Since variable $\ve{\xi}$ is independent of $\tau$ and $\ve{x}_A\left(\tau+t^{\ast}\right)$, we consider partial derivatives with respect to variable $\ve{\xi}$. 
We obtain the following relations:  
\begin{eqnarray}
P^{ij}\,\frac{\partial}{\partial \xi^j}\,
\ln \left(r^{\rm N}_A - \ve{\sigma}\cdot\ve{r}^{\rm N}_A\right)
&=& \frac{d_A^i}{r^{\rm N}_A}
\frac{1}{r^{\rm N}_A - \ve{\sigma}\cdot\ve{r}^{\rm N}_A}\,. 
\label{Relation_2}
\end{eqnarray}

\noindent
\begin{eqnarray}
P^{ij}\,\frac{\partial}{\partial \xi^j}\,
\left[r^{\rm N}_A + \ve{\sigma}\cdot\ve{r}_A^{\rm N}
\ln \left(r^{\rm N}_A - \ve{\sigma}\cdot\ve{r}^{\rm N}_A\right)\right]
&=& \frac{d_A^i}{r^{\rm N}_A - \ve{\sigma}\cdot\ve{r}^{\rm N}_A}\,. 
\nonumber\\
\label{Relation_3}
\end{eqnarray}

\noindent
\begin{eqnarray}
P^{ij}\,\frac{\partial}{\partial \xi^j}\,\frac{\ve{d}_A\cdot\ve{v}_A}{r_A^{\rm N} - \ve{\sigma}\cdot\ve{r}^{\rm N}_A} &=& 
- \frac{d_A^i\,\left(\ve{d}_A\cdot\ve{v}_A\right)}{r_A^{\rm N}\;\left(r_A^{\rm N} - \ve{\sigma}\cdot\ve{r}^{\rm N}_A\right)^2}
\nonumber\\ 
\nonumber\\ 
&& + \frac{v_A^i - \sigma^i\,\left(\ve{\sigma}\cdot\ve{v}_A\right)}{r_A^{\rm N} - \ve{\sigma}\cdot\ve{r}^{\rm N}_A}\,.  
\label{Relation_4}
\end{eqnarray}

\noindent
\begin{eqnarray}
P^{ij}\,\frac{\partial}{\partial \xi^j}\left(\ve{d}_A\cdot\ve{v}_A\right)\ln \left(r_A^{\rm N} - \ve{\sigma}\cdot\ve{r}^{\rm N}_A\right) &=&  
\frac{\left(\ve{d}_A\cdot\ve{v}_A\right)}{r_A^{\rm N} - \ve{\sigma}\cdot\ve{r}^{\rm N}_A}\,\frac{d_A^i}{r_A^{\rm N}} 
\nonumber\\
\nonumber\\
&& \hspace{-4.0cm} + \left[v_A^i - \sigma^i\,\left(\ve{\sigma}\cdot\ve{v}_A\right)\right]\,\ln \left(r_A^{\rm N} - \ve{\sigma}\cdot\ve{r}^{\rm N}_A\right)\,.  
\label{Relation_5}
\end{eqnarray}

\noindent
Notice that $\displaystyle \frac{\partial}{\partial \xi^j}\,\left(\ve{\sigma}\cdot\ve{r}^{\rm N}_A\right) = 0$.

\subsection{Time-derivatives} 

In this appendix some relations of time-derivatives are summarized which are of relevance for integrations by part.   
Since the position of massive body depends on time-variable, $\ve{x}_A\left(\tau+t^{\ast}\right)$, this vector is not independent of $\tau$. 
Therefore, in order to perform integration by parts, we need to consider the derivatives with respect to variable $\tau$. 
Taking into account $\ve{r}_A^{\rm N} = \ve{d}_A + \ve{\sigma} \left(\ve{\sigma} \cdot \ve{r}_A^{\rm N}\right)$, we find the following relations: 
\begin{eqnarray}
\frac{\partial}{\partial c\tau}\,\ln \left(r_A^{\rm N} - \ve{\sigma}\cdot\ve{r}_A^{\rm N}\right)  
&=& - \frac{1}{r_A^{\rm N}} + \frac{\ve{\sigma} \cdot \ve{v}_A}{c}\,\frac{1}{r_A^{\rm N}}  
\nonumber\\ 
\nonumber\\ 
&& \hspace{-2.0cm} - \frac{\ve{v}_A \cdot \ve{d}_A}{c\,r_A^{\rm N}}\,\frac{1}{r_A^{\rm N} - \ve{\sigma} \cdot \ve{r}_A^{\rm N}}\,.  
\label{Relation_1}
\end{eqnarray}

\noindent 
\begin{eqnarray}
\frac{\partial}{\partial c\,\tau}\,\frac{1}{r_A^{\rm N}}\,\frac{1}{r_A^{\rm N} - \ve{\sigma}\cdot\ve{r}_A^{\rm N}} &=&
\frac{1}{\left(r_A^{\rm N}\right)^3}\,
- \frac{1}{\left(r_A^{\rm N}\right)^3}\,\frac{\ve{\sigma}\cdot\ve{v}_A}{c}
\nonumber\\
\nonumber\\
&& \hspace{-3.5cm} + \frac{1}{\left(r_A^{\rm N}\right)^3}\,\frac{\ve{d}_A\cdot\ve{v}_A}{c\,\left(r_A^{\rm N} - \ve{\sigma}\cdot\ve{r}_A^{\rm N}\right)}
+ \frac{1}{\left(r_A^{\rm N}\right)^2}\,\frac{\ve{d}_A\cdot\ve{v}_A}{c\,\left(r_A^{\rm N} - \ve{\sigma}\cdot\ve{r}_A^{\rm N}\right)^2}\,. 
\nonumber\\
\label{Appendix_Integral_I1_25}
\end{eqnarray}

\noindent
\begin{eqnarray}
\frac{\partial}{\partial c\tau}\Bigg(r_A^{\rm N} + \ve{\sigma} \cdot \ve{r}^{\rm N}_A\,\ln \left(r^{\rm N}_A-\ve{\sigma} \cdot \ve{r}^{\rm N}_A\right)\Bigg)  
&=& \ln \left(r^{\rm N}_A-\ve{\sigma} \cdot \ve{r}^{\rm N}_A\right)  
\nonumber\\
\nonumber\\
&& \hspace{-6.0cm} - \frac{\ve{\sigma}\cdot\ve{v}_A}{c}\,\ln \left(r^{\rm N}_A-\ve{\sigma} \cdot \ve{r}^{\rm N}_A\right)
- \frac{\ve{d}_A \cdot \ve{v}_A}{c}\,\frac{1}{r_A^{\rm N} - \ve{\sigma} \cdot \ve{r}^{\rm N}_A}\,. 
\label{Integral_A_6}
\end{eqnarray}

\noindent
\begin{eqnarray}
\frac{\partial}{\partial c\,\tau}\,\frac{1}{r_A^{\rm N} - \ve{\sigma}\cdot\ve{r}_A^{\rm N}} &=&
\frac{1}{r_A^{\rm N}}\,\frac{1}{r_A^{\rm N} - \ve{\sigma}\cdot\ve{r}_A^{\rm N}}
\nonumber\\
\nonumber\\
&& \hspace{-3.2cm} + \frac{1}{\left(r_A^{\rm N} - \ve{\sigma}\cdot\ve{r}_A^{\rm N}\right)^2}\,\frac{\ve{v}_A\cdot\ve{d}_A}{c\;r_A^{\rm N}}
- \frac{1}{r_A^{\rm N}}\,\frac{1}{r_A^{\rm N} - \ve{\sigma}\cdot\ve{r}_A^{\rm N}}\,\frac{\ve{\sigma}\cdot\ve{v}_A}{c}\,. 
\nonumber\\
\label{Integral_A_20}
\end{eqnarray} 

\begin{eqnarray}
\frac{2}{r^{\rm N}_A - \ve{\sigma}\cdot \ve{r}^{\rm N}_A}  
&=& \frac{\partial}{\partial c \tau}\,\frac{\ve{\sigma}\cdot\ve{r}_A^{\rm N}}{r^{\rm N}_A - \ve{\sigma}\cdot \ve{r}^{\rm N}_A} 
- \frac{\partial}{\partial c \tau}\,\ln \left(r_A^{\rm N} - \ve{\sigma}\cdot\ve{r}_A^{\rm N}\right)   
\nonumber\\
\nonumber\\
&& \hspace{-2.0cm} + 2\,\frac{\ve{\sigma}\cdot\ve{v}_A}{c}\,\frac{1}{r_A^{\rm N} - \ve{\sigma}\cdot\ve{r}_A^{\rm N}} 
- \frac{\ve{v}_A\cdot\ve{d}_A}{c\,\left(r_A^{\rm N} - \ve{\sigma}\cdot\ve{r}_A^{\rm N}\right)^2}\,.  
\label{Integral_C_10_A}
\end{eqnarray}

\section{The integral $I_1$}\label{Integral_1} 

The first integration of the expression in the first line in (\ref{transformed_geodesic_equation_C1}) reads: 
\begin{eqnarray}
I_1\left(\tau+t^{\ast}\right) &=& + \frac{2\,G}{c^2}\,P^{ij}\,\frac{\partial}{\partial \xi^j}\,
\sum\limits_{l = 0}^{\infty} \frac{\left(-1\right)^l}{l!}\;
\nonumber\\
\nonumber\\
&& \hspace{-2.0cm} \times \int \limits_{- \infty}^{\tau}\,d c \tau^{\prime}\,M_{\langle L \rangle}^A\left(\tau^{\prime}+t^{\ast}\right)
\partial^{\prime}_{\langle L \rangle}\;\frac{1}{r^{\rm N}_A\left(\tau^{\prime}+t^{\ast}\right)}\,,  
\label{Appendix_Integral_I1_5}
\end{eqnarray}

\noindent
where $\partial^{\prime}_{\langle L \rangle}$ is given by Eq.~(\ref{Appendix_Partial_Derivative_10}) where $c\tau$ 
is formally replaced by the integration variable $c\tau^{\prime}$.  
In order to solve that integral, we use relation (\ref{Relation_1}) and obtain  
\begin{widetext}
\begin{eqnarray}
I_1\left(\tau+t^{\ast}\right) &=& - \frac{2\,G}{c^2}\,P^{ij}\,\frac{\partial}{\partial \xi^j}\,
\sum\limits_{l = 0}^{\infty} \frac{\left(-1\right)^l}{l!}\;
\int \limits_{- \infty}^{\tau}\,d c \tau^{\prime}\,M_{\langle L \rangle}^A\left(\tau^{\prime}+t^{\ast}\right)
\partial^{\prime}_{\langle L \rangle}\;\frac{\partial}{\partial c \tau^{\prime}}\,
\ln \left[r^{\rm N}_A\left(\tau^{\prime}+t^{\ast}\right) - \ve{\sigma}\cdot\ve{r}^{\rm N}_A\left(\tau^{\prime}+t^{\ast}\right)\right]
\nonumber\\
\nonumber\\
&& \hspace{-1.0cm} + \frac{2\,G}{c^3}\,P^{ij}\,\frac{\partial}{\partial \xi^j}\,
\sum\limits_{l = 0}^{\infty} \frac{\left(-1\right)^l}{l!}\;
\int \limits_{- \infty}^{\tau}\,d c \tau^{\prime}\,M_{\langle L \rangle}^A\left(\tau^{\prime}+t^{\ast}\right)
\partial^{\prime}_{\langle L \rangle}\;\ve{\sigma}\cdot\ve{v}_A\left(\tau^{\prime}+t^{\ast}\right)\,\frac{1}{r^{\rm N}_A\left(\tau^{\prime}+t^{\ast}\right)} 
\nonumber\\
\nonumber\\
&& \hspace{-1.0cm} - \frac{2\,G}{c^3}\,P^{ij}\,\frac{\partial}{\partial \xi^j}\,
\sum\limits_{l = 0}^{\infty} \frac{\left(-1\right)^l}{l!}\;
\int \limits_{- \infty}^{\tau}\,d c \tau^{\prime}\,M_{\langle L \rangle}^A\left(\tau^{\prime}+t^{\ast}\right)
\partial^{\prime}_{\langle L \rangle}\,
\frac{\ve{v}_A\left(\tau^{\prime}+t^{\ast}\right)\cdot\ve{d}_A\left(\tau^{\prime}+t^{\ast}\right)}
{r^{\rm N}_A\left(\tau^{\prime}+t^{\ast}\right) - \ve{\sigma}\cdot \ve{r}^{\rm N}_A\left(\tau^{\prime}+t^{\ast}\right)}\,
\frac{1}{r^{\rm N}_A\left(\tau^{\prime}+t^{\ast}\right)}\,,   
\label{Appendix_Integral_I1_10}
\end{eqnarray}
\end{widetext}

\noindent
where we also have used that $\ve{r}_A^{\rm N} = \ve{d}_A + \ve{\sigma} \left(\ve{\sigma}\cdot\ve{r}_A^{\rm N}\right)$.
Note that Eq.~(\ref{Appendix_Integral_I1_10}) is an exact expression for the integral in Eq.~(\ref{Appendix_Integral_I1_5}).  
Now the expression in the first line in (\ref{Appendix_Integral_I1_10}) will be integrated by parts. For the integral in 
the second line in (\ref{Appendix_Integral_I1_10}) we use relation (\ref{Relation_1}) again, while for the 
integral in the third line in (\ref{Appendix_Integral_I1_10}) we will use relation (\ref{Integral_A_20})   
and obtain by means of integration by parts:  
\begin{widetext}
\begin{eqnarray}
I_1\left(\tau+t^{\ast}\right) &=& - \frac{2\,G}{c^2}\,P^{ij}\,\frac{\partial}{\partial \xi^j}\,
\sum\limits_{l = 0}^{\infty} \frac{\left(-1\right)^l}{l!}\;
M_{\langle L \rangle}^A\left(\tau+t^{\ast}\right)
\partial_{\langle L \rangle}\;
\ln \left[r^{\rm N}_A\left(\tau+t^{\ast}\right) - \ve{\sigma}\cdot\ve{r}^{\rm N}_A\left(\tau+t^{\ast}\right)\right]
\nonumber\\
\nonumber\\
&& + \frac{2\,G}{c^3}\,P^{ij}\,\frac{\partial}{\partial \xi^j}\,
\sum\limits_{l = 1}^{\infty} \frac{\left(-1\right)^l}{l!}\;
\int \limits_{- \infty}^{\tau}\,d c \tau^{\prime}\,\dot{M}_{\langle L \rangle}^A\left(\tau^{\prime}+t^{\ast}\right)
\partial^{\prime}_{\langle L \rangle}\;
\ln \left[r^{\rm N}_A\left(\tau^{\prime}+t^{\ast}\right) - \ve{\sigma}\cdot\ve{r}^{\rm N}_A\left(\tau^{\prime}+t^{\ast}\right)\right]
\nonumber\\
\nonumber\\
&& - \frac{2\,G}{c^3}\,P^{ij}\,\frac{\partial}{\partial \xi^j}\,
\sum\limits_{l = 0}^{\infty} \frac{\left(-1\right)^l}{l!}\;
M_{\langle L \rangle}^A\left(\tau+t^{\ast}\right)
\partial_{\langle L \rangle}\;\ve{\sigma}\cdot\ve{v}_A\left(\tau+t^{\ast}\right)\,
\ln \left[r^{\rm N}_A\left(\tau+t^{\ast}\right) - \ve{\sigma}\cdot\ve{r}^{\rm N}_A\left(\tau+t^{\ast}\right)\right] 
\nonumber\\
\nonumber\\
&& - \frac{2\,G}{c^3}\,P^{ij}\,\frac{\partial}{\partial \xi^j}\,
\sum\limits_{l = 0}^{\infty} \frac{\left(-1\right)^l}{l!}\;
M_{\langle L \rangle}^A\left(\tau+t^{\ast}\right)
\partial_{\langle L \rangle}\,
\frac{\ve{v}_A\left(\tau+t^{\ast}\right)\cdot\ve{d}_A\left(\tau+t^{\ast}\right)}
{r^{\rm N}_A\left(\tau+t^{\ast}\right) - \ve{\sigma}\cdot \ve{r}^{\rm N}_A\left(\tau+t^{\ast}\right)} 
+ {\cal O}\left(\frac{v_A}{c} \dot{M}_L^A\right) + {\cal O}\left(\frac{v_A^2}{c^2} M_L^A\right).  
\nonumber\\
\label{Appendix_Integral_I1_15}
\end{eqnarray}
\end{widetext}

\noindent
By means of relation (C5) in \cite{Zschocke_1PN} one may show that the lower integration limit $\tau \rightarrow - \infty$ in the first line of
Eq.~(\ref{Appendix_Integral_I1_10}) vanishes. 
In order to determine the integral in the second line in (\ref{Appendix_Integral_I1_15}), we use relation (\ref{Integral_A_6}), and obtain finally: 
\begin{widetext}
\begin{eqnarray}
I_1\left(\tau+t^{\ast}\right) &=& - \frac{2\,G}{c^2}\,P^{ij}\,\frac{\partial}{\partial \xi^j}\,
\sum\limits_{l = 0}^{\infty} \frac{\left(-1\right)^l}{l!}\;
M_{\langle L \rangle}^A\left(\tau+t^{\ast}\right)
\partial_{\langle L \rangle}\;
\ln \left[r^{\rm N}_A\left(\tau+t^{\ast}\right) - \ve{\sigma}\cdot\ve{r}^{\rm N}_A\left(\tau+t^{\ast}\right)\right]
\nonumber\\
\nonumber\\
&& \hspace{-2.0cm} + \frac{2\,G}{c^3}\,P^{ij}\,\frac{\partial}{\partial \xi^j}\,
\sum\limits_{l = 1}^{\infty} \frac{\left(-1\right)^l}{l!}\;
\dot{M}_{\langle L \rangle}^A\left(\tau+t^{\ast}\right)
\partial_{\langle L \rangle}\;
\bigg[r^{\rm N}_A\left(\tau+t^{\ast}\right) + \ve{\sigma}\cdot \ve{r}^{\rm N}_A\left(\tau+t^{\ast}\right) 
\ln \left[r^{\rm N}_A\left(\tau+t^{\ast}\right) - \ve{\sigma}\cdot\ve{r}^{\rm N}_A\left(\tau+t^{\ast}\right)\right]\bigg]
\nonumber\\
\nonumber\\
&& \hspace{-2.0cm} - \frac{2\,G}{c^3}\,P^{ij}\,\frac{\partial}{\partial \xi^j}\,
\sum\limits_{l = 0}^{\infty} \frac{\left(-1\right)^l}{l!}\;
M_{\langle L \rangle}^A\left(\tau+t^{\ast}\right)
\partial_{\langle L \rangle}\;\ve{\sigma}\cdot\ve{v}_A\left(\tau+t^{\ast}\right)\,
\ln \left[r^{\rm N}_A\left(\tau+t^{\ast}\right) - \ve{\sigma}\cdot\ve{r}^{\rm N}_A\left(\tau+t^{\ast}\right)\right]
\nonumber\\
\nonumber\\
&& \hspace{-2.0cm} - \frac{2\,G}{c^3}\,P^{ij}\,\frac{\partial}{\partial \xi^j}\,
\sum\limits_{l = 0}^{\infty} \frac{\left(-1\right)^l}{l!}\;
M_{\langle L \rangle}^A\left(\tau+t^{\ast}\right)
\partial_{\langle L \rangle}\,
\frac{\ve{v}_A\left(\tau+t^{\ast}\right)\cdot\ve{d}_A\left(\tau+t^{\ast}\right)}
{r^{\rm N}_A\left(\tau+t^{\ast}\right) - \ve{\sigma}\cdot \ve{r}^{\rm N}_A\left(\tau+t^{\ast}\right)}
+ {\cal O}\left(\frac{v_A}{c}\,\dot{M}_L^A\right) + {\cal O}\left(\ddot{M}_L^A\right) + {\cal O}\left(\frac{v_A^2}{c^2} M_L^A\right). 
\nonumber\\
\label{Appendix_Integral_I1_20}
\end{eqnarray}
\end{widetext}

\noindent
In general, terms of the order $\displaystyle {\cal O}\left(\frac{v_A}{c}\,\dot{M}_L^A\right)$, ${\cal O}\left(\ddot{M}_L^A\right)$ or  
$\displaystyle {\cal O}\left(\frac{v^2_A}{c^2}\,M_L^A\right)$ have to be neglected in order to be consistent  
with the DSX metric in Eqs.~(\ref{global_metric_perturbation_A}) - (\ref{global_metric_perturbation_C}), where such terms
are absent because they would be beyond 1.5PN approximation for the lightray metric, cf. text at the end of section \ref{Solar_System_Metric}.
This fact is also valid for all subsequent calculations but will not be mentioned explicitly in what follows.

\section{The integral $I_2$}\label{Integral_2} 

The integration of the fifth term in Eq.~(\ref{transformed_geodesic_equation_C1}) reads as follows:  
\begin{widetext}
\begin{eqnarray}
I_2\left(\tau+t^{\ast}\right) &=& + \frac{4\,G}{c^3}\,\sigma^j P^{ik} \frac{\partial}{\partial \xi^k}\,
\sum\limits_{l=1}^{\infty} \frac{\left(-1\right)^l}{l!}\;\int\limits_{-\infty}^{\tau} d c \tau^{\prime}\;
\dot{M}^A_{\langle j L-1 \rangle}\left(\tau^{\prime}+t^{\ast}\right)\,
\partial^{\prime}_{\langle L-1 \rangle}\,\frac{1}{r^{\rm N}_A\left(\tau^{\prime}+t^{\ast}\right)}
\nonumber\\
\nonumber\\
&=& - \frac{4\,G}{c^3}\,\sigma^j P^{ik} \frac{\partial}{\partial \xi^k}\,
\sum\limits_{l=1}^{\infty} \frac{\left(-1\right)^l}{l!}\;\int\limits_{-\infty}^{\tau} d c \tau^{\prime}\;
\dot{M}^A_{\langle j L-1 \rangle}\left(\tau^{\prime}+t^{\ast}\right)\,
\partial^{\prime}_{\langle L-1 \rangle} \frac{\partial}{\partial c\tau^{\prime}}\,
\ln \left(r_A^{\rm N}\left(\tau^{\prime}+t^{\ast}\right) - \ve{\sigma}\cdot\ve{r}_A^{\rm N}\left(\tau^{\prime}+t^{\ast}\right)\right)  
\nonumber\\
\nonumber\\
&& + \frac{4\,G}{c^3}\,\sigma^j P^{ik} \frac{\partial}{\partial \xi^k}\,
\sum\limits_{l=1}^{\infty} \frac{\left(-1\right)^l}{l!}\;\int\limits_{-\infty}^{\tau} d c \tau^{\prime}\;
\dot{M}^A_{\langle j L-1 \rangle}\left(\tau^{\prime}+t^{\ast}\right)\,
\partial^{\prime}_{\langle L-1 \rangle}\,\frac{\ve{\sigma} \cdot \ve{v}_A\left(\tau^{\prime}+t^{\ast}\right)}{c}\,
\frac{1}{r_A^{\rm N}\left(\tau^{\prime}+t^{\ast}\right)}   
\nonumber\\
\nonumber\\
&& \hspace{-2.0cm} - \frac{4 G}{c^3} \sigma^j P^{ik} \frac{\partial}{\partial \xi^k}
\sum\limits_{l=1}^{\infty} \frac{\left(-1\right)^l}{l!} \int\limits_{-\infty}^{\tau} d c \tau^{\prime}
\dot{M}^A_{\langle j L-1 \rangle}\left(\tau^{\prime}+t^{\ast}\right)
\partial^{\prime}_{\langle L-1 \rangle}
\frac{\ve{v}_A\left(\tau^{\prime}+t^{\ast}\right) \cdot \ve{d}_A\left(\tau^{\prime}+t^{\ast}\right)}
{c\,r_A^{\rm N}\left(\tau^{\prime}+t^{\ast}\right)}\,
\frac{1}{r_A^{\rm N}\left(\tau^{\prime}+t^{\ast}\right) - \ve{\sigma} \cdot \ve{r}_A^{\rm N}\left(\tau^{\prime}+t^{\ast}\right)}\,,  
\nonumber\\ 
\label{Example_Integral_2_1} 
\end{eqnarray} 
\end{widetext}

\noindent
where we have used relation (\ref{Relation_1}). We recognize that the last two terms in (\ref{Example_Integral_2_1}) are terms of the 
order $\displaystyle {\cal O}\left(\frac{v_A}{c}\,\dot{M}_L^A\right)$, hence they are neglected. Accordingly, integration by parts of the remaining 
integral in (\ref{Example_Integral_2_1}) results in 
\begin{widetext}
\begin{eqnarray}
I_2\left(\tau+t^{\ast}\right) &=& - \frac{4\,G}{c^3} \sigma^j P^{ik} \frac{\partial}{\partial \xi^k}  
\sum\limits_{l=1}^{\infty} \frac{\left(-1\right)^l}{l!}\,\dot{M}^A_{\langle j L-1 \rangle}\left(\tau+t^{\ast}\right)  
\partial_{\langle L-1 \rangle} \ln \left(r_A^{\rm N}\left(\tau+t^{\ast}\right) - \ve{\sigma}\cdot\ve{r}_A^{\rm N}\left(\tau+t^{\ast}\right)\right)  
\nonumber\\
\nonumber\\
&& + {\cal O}\left(\frac{v_A}{c}\,\dot{M}_L^A\right) + {\cal O}\left(\ddot{M}_L^A\right) + {\cal O}\left(\frac{v_A^2}{c^2} M_L^A\right). 
\label{Example_Integral_2_2}
\end{eqnarray}
\end{widetext}
 
\noindent
Finally, by means of relation (\ref{Relation_2}), we just obtain the expression in the last line of Eq.~(\ref{First_Integration_10}).

\section{Integrals of second integration of mass-multipole terms}\label{Appendix_Integrals_Mass_Multipoles} 

By inserting (\ref{First_Integration_9}) and (\ref{First_Integration_10}) into (\ref{Second_Integration_10}) we obtain the following integrals,  
each of which will be considered separately.

\subsection{Integral $I_3$}\label{Integral_3}  

The integral $I_3$, using relation (\ref{Relation_2}), reads 
\begin{widetext}
\begin{eqnarray}
I_3\left(\tau + t^{\ast}, \tau_0 + t^{\ast}\right) &=&
- \frac{2\,G}{c^2}\,P^{ij}\,\frac{\partial}{\partial \xi^j}\,\sum\limits_{l = 0}^{\infty} \frac{\left(-1\right)^l}{l!} 
\int\limits_{\tau_0}^{\tau}\,d c\tau^{\prime}\,M_{\langle L \rangle}^A\left(\tau^{\prime}+t^{\ast}\right)  
\partial^{\prime}_{\langle L \rangle}\;
\ln \left[r^{\rm N}_A\left(\tau^{\prime}+t^{\ast}\right)-\ve{\sigma} \cdot \ve{r}^{\rm N}_A\left(\tau^{\prime}+t^{\ast}\right)\right]. 
\label{Integral_A_5}
\end{eqnarray}
\end{widetext}

\noindent
We insert relation (\ref{Integral_A_6}) into (\ref{Integral_A_5})
and obtain: 
\begin{widetext}
\begin{eqnarray}
I_3\left(\tau + t^{\ast}, \tau_0 + t^{\ast}\right) &=&
\nonumber\\
\nonumber\\
&& \hspace{-3.0cm} - \frac{2\,G}{c^2}\,P^{ij}\,\frac{\partial}{\partial \xi^j}\, 
\sum\limits_{l = 0}^{\infty} \frac{\left(-1\right)^l}{l!}\;M_{\langle L \rangle}^A\left(\tau+t^{\ast}\right)
\partial_{\langle L \rangle}\;
\Bigg(r_A^{\rm N}\left(\tau+t^{\ast}\right) + \ve{\sigma} \cdot \ve{r}^{\rm N}_A\left(\tau+t^{\ast}\right)\, 
\ln \left[r^{\rm N}_A\left(\tau+t^{\ast}\right)-\ve{\sigma} \cdot \ve{r}^{\rm N}_A\left(\tau+t^{\ast}\right)\right]\Bigg)
\nonumber\\
\nonumber\\
&& \hspace{-3.0cm} + \frac{2\,G}{c^2}\,P^{ij}\,\frac{\partial}{\partial \xi^j}\,
\sum\limits_{l = 0}^{\infty} \frac{\left(-1\right)^l}{l!}\;M_{\langle L \rangle}^A\left(\tau_0+t^{\ast}\right)
\partial_{\langle L \rangle}\;
\Bigg(r_A^{\rm N}\left(\tau_0+t^{\ast}\right) + \ve{\sigma} \cdot \ve{r}^{\rm N}_A\left(\tau_0+t^{\ast}\right)\,
\ln \left[r^{\rm N}_A\left(\tau_0+t^{\ast}\right)-\ve{\sigma} \cdot \ve{r}^{\rm N}_A\left(\tau_0+t^{\ast}\right)\right]\Bigg)
\nonumber\\
\nonumber\\
&& \hspace{-3.0cm} + \frac{2\,G}{c^3}\,\sum\limits_{l = 1}^{\infty} \frac{\left(-1\right)^l}{l!}  
\int\limits_{\tau_0}^{\tau}\,d c\tau^{\prime}\, 
\dot{M}_{\langle L \rangle}^A\left(\tau^{\prime}+t^{\ast}\right)
\partial^{\prime}_{\langle L \rangle}\;
\frac{d_A^i\left(\tau^{\prime}+t^{\ast}\right)}{r^{\rm N}_A\left(\tau^{\prime}+t^{\ast}\right)-\ve{\sigma}\cdot\ve{r}^{\rm N}_A\left(\tau^{\prime}+t^{\ast}\right)} 
\nonumber\\
\nonumber\\
&& \hspace{-3.0cm} - \frac{2\,G}{c^3}\,P^{ij}\,\frac{\partial}{\partial \xi^j}\,
\sum\limits_{l = 0}^{\infty} \frac{\left(-1\right)^l}{l!} 
\int\limits_{\tau_0}^{\tau}\,d c\tau^{\prime}
\,M_{\langle L \rangle}^A\left(\tau^{\prime}+t^{\ast}\right)
\partial^{\prime}_{\langle L \rangle}\;
\ve{\sigma}\cdot\ve{v}_A\left(\tau^{\prime}+t^{\ast}\right)\, 
\ln \left[r^{\rm N}_A\left(\tau^{\prime}+t^{\ast}\right)-\ve{\sigma} \cdot \ve{r}^{\rm N}_A\left(\tau^{\prime}+t^{\ast}\right)\right]  
\nonumber\\
\nonumber\\
&& \hspace{-3.0cm} - \frac{2\,G}{c^3}\,P^{ij}\,\frac{\partial}{\partial \xi^j}\, 
\sum\limits_{l = 0}^{\infty} \frac{\left(-1\right)^l}{l!} 
\int\limits_{\tau_0}^{\tau}\,d c\tau^{\prime}\, 
M_{\langle L \rangle}^A\left(\tau^{\prime}+t^{\ast}\right)
\partial^{\prime}_{\langle L \rangle}\;
\frac{\ve{d}_A\left(\tau^{\prime}+t^{\ast}\right) \cdot \ve{v}_A\left(\tau^{\prime}+t^{\ast}\right)}
{r^{\rm N}_A\left(\tau^{\prime}+t^{\ast}\right)-\ve{\sigma} \cdot \ve{r}^{\rm N}_A\left(\tau^{\prime}+t^{\ast}\right)}\,.  
\label{Integral_A_10}
\end{eqnarray}
\end{widetext}

\noindent
In order to get the expression in (\ref{Integral_A_10}), we have performed an integration by parts 
which results in the expressions in the first and second line. Furthermore, for the  
expression in the third line we have used relation (\ref{Relation_3}). Consequently, (\ref{Integral_A_10}) represents 
an exact expression for the integral in (\ref{Integral_A_5}).  

Now we are going to proceed with the consideration of the remaining three integrals in the third, fourth, and fifth line in (\ref{Integral_A_10}).  
For the integral in the third and fifth line we will use relation (\ref{Integral_C_10_A}) and integrate by parts; note that we also need  
relation (\ref{Relation_1}) and the facts that 
\begin{eqnarray}
\frac{\partial}{\partial c \tau}\,\ve{d}_A\left(\tau+t^{\ast}\right) &=& 
\ve{\sigma} \times \left(\ve{\sigma} \times \frac{\ve{v}_A\left(\tau+t^{\ast}\right)}{c} \right),   
\nonumber\\
\label{time_derivative_impact_vector}
\\
\nonumber\\
\frac{\partial}{\partial c \tau}\,M_{\langle L \rangle}^A\left(\tau+t^{\ast}\right) &=& \frac{\dot{M}_{\langle L \rangle}^A\left(\tau+t^{\ast}\right)}{c}\,. 
\label{time_derivative_mass_multipoles}
\end{eqnarray}

\noindent
For the integral in the fourth line we use relation (\ref{Integral_A_6})  
and integrate by parts and afterwards we apply relation (\ref{Relation_3}). Altogether, we obtain:  
\begin{widetext}
\begin{eqnarray}
I_3\left(\tau+t^{\ast},\tau_0+t^{\ast}\right) &=& I_3\left(\tau+t^{\ast}\right) - I_3\left(\tau_0+t^{\ast}\right), \quad {\rm where}  
\nonumber\\
\nonumber\\
I_3\left(\tau+t^{\ast}\right) &=&
- \frac{2\,G}{c^2}\,
\sum\limits_{l = 0}^{\infty} \frac{\left(-1\right)^l}{l!}\;M_{\langle L \rangle}^A\left(\tau+t^{\ast}\right)
\partial_{\langle L \rangle}\;
\frac{d^i_A\left(\tau+t^{\ast}\right)}{r^{\rm N}_A\left(\tau+t^{\ast}\right)-\ve{\sigma}\cdot\ve{r}^{\rm N}_A\left(\tau+t^{\ast}\right)}
\nonumber\\
\nonumber\\
&& \hspace{-2.0cm} + \frac{G}{c^3}\,
\sum\limits_{l = 1}^{\infty} \frac{\left(-1\right)^l}{l!}\;\dot{M}_{\langle L \rangle}^A\left(\tau+t^{\ast}\right)
\partial_{\langle L \rangle}\;
d_A^i\left(\tau+t^{\ast}\right)\,
\frac{\ve{\sigma}\cdot\ve{r}^{\rm N}_A\left(\tau+t^{\ast}\right)}{r^{\rm N}_A\left(\tau+t^{\ast}\right)-\ve{\sigma}\cdot\ve{r}^{\rm N}_A\left(\tau+t^{\ast}\right)}
\nonumber\\
\nonumber\\
&& \hspace{-2.0cm} - \frac{G}{c^3}\,
\sum\limits_{l = 1}^{\infty} \frac{\left(-1\right)^l}{l!}\;\dot{M}_{\langle L \rangle}^A\left(\tau+t^{\ast}\right)
\partial_{\langle L \rangle}\;
d_A^i\left(\tau+t^{\ast}\right)\,
\ln \left[r^{\rm N}_A\left(\tau+t^{\ast}\right)-\ve{\sigma}\cdot\ve{r}^{\rm N}_A\left(\tau+t^{\ast}\right)\right] 
\nonumber\\
\nonumber\\
&& \hspace{-2.0cm} - \frac{2\,G}{c^3}\,
\sum\limits_{l = 0}^{\infty} \frac{\left(-1\right)^l}{l!}\;M_{\langle L \rangle}^A\left(\tau+t^{\ast}\right)
\partial_{\langle L \rangle}\;
\ve{\sigma}\cdot\ve{v}_A\left(\tau+t^{\ast}\right)\,
\frac{d_A^i\left(\tau+t^{\ast}\right)}{r^{\rm N}_A\left(\tau+t^{\ast}\right)-\ve{\sigma} \cdot \ve{r}^{\rm N}_A\left(\tau+t^{\ast}\right)}
\nonumber\\
\nonumber\\
&& \hspace{-2.0cm} - \frac{G}{c^3}\,P^{ij}\,\frac{\partial}{\partial \xi^j}\, 
\sum\limits_{l = 0}^{\infty} \frac{\left(-1\right)^l}{l!}\;M_{\langle L \rangle}^A\left(\tau+t^{\ast}\right)
\partial_{\langle L \rangle}\;
\ve{d}_A\left(\tau+t^{\ast}\right) \cdot \ve{v}_A\left(\tau+t^{\ast}\right)\, 
\frac{\ve{\sigma}\cdot\ve{r}^{\rm N}_A\left(\tau+t^{\ast}\right)}{r^{\rm N}_A\left(\tau+t^{\ast}\right)-\ve{\sigma} \cdot \ve{r}^{\rm N}_A\left(\tau+t^{\ast}\right)}
\nonumber\\
\nonumber\\
&& \hspace{-2.0cm} + \frac{G}{c^3}\,P^{ij}\,\frac{\partial}{\partial \xi^j}\,
\sum\limits_{l = 0}^{\infty} \frac{\left(-1\right)^l}{l!}\;M_{\langle L \rangle}^A\left(\tau+t^{\ast}\right)
\partial_{\langle L \rangle}\;
\ve{d}_A\left(\tau+t^{\ast}\right) \cdot \ve{v}_A\left(\tau+t^{\ast}\right)\,  
\ln \left[r^{\rm N}_A\left(\tau+t^{\ast}\right)-\ve{\sigma} \cdot \ve{r}^{\rm N}_A\left(\tau+t^{\ast}\right)\right] 
\nonumber\\
\nonumber\\
&& \hspace{-2.0cm} + {\cal O}\left(\frac{v_A}{c}\,\dot{M}_L^A\right) + {\cal O}\left(\ddot{M}_L^A\right) + {\cal O}\left(\frac{v_A^2}{c^2} M_L^A\right),  
\label{Integral_A_15}
\end{eqnarray}
\end{widetext}

\noindent
where in the first line we have used relation (\ref{Relation_3}).

\subsection{Integral $I_4$}\label{Integral_4}  

The integral $I_4$ reads 
\begin{eqnarray}
I_4\left(\tau+t^{\ast}, \tau_0+t^{\ast}\right) &=&
\nonumber\\
\nonumber\\
&& \hspace{-3.5cm} - \frac{2\,G}{c^2} \sigma^i \sum\limits_{l = 0}^{\infty} \frac{\left(-1\right)^l}{l!}  
\int\limits_{\tau_0}^{\tau}\,d c\tau^{\prime} 
M_{\langle L \rangle}^A\left(\tau^{\prime}+t^{\ast}\right)
\partial^{\prime}_{\langle L \rangle}\;\frac{1}{r^{\rm N}_A\left(\tau^{\prime}+t^{\ast}\right)}\,. 
\nonumber\\
\label{Integral_B_5}
\end{eqnarray}

\noindent
The evaluation of that integral goes very similar to the determination of the integral $I_1$ as given in appendix \ref{Integral_1}.  
Accordingly we obtain:
\begin{widetext}
\begin{eqnarray}
I_4\left(\tau+t^{\ast},\tau_0+t^{\ast}\right) &=& I_4\left(\tau+t^{\ast}\right) - I_4\left(\tau_0+t^{\ast}\right), \quad {\rm where} 
\nonumber\\
\nonumber\\
I_4\left(\tau+t^{\ast}\right) &=& + \frac{2\,G}{c^2}\,\sigma^i\,
\sum\limits_{l = 0}^{\infty} \frac{\left(-1\right)^l}{l!}\;
M_{\langle L \rangle}^A\left(\tau+t^{\ast}\right)
\partial_{\langle L \rangle}\;
\ln \left[r^{\rm N}_A\left(\tau+t^{\ast}\right) - \ve{\sigma}\cdot\ve{r}^{\rm N}_A\left(\tau+t^{\ast}\right)\right]
\nonumber\\
\nonumber\\
&& \hspace{-3.0cm} - \frac{2\,G}{c^3}\,\sigma^i\,
\sum\limits_{l = 1}^{\infty} \frac{\left(-1\right)^l}{l!}\;
\dot{M}_{\langle L \rangle}^A\left(\tau+t^{\ast}\right)
\partial_{\langle L \rangle}\;
\bigg[r^{\rm N}_A\left(\tau+t^{\ast}\right) + \ve{\sigma}\cdot \ve{r}^{\rm N}_A\left(\tau+t^{\ast}\right)
\ln \left[r^{\rm N}_A\left(\tau+t^{\ast}\right) - \ve{\sigma}\cdot\ve{r}^{\rm N}_A\left(\tau+t^{\ast}\right)\right]\bigg]
\nonumber\\
\nonumber\\
&& \hspace{-3.0cm} + \frac{2\,G}{c^3}\,\sigma^i\,
\sum\limits_{l = 0}^{\infty} \frac{\left(-1\right)^l}{l!}\;
M_{\langle L \rangle}^A\left(\tau+t^{\ast}\right)
\partial_{\langle L \rangle}\;\ve{\sigma}\cdot\ve{v}_A\left(\tau+t^{\ast}\right)\,
\ln \left[r^{\rm N}_A\left(\tau+t^{\ast}\right) - \ve{\sigma}\cdot\ve{r}^{\rm N}_A\left(\tau+t^{\ast}\right)\right]
\nonumber\\
\nonumber\\
&& \hspace{-3.0cm} + \frac{2\,G}{c^3}\,\sigma^i\,
\sum\limits_{l = 0}^{\infty} \frac{\left(-1\right)^l}{l!}\;
M_{\langle L \rangle}^A\left(\tau+t^{\ast}\right)
\partial_{\langle L \rangle}\,
\frac{\ve{v}_A\left(\tau+t^{\ast}\right)\cdot\ve{d}_A\left(\tau+t^{\ast}\right)}
{r^{\rm N}_A\left(\tau+t^{\ast}\right) - \ve{\sigma}\cdot \ve{r}^{\rm N}_A\left(\tau+t^{\ast}\right)}
+ {\cal O}\left(\frac{v_A}{c}\,\dot{M}_L^A\right) + {\cal O}\left(\ddot{M}_L^A\right) + {\cal O}\left(\frac{v_A^2}{c^2} M_L^A\right).   
\nonumber\\ 
\label{Integral_B_10}
\end{eqnarray}
\end{widetext}

\subsection{Integral $I_5$}\label{Integral_5}  

The integral $I_5$ reads 
\begin{eqnarray}
I_5\left(\tau+t^{\ast},\tau_0+t^{\ast}\right) &=& 
\nonumber\\
\nonumber\\
&& \hspace{-3.0cm} + \frac{2 G}{c^3} \sum\limits_{l = 1}^{\infty} \frac{\left(-1\right)^l}{l!}  
\int\limits_{\tau_0}^{\tau} d c\tau^{\prime} 
\,\dot{M}_{\langle L \rangle}^A\left(\tau^{\prime}+t^{\ast}\right)
\nonumber\\
\nonumber\\
&& \hspace{-3.0cm} \times\;\partial^{\prime}_{\langle L \rangle}
\frac{d_A^i\left(\tau^{\prime}+t^{\ast}\right)}{r^{\rm N}_A\left(\tau^{\prime}+t^{\ast}\right)-\ve{\sigma}\cdot \ve{r}^{\rm N}_A\left(\tau^{\prime}+t^{\ast}\right)}
\,. 
\label{Integral_C_5}
\end{eqnarray}

\noindent
In order to perform that integral we need relation (\ref{Integral_C_10_A}).  
Integration by parts and by inspection of relations (\ref{time_derivative_impact_vector}) and (\ref{time_derivative_mass_multipoles}) one obtains:  
\begin{widetext}
\begin{eqnarray}
I_5\left(\tau+t^{\ast},\tau_0+t^{\ast} \right) &=& I_5\left(\tau+t^{\ast}\right) - I_5\left(\tau_0+t^{\ast}\right), \quad {\rm where} 
\nonumber\\
\nonumber\\
I_5\left(\tau+t^{\ast}\right) &=& 
+ \frac{G}{c^3}\sum\limits_{l = 1}^{\infty} \frac{\left(-1\right)^l}{l!} \dot{M}_{\langle L \rangle}^A\left(\tau+t^{\ast}\right) 
\partial_{\langle L \rangle} \;d_A^i\left(\tau+t^{\ast}\right) 
\frac{\ve{\sigma}\cdot\ve{r}_A^{\rm N}\left(\tau+t^{\ast}\right)}{r^{\rm N}_A\left(\tau+t^{\ast}\right)-\ve{\sigma}\cdot \ve{r}^{\rm N}_A\left(\tau+t^{\ast}\right)} 
\nonumber\\
\nonumber\\
&& \hspace{-2.0cm} - \frac{G}{c^3}\sum\limits_{l = 1}^{\infty} \frac{\left(-1\right)^l}{l!} \dot{M}_{\langle L \rangle}^A\left(\tau+t^{\ast}\right) 
\partial_{\langle L \rangle} \;d_A^i\left(\tau+t^{\ast}\right) 
\ln \left[ r^{\rm N}_A\left(\tau+t^{\ast}\right)-\ve{\sigma}\cdot \ve{r}^{\rm N}_A\left(\tau+t^{\ast}\right)\right] 
+ {\cal O}\left(\frac{v_A}{c}\,\dot{M}_L^A\right) + {\cal O}\left(\ddot{M}_L^A\right).  
\nonumber\\
\label{Integral_C_15}
\end{eqnarray}
\end{widetext}

\subsection{Integral $I_6$}\label{Integral_6}  

The integral $I_6$, using relation (\ref{Relation_2}), reads 
\begin{eqnarray}
I_6\left(\tau+t^{\ast},\tau_0+t^{\ast}\right) &=&
\nonumber\\
\nonumber\\
&& \hspace{-3.0cm} + \frac{2 G}{c^3}\,P^{ij} \frac{\partial}{\partial \xi^j} 
\sum\limits_{l = 0}^{\infty} \frac{\left(-1\right)^l}{l!} 
\int\limits_{\tau_0}^{\tau}\,d c\tau^{\prime}\,  
M_{\langle L \rangle}^A\left(\tau^{\prime}+t^{\ast}\right)
\nonumber\\
\nonumber\\
&& \hspace{-3.0cm} \times\;\ve{\sigma} \cdot \ve{v}_A\left(\tau^{\prime}+t^{\ast}\right)
\partial^{\prime}_{\langle L \rangle}
\ln \left[r^{\rm N}_A\left(\tau^{\prime}+t^{\ast}\right) - \ve{\sigma} \cdot \ve{r}^{\rm N}_A\left(\tau^{\prime}+t^{\ast}\right)\right]. 
\nonumber\\
\label{Integral_D_5}
\end{eqnarray}

\noindent
Integration by parts using relation (\ref{Integral_A_6}), and recalling relation (\ref{time_derivative_mass_multipoles}), yields: 
\begin{widetext}
\begin{eqnarray}
I_6\left(\tau+t^{\ast},\tau_0+t^{\ast}\right) &=& I_6\left(\tau+t^{\ast}\right) - I_6 \left(\tau_0+t^{\ast}\right), \quad {\rm where} 
\nonumber\\
\nonumber\\
I_6\left(\tau+t^{\ast}\right) &=&
+ \frac{2 G}{c^3}
\sum\limits_{l = 0}^{\infty} \frac{\left(-1\right)^l}{l!} M_{\langle L \rangle}^A\left(\tau+t^{\ast}\right)
\ve{\sigma} \cdot \ve{v}_A\left(\tau+t^{\ast}\right)
\partial_{\langle L \rangle}
\frac{d_A^i\left(\tau+t^{\ast}\right)}{r_A^{\rm N}\left(\tau+t^{\ast}\right) - \ve{\sigma}\cdot \ve{r}_A^{\rm N}\left(\tau+t^{\ast}\right)}
\nonumber\\
\nonumber\\
&& + {\cal O}\left(\frac{v_A}{c}\,\dot{M}_L^A\right) + {\cal O}\left(\frac{v_A^2}{c^2} M_L^A\right). 
\label{Integral_D_10}
\end{eqnarray}
\end{widetext}

\noindent
In (\ref{Integral_D_10}) we have also used relation (\ref{Relation_3}).

\subsection{Integral $I_7$}\label{Integral_7}  

The integral $I_7$ reads 
\begin{eqnarray}
I_7\left(\tau+t^{\ast},\tau_0+t^{\ast}\right) &=&
\nonumber\\
\nonumber\\
&& \hspace{-3.0cm} - \frac{2 G}{c^3} P^{ij}\,\frac{\partial}{\partial \xi^j} 
\sum\limits_{l = 0}^{\infty} \frac{\left(-1\right)^l}{l!} 
\int\limits_{\tau_0}^{\tau} d c\tau^{\prime}\,  
M_{\langle L \rangle}^A\left(\tau^{\prime}+t^{\ast}\right)
\nonumber\\
\nonumber\\
&& \hspace{-3.0cm} \times\;\partial^{\prime}_{\langle L \rangle}
\frac{\ve{v}_A\left(\tau^{\prime}+t^{\ast}\right) \cdot \ve{d}_A\left(\tau^{\prime}+t^{\ast}\right)}
{r^{\rm N}_A\left(\tau^{\prime}+t^{\ast}\right) - \ve{\sigma} \cdot \ve{r}^{\rm N}_A\left(\tau^{\prime}+t^{\ast}\right)}\,. 
\label{Integral_E_5}
\end{eqnarray}

\noindent
Inserting relation (\ref{Integral_C_10_A}) and integration by parts, recalling relations (\ref{time_derivative_impact_vector}) 
and (\ref{time_derivative_mass_multipoles}), yields:  
\begin{widetext}
\begin{eqnarray}
I_7\left(\tau+t^{\ast},\tau_0+t^{\ast}\right) &=& I_7\left(\tau+t^{\ast}\right) - I_7\left(\tau_0+t^{\ast}\right), \quad {\rm where} 
\nonumber\\
\nonumber\\
I_7\left(\tau+t^{\ast}\right) &=&
- \frac{G}{c^3} P^{ij}\,\frac{\partial}{\partial \xi^j} 
\sum\limits_{l = 0}^{\infty} \frac{\left(-1\right)^l}{l!} M_{\langle L \rangle}^A\left(\tau+t^{\ast}\right)
\partial_{\langle L \rangle}
\ve{v}_A\left(\tau+t^{\ast}\right) \cdot \ve{d}_A\left(\tau+t^{\ast}\right)
\frac{\ve{\sigma}\cdot\ve{r}^{\rm N}_A\left(\tau+t^{\ast}\right)}{r^{\rm N}_A\left(\tau+t^{\ast}\right)-\ve{\sigma}\cdot\ve{r}^{\rm N}_A\left(\tau+t^{\ast}\right)}
\nonumber\\
\nonumber\\
&& \hspace{-3.0cm} + \frac{G}{c^3} P^{ij}\,\frac{\partial}{\partial \xi^j}   
\sum\limits_{l = 0}^{\infty} \frac{\left(-1\right)^l}{l!} M_{\langle L \rangle}^A\left(\tau+t^{\ast}\right)
\partial_{\langle L \rangle} \ve{v}_A\left(\tau+t^{\ast}\right) \cdot \ve{d}_A\left(\tau+t^{\ast}\right)  
\ln \left[r^{\rm N}_A\left(\tau+t^{\ast}\right)-\ve{\sigma}\cdot\ve{r}^{\rm N}_A\left(\tau+t^{\ast}\right)\right] 
\nonumber\\
\nonumber\\
&& \hspace{-3.0cm} + {\cal O}\left(\frac{v_A}{c}\,\dot{M}_L^A\right) + {\cal O}\left(\frac{v_A^2}{c^2} M_L^A\right).  
\label{Integral_E_10}
\end{eqnarray}
\end{widetext}

\subsection{Integral $I_8$}\label{Integral_8}  

The integral $I_8$ reads 
\begin{eqnarray}
I_8\left(\tau+t^{\ast},\tau_0+t^{\ast}\right) &=&
\nonumber\\
\nonumber\\
&& \hspace{-3.0cm} - \frac{4\,G}{c^3}\sum\limits_{l=1}^{\infty} \frac{\left(-1\right)^l}{l!} \int\limits_{\tau_0}^{\tau}\,d c\tau^{\prime}\, 
\dot{M}^A_{\langle i L-1 \rangle}\left(\tau^{\prime}+t^{\ast}\right)\,
\nonumber\\
\nonumber\\
&& \hspace{-3.0cm} \times \partial^{\prime}_{\langle L-1 \rangle} \frac{1}{r^{\rm N}_A\left(\tau^{\prime}+t^{\ast}\right)}\,. 
\label{Integral_F_5}
\end{eqnarray}

\noindent
Integration by parts using relation (\ref{Relation_1}), and recalling relation (\ref{time_derivative_mass_multipoles}), yields: 
\begin{widetext}
\begin{eqnarray}
I_8\left(\tau+t^{\ast},\tau_0+t^{\ast}\right) &=& I_8\left(\tau+t^{\ast}\right) - I_8\left(\tau_0+t^{\ast}\right), \quad {\rm where}  
\nonumber\\
\nonumber\\
I_8\left(\tau+t^{\ast}\right) &=&
+ \frac{4\,G}{c^3}
\sum\limits_{l=1}^{\infty} \frac{\left(-1\right)^l}{l!}\;\dot{M}^A_{\langle i L-1 \rangle}\left(\tau+t^{\ast}\right)
\partial_{\langle L-1 \rangle} 
\ln \left[r^{\rm N}_A\left(\tau + t^{\ast}\right) - \ve{\sigma}\cdot \ve{r}^{\rm N}_A\left(\tau+t^{\ast}\right)\right] 
+ {\cal O}\left(\ddot{M}_L^A\right).  
\label{Integral_F_10}
\end{eqnarray}
\end{widetext}

\subsection{Integral $I_9$}\label{Integral_9}   

The integral $I_9$ reads 
\begin{eqnarray}
I_9\left(\tau+t^{\ast},\tau_0+t^{\ast}\right) &=&
\nonumber\\
\nonumber\\
&& \hspace{-3.0cm} + \frac{4\,G}{c^3}\sum\limits_{l = 0}^{\infty} \frac{\left(-1\right)^l}{l!}  
\int\limits_{\tau_0}^{\tau}\,d c\tau^{\prime}\,v_A^i\left(\tau^{\prime}+t^{\ast}\right)\, 
M_{\langle L \rangle}^A\left(\tau^{\prime}+t^{\ast}\right)
\nonumber\\
\nonumber\\
&& \hspace{-3.0cm} \times\;\partial^{\prime}_{\langle L \rangle}\,
\frac{1}{r^{\rm N}_A\left(\tau^{\prime}+t^{\ast}\right)}\,. 
\label{Integral_G_5}
\end{eqnarray}

\noindent
Integration by parts using relation (\ref{Relation_1}), and recalling relation (\ref{time_derivative_mass_multipoles}) yields: 
\begin{widetext}
\begin{eqnarray}
I_9\left(\tau+t^{\ast},\tau_0+t^{\ast}\right) &=& I_9\left(\tau+t^{\ast}\right) - I_9\left(\tau_0+t^{\ast}\right) , \quad {\rm where} 
\nonumber\\
\nonumber\\
I_9\left(\tau+t^{\ast}\right) &=&
- \frac{4\,G}{c^3}\,v_A^i\left(\tau+t^{\ast}\right)\sum\limits_{l = 0}^{\infty} \frac{\left(-1\right)^l}{l!}
M_{\langle L \rangle}^A\left(\tau+t^{\ast}\right)\,\partial_{\langle L \rangle}\,
\ln \left[r^{\rm N}_A\left(\tau + t^{\ast}\right) - \ve{\sigma}\cdot \ve{r}^{\rm N}_A\left(\tau+t^{\ast}\right)\right] 
\nonumber\\
\nonumber\\
&& + {\cal O}\left(\frac{v_A}{c}\,\dot{M}_L^A\right) + {\cal O}\left(\frac{v_A^2}{c^2} M_L^A\right). 
\label{Integral_G_10}
\end{eqnarray}
\end{widetext}

\subsection{Integral $I_{10}$}\label{Integral_10}  

The integral $I_{10}$, using relation (\ref{Relation_2}), reads 
\begin{eqnarray}
I_{10}\left(\tau+t^{\ast},\tau_0+t^{\ast}\right) &=&
\nonumber\\
\nonumber\\
&& \hspace{-3.0cm} - \frac{4\,G}{c^3} \sigma^j P^{ik} \frac{\partial}{\partial \xi^k}\,
\sum\limits_{l=1}^{\infty} \frac{\left(-1\right)^l}{l!} 
\int\limits_{\tau_0}^{\tau} d c\tau^{\prime}\, 
\dot{M}^A_{\langle j L-1 \rangle}\left(\tau^{\prime}+t^{\ast}\right)
\nonumber\\
\nonumber\\
&& \hspace{-3.0cm} \times\;\partial^{\prime}_{\langle L-1 \rangle} 
\ln \left[r^{\rm N}_A\left(\tau^{\prime}+t^{\ast}\right) - \ve{\sigma} \cdot \ve{r}^{\rm N}_A\left(\tau^{\prime}+t^{\ast}\right)\right].  
\label{Integral_H_5}
\end{eqnarray}

\noindent
Integration by parts using relation (\ref{Integral_A_6}) and (\ref{time_derivative_mass_multipoles}) yields, 
with the aid of relation (\ref{Relation_3}):  
\begin{widetext}
\begin{eqnarray}
I_{10} \left(\tau+t^{\ast},\tau_0+t^{\ast}\right) &=& I_{10} \left(\tau+t^{\ast}\right) - I_{10}\left(\tau_0+t^{\ast}\right), \quad {\rm where} 
\nonumber\\
\nonumber\\
I_{10}\left(\tau+t^{\ast}\right) &=&
- \frac{4\,G}{c^3}\,\sigma^j\, 
\sum\limits_{l=1}^{\infty} \frac{\left(-1\right)^l}{l!}\;\dot{M}^A_{\langle j L-1 \rangle}\left(\tau+t^{\ast}\right)\,
\partial_{\langle L-1 \rangle}
\frac{d_A^i\left(\tau+t^{\ast}\right)}{r_A^{\rm N}\left(\tau+t^{\ast}\right) - \ve{\sigma}\cdot \ve{r}_A^{\rm N}\left(\tau+t^{\ast}\right)}
\nonumber\\
\nonumber\\
&& + {\cal O}\left(\ddot{M}_L^A\right) + {\cal O}\left(\frac{v_A}{c}\,\dot{M}_L^A\right). 
\label{Integral_H_10}
\end{eqnarray}
\end{widetext}

\newpage

\section{Light trajectory in the field of spin-multipoles at rest}\label{Appendix_Spin} 

\subsection{First integration}\label{Appendix_Spin1}  

The contribution of the spin-multipoles in the first integration of geodesic equation for the light trajectory in the field of arbitrarily moving bodies 
with time-dependent spin-multipoles is given by Eq.~(\ref{First_Integration_15}). In \cite{Kopeikin1997} 
the light trajectory has been determined in the field of motionless bodies located at the origin of coordinate system 
($\ve{x}_A=\ve{0}$) and with time-independent mass-multipoles and spin-multipoles. 
Accordingly, in order to compare our results with \cite{Kopeikin1997}, we have to consider the following limits in our solution: 
\begin{eqnarray}
S^A_{\langle L \rangle}\left(\tau+t^{\ast}\right) &\rightarrow& S^A_{\langle L \rangle}\,,
\label{Appendix_First_Integration_1} 
\\
\nonumber\\
\ve{d}_A\left(\tau+t^{\ast}\right) &\rightarrow& \ve{\xi}\,,
\label{Appendix_First_Integration_2}
\\
\nonumber\\
d_A\left(\tau+t^{\ast}\right) &\rightarrow& d = \left|\ve{\xi}\right|\,,
\label{Appendix_First_Integration_3}
\\
\nonumber\\
\ve{r}^{\rm N}_A\left(\tau+t^{\ast}\right) &\rightarrow& \ve{r} = \ve{\xi} + c\,\tau\,\ve{\sigma}\,, 
\label{Appendix_First_Integration_4}
\\
\nonumber\\
r^{\rm N}_A\left(\tau+t^{\ast}\right) &\rightarrow& r = \sqrt{d^2 + c^2 \tau^2}\,,  
\label{Appendix_First_Integration_5}
\end{eqnarray}

\noindent
where  
\begin{widetext}
\begin{eqnarray}
r^{\rm N}_A\left(\tau+t^{\ast}\right) = \sqrt{{\xi}^2 + c^2 \tau^2 + x_A^2\left(\tau+t^{\ast}\right) - 2\,c\,\tau\,\ve{\sigma}\cdot\ve{x}_A 
\left(\tau+t^{\ast}\right) - 2\,\ve{\xi}\cdot\ve{x}_A\left(\tau+t^{\ast}\right)}\,. 
\label{Appendix_First_Integration_6}
\end{eqnarray}
\end{widetext}

\noindent  
In these limits the expression in Eq.~(\ref{First_Integration_15}) simplifies to  
\begin{eqnarray}
\frac{\Delta \dot{x}^{i\;{\cal S}}_A \left(\tau\right)}{c} &=&
- \frac{4\,G}{c^3}\,\sum\limits_{l=1}^{\infty} \frac{\left(-1\right)^l\;l}{\left(l + 1 \right)!}\,
\epsilon_{iab}\,S^A_{\langle b L-1 \rangle}\;\partial_{\langle a L-1 \rangle}\,\frac{1}{r} 
\nonumber\\
\nonumber\\
&& \hspace{-2.0cm} - \frac{4\,G}{c^3}\,\sigma^j
\sum\limits_{l=1}^{\infty} \frac{\left(-1\right)^l\;l}{\left(l + 1 \right)!}\,
\epsilon_{jab}\,S^A_{\langle b L-1 \rangle}\;\partial_{\langle a L-1 \rangle}\,
\frac{\xi^i}{d^2}\,\left(1+ \frac{c\,\tau}{r}\right),  
\nonumber\\
\label{Appendix_First_Integration_10}
\end{eqnarray}

\noindent
up to terms ${\cal O}\left(c^{-4}\right)$. In (\ref{Appendix_First_Integration_10}) we have used 
$\ve{\sigma}\cdot\ve{r}=c\,\tau$ and $\displaystyle \frac{1}{r}\,\frac{1}{r-\ve{\sigma}\cdot\ve{r}} = \frac{1}{d^2}\left(1+\frac{c\,\tau}{r}\right)$.  
The derivative operator has been given by Eq.~(\ref{Appendix_Partial_Derivative_10}) 
and simplifies as follows:  
\begin{widetext}
\begin{eqnarray}
\partial_{\langle a L-1 \rangle} &=& \underset{a i_1 ... i_{l-1}}{\rm STF}\;   
P^{a b}\;P^{i_1\,j_1}\;...\;P^{i_{l-1}\,j_{l-1}}\;\frac{\partial}{\partial \xi^b}\;\frac{\partial}{\partial \xi^{j_1}}\;...\;
\frac{\partial}{\partial \xi^{j_{l-1}}}\;
\nonumber\\
\nonumber\\
&+& \underset{a i_1 ... i_{l-1}}{\rm STF}\,\sum\limits_{p=1}^{l} \frac{l!}{\left(l-p\right)!\;p!}\;
\sigma^{i_1}\,...\,\sigma^{i_p}\;P^{a b}\;P^{i_{p+1}\,j_{p+1}}\;...\;P^{i_{l-1}\,j_{l-1}}\; 
\frac{\partial}{\partial \xi^{b}}\;\frac{\partial}{\partial \xi^{j_{p+1}}}\;...\;
\frac{\partial}{\partial \xi^{j_{l-1}}}\;
\left(\frac{\partial}{\partial c\,\tau}\right)^{p}\,, 
\label{Appendix_First_Integration_15}
\end{eqnarray}
\end{widetext}

\noindent
because there is no dependence on variable $t^{\ast}$ any longer, and the expression in (\ref{Appendix_First_Integration_15}) has been 
subdivided into one summand $p=0$ and all other terms with $p \ge 1$. By inserting the expression (\ref{Appendix_First_Integration_15})  
into (\ref{Appendix_First_Integration_10}) we confirm an agreement with Eq.~(37) in \cite{Kopeikin1997}, up to an  
overall sign which has been clarified by private communication \cite{Communication1}.  
For such a comparison it may be useful to note the relations 
\begin{eqnarray}
\left(\frac{\partial}{\partial c \tau}\right)^p \frac{1}{r} &=& - \left(\frac{\partial}{\partial c \tau}\right)^{p-1} \frac{c\,\tau}{r^3}\,, 
\label{Appendix_First_Integration_20}
\\
\left(\frac{\partial}{\partial c \tau}\right)^p \left(1 + \frac{c\,\tau}{r} \right) &=& + \left(\frac{\partial}{\partial c \tau}\right)^{p-1} 
\frac{d^2}{r^3}\,,  
\label{Appendix_First_Integration_25}
\end{eqnarray}
 
\noindent
while $\ve{d}$ is here time-independent.

\subsection{Second integration}\label{Appendix_Spin2} 

The contribution of the spin-multipoles in the second integration of geodesic equation for the light trajectory in the field of arbitrarily moving bodies        
with time-dependent spin-multipoles is given by Eq.~(\ref{Second_Integration_Spin_5}). In \cite{Kopeikin1997}          
the light trajectory has been determined in the field of motionless bodies located at the origin of coordinate system
($\ve{x}_A=\ve{0}$) and with time-independent mass-multipoles and spin-multipoles. Accordingly, we consider the limits  
(\ref{Appendix_First_Integration_1}) - (\ref{Appendix_First_Integration_5}) in our solution (\ref{Second_Integration_Spin_5}) and  
obtain 
\begin{eqnarray}
\Delta \ve{x}^{{\cal S}}_A\left(\tau\,,\,\tau_0\right) &=& \Delta \ve{x}^{{\cal S}}_A\left(\tau\right) - \Delta \ve{x}^{{\cal S}}_A\left(\tau_0\right),  
\label{Appendix_Second_Integration_1}
\end{eqnarray}

\noindent
with 
\begin{eqnarray}
\Delta x^{i\,{\cal S}}_A\left(\tau\right) \!\!&=& \!\!  
\frac{4\,G}{c^3}\,\sum\limits_{l=1}^{\infty} \frac{\left(-1\right)^l\;l}{\left(l + 1 \right)!}
\epsilon_{iab} S^A_{\langle b L-1 \rangle} \partial_{\langle a L-1 \rangle}
\ln \left(r - c \tau\right) 
\nonumber\\
\nonumber\\
&& \hspace{-1.5cm} - \frac{4\,G}{c^3}\,\sigma^j
\sum\limits_{l=1}^{\infty} \frac{\left(-1\right)^l\;l}{\left(l + 1 \right)!}\,
\epsilon_{jab}\,S^A_{\langle b L-1 \rangle}\;\partial_{\langle a L-1 \rangle}\,
\frac{\xi^i}{d^2}\left(r+c\,\tau\right), 
\nonumber\\
\label{Appendix_Second_Integration_5}
\end{eqnarray}

\noindent
up to terms of the order ${\cal O}\left(c^{-4}\right)$, and the derivative operator is given by 
\begin{widetext}
\begin{eqnarray}
\partial_{\langle a L-1 \rangle} &=& \underset{a i_1 ... i_{l-1}}{\rm STF}\;
P^{a b}\;P^{i_1\,j_1}\;...\;P^{i_{l-1}\,j_{l-1}}\;\frac{\partial}{\partial \xi^b}\;\frac{\partial}{\partial \xi^{j_1}}\;...\;
\frac{\partial}{\partial \xi^{j_{l-1}}}\;
\nonumber\\
\nonumber\\
&+& \underset{a i_1 ... i_{l-1}}{\rm STF}\;l\;  
\sigma^{i_1}\;P^{a b}\;P^{i_2\,j_2}\;...\;P^{i_{l-1}\,j_{l-1}}\;
\frac{\partial}{\partial \xi^{b}}\;\frac{\partial}{\partial \xi^{j_{p+1}}}\;...\;
\frac{\partial}{\partial \xi^{j_{l-1}}}\;
\frac{\partial}{\partial c\,\tau}
\nonumber\\
\nonumber\\
&+& \underset{a i_1 ... i_{l-1}}{\rm STF}\,\sum\limits_{p=2}^{l} \frac{l!}{\left(l-p\right)!\;p!}\;
\sigma^{i_1}\,...\,\sigma^{i_p}\;P^{a b}\;P^{i_{p+1}\,j_{p+1}}\;...\;P^{i_{l-1}\,j_{l-1}}\;
\frac{\partial}{\partial \xi^{b}}\;\frac{\partial}{\partial \xi^{j_{p+1}}}\;...\;
\frac{\partial}{\partial \xi^{j_{l-1}}}\;
\left(\frac{\partial}{\partial c\,\tau}\right)^{p},
\label{Appendix_Second_Integration_10}
\end{eqnarray}
\end{widetext}

\noindent
where the expression has been subdivided into three pieces: one term $p=0$, one term $p=1$, and all other terms with $p \ge 2$. 
By inserting (\ref{Appendix_Second_Integration_10}) into (\ref{Appendix_Second_Integration_5}), we have found an agreement with 
Eq.~(38) in \cite{Kopeikin1997}, up to an overall sign which has been clarified by private communication \cite{Communication1}.  
For such comparison, it might be useful to recall  
$\displaystyle \ln \frac{r - c\tau}{r_0 - c\tau_0} = - \ln \frac{r + c\tau}{r_0 + c\tau_0}$ and to note the following relations:    
\begin{eqnarray}
\frac{\partial}{\partial c \tau}\, \ln \left(r - c\,\tau\right) &=& - \frac{1}{r}\,,
\label{Appendix_Second_Integration_15}
\\
\frac{\partial}{\partial c \tau}\, \left(r + c\,\tau\right) &=& 1 + \frac{c\,\tau}{r}\,, 
\label{Appendix_Second_Integration_20}
\end{eqnarray}

\noindent 
as well as 
\begin{eqnarray}
\left(\frac{\partial}{\partial c \tau}\right)^p \ln \left(r - c\,\tau\right) &=& \left(\frac{\partial}{\partial c \tau}\right)^{p-2}\frac{c\,\tau}{r^3}\,,
\label{Appendix_Second_Integration_25}
\\
\left(\frac{\partial}{\partial c \tau}\right)^p \left(r + c\,\tau\right) &=& \left(\frac{\partial}{\partial c \tau}\right)^{p-2}\frac{d^2}{r^3}\,.  
\label{Appendix_Second_Integration_30}
\end{eqnarray}

\noindent
We also note that time-independent terms cancel each other in view of relation (\ref{Appendix_Second_Integration_1}).

\section{Light propagation in the field of arbitrarily moving bodies in 1PM approximation}\label{Solution_Kopeikin_Mashhoon} 

In \cite{KopeikinMashhoon2002} the light trajectory in the field of $N$ bodies with spin-dipole in post-Minkowskian approximation 
has been determined. That solution is given by Eq.~(39) in \cite{KopeikinMashhoon2002} and reads: 
\begin{eqnarray}
\ve{x}_{\rm S} \left(\tau + t^{\ast} \right) &=& \ve{\xi} + c\,\tau\,\ve{\sigma} + \ve{\Xi} \left(\tau\right) - \ve{\Xi} \left(\tau_0\right),  
\label{Appendix_Spin_PM_5}
\end{eqnarray}

\noindent
where according to Eq.~(41) in \cite{KopeikinMashhoon2002}  
\begin{eqnarray}
\Xi^i \left(\tau\right) &=& + \frac{1}{2}\,\sigma_{\alpha}\,\sigma_{\beta}\,\hat{\partial}_i\,D_S^{\alpha \beta}\left(\tau\right) 
- \sigma_{\alpha}\,B_S^{\alpha i}\left(\tau\right) 
\nonumber\\
\nonumber\\
&& - \frac{1}{2}\,\sigma^i\,B_S^{00}\left(\tau\right) 
+ \frac{1}{2}\,\sigma^i\,\sigma_p\,\sigma_q\,B_S^{p q}\left(\tau\right),  
\label{Appendix_Spin_PM_10}
\end{eqnarray}

\noindent
with $\sigma_{\alpha} = \left(-1,\sigma_i\right)$ and $\sigma_i = \sigma^i$. The expressions for $B_S^{\alpha \beta}$ and $\hat{\partial}_i\,D_S^{\alpha \beta}$  
were given by Eqs.~(C16) and (C17) in \cite{KopeikinMashhoon2002}, respectively 
(note a missing factor $4$ in the last term in Eq.~(C17) in \cite{KopeikinMashhoon2002}). Inserting these expressions into (\ref{Appendix_Spin_PM_10}) yields  
\begin{eqnarray}
\Xi^i \left(\tau\right) &=& - \frac{2 G}{c^4}\,\frac{P_{ij}\,r^j}{r - \ve{v}\cdot\ve{r}/c}\,
\sigma_{\alpha}\,\sigma_{\beta}\,\frac{r_{\gamma}\,S^{\gamma \,(\,\alpha}\,u^{\beta\,)}}{\left(r - \ve{\sigma}\cdot\ve{r}\right)^2}
\nonumber\\
\nonumber\\
&& + \frac{2 G}{c^4}\,\frac{P_{ij}\,r^j}{1 - \ve{\sigma}\cdot\ve{v}/c}\,\sigma_{\alpha}\,\sigma_{\beta}\,
\frac{\sigma_{\gamma}\,S^{\gamma \,(\,\alpha}\,u^{\beta\,)}}{\left(r - \ve{\sigma}\cdot\ve{r}\right)^2}
\nonumber\\
\nonumber\\
&& + \frac{2 G}{c^4}\,\frac{P_{ij}\,v^j/c}{\left(1 - \ve{\sigma}\cdot\ve{v}/c\right)^2}\,\sigma_{\alpha}\,\sigma_{\beta}\,
\frac{\sigma_{\gamma}\,S^{\gamma \,(\,\alpha}\,u^{\beta\,)}}{r - \ve{\sigma}\cdot\ve{r}}
\nonumber\\
\nonumber\\
&& + \frac{2 G}{c^4}\,\frac{P_{ij}\,\sigma_{\alpha}\,\sigma_{\beta}\,
S^{j\,(\,\alpha}\,u^{\beta\,)}}{1 - \ve{\sigma}\cdot\ve{v}/c}\,\frac{1}{r - \ve{\sigma}\cdot\ve{r}}
\nonumber\\
\nonumber\\
&& - \frac{4 G}{c^4}\,\sigma_{\alpha}\,
\frac{r_{\gamma}\,S^{\gamma \,(\,\alpha}\,u^{i\,)}}{r - \ve{\sigma}\cdot\ve{r}}\,\frac{1}{r - \ve{v}\cdot\ve{r}/c}
\nonumber\\
\nonumber\\
&& + \frac{4 G}{c^4}\,\sigma_{\alpha}\,
\frac{\sigma_{\gamma}\,S^{\gamma \,(\,\alpha}\,u^{i\,)}}{r - \ve{\sigma}\cdot\ve{r}}\,\frac{1}{1 - \ve{\sigma}\cdot\ve{v}/c}
\nonumber\\
\nonumber\\
&& - \frac{2 G}{c^4}\,\sigma_i\,
\frac{r_{\gamma}\,S^{\gamma \,(\,0}\,u^{0\,)}}{r - \ve{\sigma}\cdot\ve{r}}\,\frac{1}{r - \ve{v}\cdot\ve{r}/c}
\nonumber\\
\nonumber\\
&& + \frac{2 G}{c^4}\,\sigma_i\,\frac{\sigma_{\gamma}\,S^{\gamma \,(\,0}\,u^{0\,)}}{r - \ve{\sigma}\cdot\ve{r}}\,\frac{1}{1 - \ve{\sigma}\cdot\ve{v}/c}
\nonumber\\
\nonumber\\
&& + \frac{2 G}{c^4}\,\sigma_i\,\sigma_p\,\sigma_q\,
\frac{r_{\gamma}\,S^{\gamma \,(\,p}\,u^{q\,)}}{r - \ve{\sigma}\cdot\ve{r}}\,\frac{1}{r - \ve{v}\cdot\ve{r}/c}
\nonumber\\
\nonumber\\
&& - \frac{2 G}{c^4}\,\sigma_i \sigma_p \sigma_q  
\frac{\sigma_{\gamma}\,S^{\gamma\,(\,p}\,u^{q\,)}}{r - \ve{\sigma}\cdot\ve{r}}\,\frac{1}{1 - \ve{\sigma}\cdot\ve{v}/c}\,, 
\label{Appendix_Spin_PM_15}
\end{eqnarray}

\noindent
where $S^{\gamma\,(\,\alpha}\,u^{\beta\,)} = \left(S^{\gamma\,\alpha}\,u^{\beta} + S^{\gamma\,\beta}\,u^{\alpha}\right)/\,2\;$ means the symmetrization with 
respect to the indices $\alpha$ and $\beta$.   
Thereby, $u^{\beta}=\gamma_v\,\left(c\,,\,\ve{v}\right)$ where $\gamma_v^{-1}=\sqrt{1-\ve{v}^2/c^2}$ is the Lorentz factor, and all time-dependent 
quantities depend on the retarded time-variable $\tau_{\rm ret}$, that means for the global spin-tensor  
$S^{\alpha \beta} = S^{\alpha \beta}\left(\tau_{\rm ret}\right)$, for the four-velocity $u^{\alpha} = u^{\alpha}\left(\tau_{\rm ret}\right)$ and for the  
three-velocity $\ve{v} = \ve{v}\left(\tau_{\rm ret}\right)$.   
Furthermore, $r_{\alpha}=\left(-r\,,\,\ve{r}\right)$  
with $r = \left|\ve{r}\right|$ and $\ve{r}$ being the vector pointing from the spatial position of the body at retarded time, $\ve{x}_A\left(t_{\rm ret}\right)$,  
toward the spatial position of the photon at global coordinate-time, $\ve{x}\left(t\right)$. That means, in (\ref{Appendix_Spin_PM_15}) we may replace 
the new variables $\ve{\xi}, \tau_{\rm ret}$ by the old variables $\ve{x}, t_{\rm ret}$ (see also text below Eq.~(\ref{arbitrarily_moving_spin_dipole_35})):  
\begin{eqnarray}
\ve{r} &\equiv& \ve{r}\left(t,t_{\rm ret}\right) = \ve{x}_0 + c\,\left(t-t_0\right)\,\ve{\sigma} - \ve{x}_A\left(t_{\rm ret}\right), 
\label{Appendix_Spin_PM_25}
\end{eqnarray}

\noindent
where the retarded time in terms of the old variables is given by Eq.~(\ref{Retarded_Time}); see also Eq.~(11) in \cite{KopeikinSchaefer1999} 
or Eqs.~(12) in \cite{KopeikinMashhoon2002}.  
The solution in (\ref{Appendix_Spin_PM_15}) is valid for the light trajectory in post-Minkowskian approximation 
in the gravitational field of an arbitrarily moving point-like body carrying a spin-dipole, while our result in Eq.~(\ref{arbitrarily_moving_spin_dipole_5}) 
and Eq.~(\ref{arbitrarily_moving_spin_dipole_35})  
is valid for the light trajectory in post-Newtonian approximation in the gravitational field of an arbitrarily moving extended body carrying a spin-dipole. 
In order to compare both results we have to expand all expressions in (\ref{Appendix_Spin_PM_15}) with respect to  
variable $v/c \ll 1$ and neglect all terms of the order ${\cal O}\left(c^{-4}\right)$, and afterwards we have to express the global spin-tensor $S^{\alpha \beta}$ 
in terms of the intrinsic spin-vector $\ve{S}$. Especially, we find 
\begin{eqnarray}
\ve{r}\left(t,t_{\rm ret}\right) &=& \ve{r}_A^{\rm N}\left(t\right) + {\cal O}\left(c^{-1}\right),  
\label{Appendix_Spin_PM_32}
\\
\nonumber\\
\ve{v}\left(t_{\rm ret}\right) &=& \ve{v}_A\left(t\right) + {\cal O}\left(c^{-1}\right),  
\label{Appendix_Spin_PM_33}
\\
\nonumber\\ 
S^{\alpha \beta}\left(t_{\rm ret}\right) &=& S^{\alpha \beta}\left(t\right) + {\cal O}\left(c^{-1}\right),  
\label{Appendix_Spin_PM_34}
\end{eqnarray}

\noindent
where in (\ref{Appendix_Spin_PM_32}) and (\ref{Appendix_Spin_PM_33}) we have attached an index $A$ in order to indicate  
that actually the body $A$ is meant here, while the spin-tensor in (\ref{Appendix_Spin_PM_34}) describes still the global spin, besides the fact that 
this spin-tensor originates from the intrinsic spin $\ve{S}_A$ of that single body.  
Let us consider one specific example by performing a series-expansion of the first term in (\ref{Appendix_Spin_PM_15}), for which we obtain: 
\begin{eqnarray}
&& - \frac{2 G}{c^4}\,\frac{P_{ij}\,r^j}{r - \ve{v}\cdot\ve{r}/c}\,
\sigma_{\alpha}\,\sigma_{\beta}\,\frac{r_{\gamma}\,S^{\gamma \,(\,\alpha}\,u^{\beta\,)}}{\left(r - \ve{\sigma}\cdot\ve{r}\right)^2}
\nonumber\\
\nonumber\\
&& = - \frac{2 G}{c^4}\,\frac{P_{ij}\,r^j}{r - \ve{v}\cdot\ve{r}/c}\,
\sigma_{\alpha}\,\sigma_{\beta}\,\frac{r_{\gamma}\,S^{\gamma \alpha}\,u^{\beta}}{\left(r - \ve{\sigma}\cdot\ve{r}\right)^2}
\nonumber\\
\nonumber\\
&& = + \frac{2 G}{c^3}\,\frac{d_A^i\left(t\right)}{r_A^{\rm N}\left(t\right)}\,
\sigma_{\alpha}\,\frac{r^{\rm N}_{\gamma}\left(t\right)\,S^{\gamma \alpha}\left(t\right)}
{\left(r_A^{\rm N}\left(t\right) - \ve{\sigma}\cdot\ve{r}_A^{\rm N}\left(t\right)\right)^2} + {\cal O}\left(c^{-1}\right),  
\nonumber\\
\label{Appendix_Spin_PM_36}
\end{eqnarray}

\noindent
where in the second line we have determined the symmetrization, while in the third line we have used (\ref{Appendix_Spin_PM_32}) - (\ref{Appendix_Spin_PM_34}) 
and $P_{ij}\,r^j = d_A^i\left(t\right)$. Very similar steps for the other terms in (\ref{Appendix_Spin_PM_15}) yield the following 
expression:  
\begin{eqnarray}
\Xi^i \left(t\right) &=& + \frac{2 G}{c^3}\,\frac{d_A^i\left(t\right)}{r_A^{\rm N}\left(t\right)}\,
\sigma_{\alpha}\,\frac{r^{\rm N}_{\gamma}\left(t\right)\,S^{\gamma \alpha}\left(t\right)}
{\left(r_A^{\rm N}\left(t\right) - \ve{\sigma}\cdot\ve{r}_A^{\rm N}\left(t\right)\right)^2}
\nonumber\\
\nonumber\\
&& - \frac{2 G}{c^3}\,P_{ij}\,\sigma_{\alpha}\,
\frac{S^{j \alpha}\left(t\right)}{r_A^{\rm N}\left(t\right) - \ve{\sigma}\cdot\ve{r}_A^{\rm N}\left(t\right)}
\nonumber\\
\nonumber\\
&& + \frac{2 G}{c^3}\,
\frac{r^{\rm N}_{\gamma}\left(t\right)\,S^{\gamma i}\left(t\right)}
{r_A^{\rm N}\left(t\right)-\ve{\sigma}\cdot\ve{r}_A^{\rm N}\left(t\right)}\,\frac{1}{r_A^{\rm N}\left(t\right)}
\nonumber\\
\nonumber\\
&& - \frac{2 G}{c^3}\,
\frac{\sigma_{\gamma}\,S^{\gamma i}\left(t\right)}{r_A^{\rm N}\left(t\right) - \ve{\sigma}\cdot\ve{r}_A^{\rm N}\left(t\right)}
\nonumber\\
\nonumber\\
&& - \frac{2 G}{c^3}\,\sigma_i\,
\frac{r^{\rm N}_{\gamma}\left(t\right)\,S^{\gamma 0}\left(t\right)}
{r_A^{\rm N}\left(t\right)-\ve{\sigma}\cdot\ve{r}_A^{\rm N}\left(t\right)}\,\frac{1}{r_A^{\rm N}\left(t\right)}
\nonumber\\
\nonumber\\
&& + \frac{2 G}{c^3}\,\sigma_i\,\frac{\sigma_{\gamma}\,S^{\gamma 0}\left(t\right)}
{r_A^{\rm N}\left(t\right) - \ve{\sigma}\cdot\ve{r}_A^{\rm N}\left(t\right)}
+ {\cal O}\left(c^{-4}\right),  
\nonumber\\ 
\label{Appendix_Spin_PM_40}
\end{eqnarray}

\noindent
where we have used $\sigma_{\alpha}\,\sigma_{\gamma}\,S^{\gamma \alpha} = 0$ because the spin-tensor is anti-symmetric,  
$S^{\gamma \alpha} = - S^{\alpha \gamma}$, and we have introduced 
$r^{\rm N}_{\gamma}\left(t\right) = \left(- r_A^{\rm N}\left(t\right), \ve{r}_A^{\rm N}\left(t\right)\right)$. 
In order to compare (\ref{Appendix_Spin_PM_40}) with our result in (\ref{arbitrarily_moving_spin_dipole_35}) we have to express  
the global spin-tensor $S^{\alpha \beta}$ in (\ref{Appendix_Spin_PM_40}) in terms of the intrinsic spin-vector $\ve{S}_A$, where the index  
refers to body $A$. Recalling relations (24) and (C.10) in \cite{Zschocke_Soffel} we have  
\begin{eqnarray}
S^{i 0} \left(t\right) &=& {\cal O}\left(c^{-1}\right),  
\label{Appendix_Spin_PM_41}
\\
\nonumber\\
S^{ij} \left(t\right) &=& \epsilon_{ijk}\,S_A^k\left(t\right) + {\cal O}\left(c^{-1}\right).  
\label{Appendix_Spin_PM_42}
\end{eqnarray}
 
\noindent
By inserting (\ref{Appendix_Spin_PM_41}) - (\ref{Appendix_Spin_PM_42}) into (\ref{Appendix_Spin_PM_40}) we arrive at 
\begin{eqnarray}
\Xi^i\left(t\right) &=& + \frac{2 G}{c^3}\,\frac{d_A^i\left(t\right)}{r_A^{\rm N}\left(t\right)}\,
\sigma_{j}\,\frac{r^{\rm N}_{k}\left(t\right)\,\epsilon_{k j l}\,S_A^{l}\left(t\right)}
{\left(r_A^{\rm N}\left(t\right) - \ve{\sigma}\cdot\ve{r}_A^{\rm N}\left(t\right)\right)^2}
\nonumber\\
\nonumber\\
&& \hspace{-1.5cm} - \frac{2 G}{c^3}\,\sigma_{k}\,\epsilon_{i k l}\, 
\frac{S_A^{l}\left(t\right)}{r_A^{\rm N}\left(t\right) - \ve{\sigma}\cdot\ve{r}_A^{\rm N}\left(t\right)}
\nonumber\\
\nonumber\\
&& \hspace{-1.5cm} + \frac{2 G}{c^3}\,
\frac{r^{\rm N}_{j}\left(t\right)\,\epsilon_{j i l}\,S_A^{l}\left(t\right)}
{r_A^{\rm N}\left(t\right)-\ve{\sigma}\cdot\ve{r}_A^{\rm N}\left(t\right)}\,\frac{1}{r_A^{\rm N}\left(t\right)}
\nonumber\\
\nonumber\\
&& \hspace{-1.5cm} - \frac{2 G}{c^3}\,\sigma_j\,\epsilon_{j i l}\,\frac{S_A^{l}\left(t\right)}
{r_A^{\rm N}\left(t\right) - \ve{\sigma}\cdot\ve{r}_A^{\rm N}\left(t\right)}
+ {\cal O}\left(c^{-4}\right). 
\label{Appendix_Spin_PM_50}
\end{eqnarray}

\noindent
Finally, using $\ve{r}_A^{\rm N}\left(t\right) = \ve{d}_A\left(t\right) + \ve{\sigma} \left(\ve{\sigma}\cdot\ve{r}_A^{\rm N}\left(t\right)\right)$ we obtain  
\begin{eqnarray}
\ve{\Xi}\left(t\right) &=&
- \frac{2\,G}{c^3}\,
\frac{\ve{\sigma} \cdot \left(\ve{d}_A\left(t\right) \times \ve{S}_A\left(t\right)\right)}
{\left(r_{\rm A}^{\rm N}\left(t\right) - \ve{\sigma}\cdot\ve{r}_{\rm A}^{\rm N}\left(t\right)\right)^2}\,
\frac{\ve{d}_A\left(t\right)}{r_{\rm A}^{\rm N}\left(t\right)}
\nonumber\\
\nonumber\\
&& - \frac{2\,G}{c^3}\,
\frac{\ve{\sigma} \times \ve{S}_A\left(t\right)}
{r_{\rm A}^{\rm N}\left(t\right) - \ve{\sigma}\cdot \ve{r}_{\rm A}^{\rm N}\left(t\right)}
\nonumber\\
\nonumber\\
&& + \frac{2\,G}{c^3}\,
\,\frac{\ve{S}_A\left(t\right) \times \ve{d}_A\left(t\right)}
{r^{\rm N}_A\left(t\right) - \ve{\sigma}\cdot \ve{r}^{\rm N}_A\left(t\right)}\;
\frac{1}{r^{\rm N}_A\left(t\right)}
\nonumber\\
\nonumber\\
&& + \frac{2\,G}{c^3}\,\frac{\ve{\sigma} \times \ve{S}_A\left(t\right)}{r^{\rm N}_A\left(t\right)}\,,  
\label{Appendix_Spin_PM_55}
\end{eqnarray}

\noindent
where the last two terms in (\ref{Appendix_Spin_PM_55}) comprise the last two terms in (\ref{Appendix_Spin_PM_50}). The expression in (\ref{Appendix_Spin_PM_55})  
agrees with our result in (\ref{arbitrarily_moving_spin_dipole_35}); note that all derivatives according to (\ref{arbitrarily_moving_spin_dipole_15}) have been   
performed, hence the replacement $\tau+t^{\ast} \rightarrow t$ in (\ref{arbitrarily_moving_spin_dipole_35}) is possible.

\newpage

\section{Light trajectory in post-post-Newtonian approximation for monopoles at rest}\label{Appendix_2PN}  

In this appendix we briefly summarize the 2PN solution for the lightray (in harmonic gauge) in the field of one monopole at rest,  
located at $\ve{x}_A = {\rm const}$.  
The 2PN metric for one monopole at rest reads \cite{Brumberg1987,Brumberg1991,Zschocke_Klioner,Deng_Xie}:  
\begin{eqnarray}
h^{(2)}_{00}\left(\ve{x}\right) &=& + \frac{2\,m_A}{r_A}\,,
\label{body_at_rest_Metric_1}
\\
\nonumber\\
h^{(2)}_{ij}\left(\ve{x}\right) &=& + \frac{2\,m_A}{r_A}\,\delta_{ij}\,,
\label{body_at_rest_Metric_2}
\\
\nonumber\\
h^{(4)}_{00}\left(\ve{x}\right) &=& - \frac{2\,m_A^2}{r^2_A}\,,
\label{body_at_rest_Metric_3}
\\
\nonumber\\
h^{(4)}_{ij}\left(\ve{x}\right) &=& + \frac{m_A^2}{r^2_A}\,\delta_{ij}
+ \frac{m_A^2}{r^4_A}\,r_A^i\,r_A^j\,,
\label{body_at_rest_Metric_4}
\end{eqnarray}

\noindent
where $m_A=G M_A/c^2$ is the Schwarzschild radius of body $A$ and $\ve{r}_A = \ve{x} - \ve{x}_A$.  
Using the constraint for lightrays, $ds=0$, the geodesic equation can be written in the following form \cite{Brumberg1987,Brumberg1991,Klioner_Zschocke,Deng_Xie}:  
\begin{eqnarray}
\ddot{\ve{x}}_{\rm 2PN} &=& -\,2\,m_A\,c^2 \frac{\ve{r}_A}{r_A^3} + 4\,m_A\,\dot{\ve{x}}
\frac{\ve{r}_A \cdot \dot{\ve{x}}}{r_A^3}
\nonumber\\
\nonumber\\
&& \hspace{-1.5cm} - 2\,m_A^2\,\dot{\ve{x}}\,\frac{\ve{r}_A \cdot \dot{\ve{x}}}{r_A^4} +\,8 m_A^2 c^2 \frac{\ve{r}_A}{r_A^4}
+ 2\,m_A^2 \ve{r}_A \frac{\left(\ve{r}_A \cdot \dot{\ve{x}}\right)^2}{r_A^6}\,.
\nonumber\\
\label{body_at_rest_geodesic_equation_4}
\end{eqnarray}

\noindent
The solution of geodesic equation (\ref{body_at_rest_geodesic_equation_4}) has been found at the first time in \cite{Brumberg1987,Brumberg1991}.  
This solution has been confirmed within several investigations, e.g. \cite{KlionerKopeikin1992,Klioner_Zschocke,Deng_Xie} and has also been recalculated 
in this work.

\subsection{Light trajectory in Newtonian approximation for monopole at rest}

The light trajectory in Newtonian approximation (N) reads: 
\begin{eqnarray}
\ve{x}_{\rm N}\left(t\right) &=& \ve{x}_0 + c\left(t - t_0\right)\ve{\sigma}\,. 
\end{eqnarray}

\subsection{Light trajectory in 1PN approximation for monopole at rest}

The light trajectory in post-Newtonian approximation (1PN) reads: 
\begin{eqnarray}
\ve{x}_{\rm 1PN}\left(t\right) &=& \ve{x}_0 + c\left(t - t_0\right)\ve{\sigma}  
\nonumber\\
\nonumber\\
&& - 2 m_A\,\left(\frac{\ve{d}_A}{r^{\rm N}_A-\ve{\sigma}\cdot\ve{r}^{\rm N}_A}-\frac{\ve{d}_A}{r^0_A-\ve{\sigma}\cdot\ve{r}^0_A}\right)  
\nonumber\\
\nonumber\\
&& + 2 m_A\,\ve{\sigma}\,\ln \frac{r^{\rm N}_A - \ve{\sigma} \cdot \ve{r}^{\rm N}_A}{r^0_A - \ve{\sigma} \cdot \ve{r}^0_A}\,,  
\label{1PN_Solution_A} 
\end{eqnarray}

\noindent
where $\ve{r}_A^{\rm N} = \ve{x}_{\rm N}\left(t\right)  - \ve{x}_A$ and $r^0_A = \ve{x}\left(t_0\right)  - \ve{x}_A$. 
The expression in (\ref{1PN_Solution_A}) corrects some typos in Eq.~(B6) in \cite{Zschocke_1PN}.  

\subsection{Light trajectory in 2PN approximation for monopole at rest}

The light trajectory in post-post-Newtonian approximation (2PN) reads: 
\begin{eqnarray}
\ve{x}_{\rm 2PN}\left(t\right) &=& \ve{x}_0 + c \left(t-t_0\right) \ve{\sigma} 
\nonumber\\
\nonumber\\
&& + m_A\left[\ve{B}_1\left(\ve{r}_A^{\rm 1PN}\right) - \ve{B}_1\left(\ve{r}^0_A\right) \right]
\nonumber\\
\nonumber\\
&& + m^2_A \left[\ve{B}_2\left(\ve{r}_A^{\rm N}\right) - \ve{B}_2\left(\ve{r}^0_A\right) \right],
\nonumber\\
\label{2PN_Solution_Brumberg_2}
\end{eqnarray}

\noindent
where $\ve{r}_A^{\rm 1PN} = \ve{x}_{\rm 1PN}\left(t\right) - \ve{x}_A$. The vectorial functions $\ve{B}_1$ and $\ve{B}_2$ are given by  
(cf. Eqs.~(3.2.41) and (3.2.42) in \cite{Brumberg1991} or Eqs.~(50) and (51) in \cite{Klioner_Zschocke}):
\begin{widetext}
\begin{eqnarray}
\ve{B}_1(\ve{r}_A^{\rm 1PN}) &=&
- 2 \, \frac{\ve{\sigma}\times(\ve{r}_A^{\rm 1PN} \times \ve{\sigma})}{r_A^{\rm 1PN} - \ve{\sigma}\cdot\ve{r}_A^{\rm 1PN}}
+2\, \ve{\sigma}\,\ln \left(r_A^{\rm 1PN}-\ve{\sigma}\cdot\ve{r}_A^{\rm 1PN}\right)\,,
\label{Vectorial_Function_B1}
\\
\nonumber\\
\ve{B}_2(\ve{r}_A^{\rm N}) &=&
+ 4\,\frac{\ve{\sigma}}{r_A^{\rm N} - \ve{\sigma}\cdot \ve{r}_A^{\rm N}}
+ 4\,\frac{\ve{d}_A}{\left(r_A^{\rm N} - \ve{\sigma}\cdot \ve{r}_A^{\rm N}\right)^2}
+ \frac{1}{4}\,\frac{\ve{r}_A^{\rm N}}{\left(r_A^{\rm N}\right)^2}
- \frac{15}{4}\,\frac{\ve{\sigma}}{d_A}\,\arctan \left(\frac{\ve{\sigma}\cdot\ve{r}_A^{\rm N}}{d_A}\right)
\nonumber\\
&& - \frac{15}{4}\,\ve{d}_A\,\frac{\ve{\sigma} \cdot \ve{r}_A^{\rm N}}{d_A^3}\,
\left[\frac{\pi}{2} + \arctan \left(\frac{\ve{\sigma}\cdot\ve{r}_A^{\rm N}}{d_A}\right)\right]. 
\label{Vectorial_Function_B2}
\end{eqnarray}
\end{widetext}

\noindent
It should be mentioned that in $\ve{B}_1$ the coordinate of the photon in 1PN approximation, $\ve{x}_{\rm 1PN}$, can be replaced by the exact  
coordinate $\ve{x}$ of the photon, and in $\ve{B}_2$ the coordinate of the photon in Newtonian approximation, $\ve{x}_{\rm N}$, can be replaced 
by the exact coordinate $\ve{x}$, because such replacements are correct up to terms of the order ${\cal O}\left(c^{-6}\right)$.

Sometimes it is useful to perform a series-expansion of the vectorial function $\ve{B}_1$ in terms of the small parameter $m_A$ and to express 
the 2PN solution (\ref{2PN_Solution_Brumberg_2}) in terms of unperturbed lightray as follows:  
\begin{widetext}
\begin{eqnarray}
\ve{x}_{\rm 2PN}\left(t\right) &=& \ve{x}_0 + c \left(t-t_0\right) \ve{\sigma}   
\nonumber\\
\nonumber\\
&& \hspace{-1.5cm} - 2\,m_A\left(\frac{\ve{d}_A}{r_A^{\rm N} - \ve{\sigma} \cdot \ve{r}_A^{\rm N}} - \frac{\ve{d}_A}{r_A^0 - \ve{\sigma}\cdot\ve{r}_A^0}\right)
+ 2\,m_A\,\ve{\sigma}\,\ln \frac{r_A^{\rm N} - \ve{\sigma} \cdot \ve{r}_A^{\rm N}}{r_A^0 - \ve{\sigma}\cdot\ve{r}_A^0}
\nonumber\\
\nonumber\\
&& \hspace{-1.5cm} - \frac{15}{4}\,\frac{m_A^2}{d_A^3}\,\ve{d}_A \left[
\left(\ve{\sigma} \cdot \ve{r}_A^{\rm N}\right)\,\left(\frac{\pi}{2} + \arctan \frac{\ve{\sigma} \cdot \ve{r}_A^{\rm N}}{d_A} \right)
- \left(\ve{\sigma} \cdot \ve{r}_A^0\right)\,\left(\frac{\pi}{2} + \arctan \frac{\ve{\sigma} \cdot \ve{r}_A^0}{d_A} \right)\right]
\nonumber\\
\nonumber\\
&& \hspace{-1.5cm}
- \frac{15}{4}\,\frac{m_A^2}{d_A}\,\ve{\sigma}\left[\arctan \frac{\ve{\sigma}\cdot\ve{r}_A^{\rm N}}{d_A} - \arctan\frac{\ve{\sigma}\cdot\ve{r}_A^0}{d_A}\right]
- 4\,\frac{m_A^2}{r_A^{\rm N}}\,\frac{\ve{d}_A}{r_A^{\rm N} - \ve{\sigma} \cdot \ve{r}_A^{\rm N}}\,
\ln \frac{r_A^{\rm N} - \ve{\sigma} \cdot \ve{r}_A^{\rm N}}{r_A^0 - \ve{\sigma}\cdot\ve{r}_A^0}
- 4\,\frac{m_A^2}{r_A^{\rm N}}\,\ve{\sigma}\,\ln \frac{r_A^{\rm N} - \ve{\sigma} \cdot \ve{r}_A^{\rm N}}{r_A^0 - \ve{\sigma}\cdot\ve{r}_A^0}
\nonumber\\
\nonumber\\
&& \hspace{-1.5cm} + \frac{1}{4}\,m_A^2\,\frac{\ve{d}_A}{\left(r_A^{\rm N}\right)^2} - \frac{1}{4}\,m_A^2\,\frac{\ve{d}_A}{\left(r_A^0\right)^2}
+ \frac{1}{4}\,m_A^2\,\ve{\sigma}\,\frac{\ve{\sigma}\cdot\ve{r}_A^{\rm N}}{\left(r_A^{\rm N}\right)^2}
- \frac{1}{4}\,m_A^2\,\ve{\sigma}\,\frac{\ve{\sigma}\cdot\ve{r}_A^0}{\left(r_A^0\right)^2}
- 4\,m_A^2\,\frac{\ve{\sigma}\cdot\ve{r}_A^{\rm N}}{r_A^{\rm N}}\,\ve{\sigma}
\left(\frac{1}{r_A^{\rm N} - \ve{\sigma}\cdot\ve{r}_A^{\rm N}} - \frac{1}{r_A^0 - \ve{\sigma}\cdot\ve{r}_A^0}\right)
\nonumber\\
\nonumber\\
&& \hspace{-1.5cm} + 4\,m_A^2\,\ve{d}_A \left(\frac{1}{r_A^{\rm N}} + \frac{1}{r_A^0 - \ve{\sigma}\cdot\ve{r}_A^0} \right)
\left(\frac{1}{r_A^{\rm N} - \ve{\sigma}\cdot\ve{r}_A^{\rm N}} - \frac{1}{r_A^0 - \ve{\sigma}\cdot\ve{r}_A^0} \right),
\label{2PN_Solution_A} 
\end{eqnarray}
\end{widetext}

\noindent
where we recall that $\ve{r}_A^{\rm N} = \ve{d}_A + \ve{\sigma}\left(\ve{\sigma}\cdot\ve{r}_A^{\rm N}\right)$ and
$\ve{r}_A^0 = \ve{d}_A + \ve{\sigma}\left(\ve{\sigma}\cdot\ve{r}_A^0\right)$.
The expression in (\ref{2PN_Solution_A}) corrects some typos in Eq.~(37) in \cite{Zschocke_1PN}.  
Let us notice that in the second line in (\ref{2PN_Solution_A}) it is not allowed to replace the coordinate of the photon in 
Newtonian approximation, $\ve{x}_{\rm N}$, by the exact coordinate of the photon, $\ve{x}$, because such a relacement would cause  
an error of the order ${\cal O}\left(c^{-4}\right)$. This is the reason for the fact that the form in (\ref{2PN_Solution_A}) is usually not 
in use in favor of the expression in Eqs.~(\ref{2PN_Solution_Brumberg_2}) - (\ref{Vectorial_Function_B2}).


\end{document}